\documentclass[preprint,aps,preprintnumbers,nofootinbib,superscriptaddress,showpacs]{myrevtex4}
\usepackage{amsmath,amssymb,amsbsy}
\usepackage{graphicx,multirow}
\usepackage{color}

\usepackage{ulem}
\preprint{CU-TP-1192, Edinburgh 2010/1}
\preprint{RBRC--827, SHEP 0928}

                               {\end{figure}}

\newenvironment{CTable}[1][tbp]{\begin{table}[#1]\centering}%
                               {\end{table}}

\def\CO{{\cal{O}}}

\def\calC{{\cal{C}}}
\def\cM{{\cal{M}}}

\def\bar{\overline}
\def\hat{\widehat}
\def\tilde{\widetilde}

\def\bea{\begin{eqnarray}}
\def\eea{\end{eqnarray}}
\def\beq{\begin{equation}}
\def\eeq{\end{equation}}

\def\spose#1{\hbox to 0pt{#1\hss}}
\def\ltapprox{\mathrel{\spose{\lower 3pt\hbox{$\mathchar"218$}}
\raise 2.0pt\hbox{$\mathchar"13C$}}}
\def\gtapprox{\mathrel{\spose{\lower 3pt\hbox{$\mathchar"218$}}
\raise 2.0pt\hbox{$\mathchar"13E$}}}
\def\inapprox{\mathrel{\spose{\lower 3pt\hbox{$\mathchar"218$}}
\raise 2.0pt\hbox{$\mathchar"232$}}}

\begin{document}

%\vphantom{}

\title{Neutral $B$-meson mixing from unquenched lattice QCD with domain-wall light quarks and static $b$-quarks}

\newcommand\riken{RIKEN-BNL Research Center, Brookhaven National
Laboratory, Upton, NY 11973, USA}
\newcommand\bnlaf{Physics Department, Brookhaven
National Laboratory, Upton, NY 11973, USA}
\newcommand\edinb{SUPA,
School of Physics, The University of Edinburgh, Edinburgh EH9 3JZ, UK}
\newcommand\epcca{EPCC, School of Physics, The University of
Edinburgh, Edinburgh EH9 3JZ, UK}
\newcommand\cuaff{Physics Department, Columbia University, New York, NY 10027, USA}
\newcommand\soton{School of Physics and Astronomy, University of
Southampton, Southampton SO17 1BJ, UK}
\newcommand\uconn{Physics Department, University of Connecticut, Storrs, CT 06269-3046, USA}
\newcommand\salamanca{Departamento de F\'isica Fundamental, Facultad de Ciencias, Universidad de Salamanca, Plaza de la Merced s/n, 37008, Salamanca, Spain}
\newcommand\valencia{Instituto de F\'isica Corpuscular, CSIC-Universidad de Valencia, P.O. Box 22085, 46071 Valencia, Spain}
\newcommand\princeton{Department of Physics, Princeton University, Princeton, NJ 08544, USA}

\author{C.~Albertus\footnote{Present address: \salamanca}}\affiliation\soton
\author{Y.~Aoki}\affiliation\riken
\author{P.~A.~Boyle}\affiliation\edinb
\author{N.~H.~Christ}\affiliation\cuaff
\author{T.~T.~Dumitrescu\footnote{Present address: \princeton}}\affiliation\cuaff
\author{J.~M.~Flynn}\affiliation\soton
\author{T.~Ishikawa\footnote{Present address: \uconn}}\affiliation\riken
\author{T.~Izubuchi}\affiliation\riken\affiliation\bnlaf
\author{O.~Loktik}\affiliation\cuaff
\author{C.~T.~Sachrajda}\affiliation\soton
\author{A.~Soni}\affiliation\bnlaf
\author{R.~S.~Van~de~Water}\affiliation\bnlaf
\author{J.~Wennekers\footnote{Also: \valencia}}\affiliation\edinb
\author{O.~Witzel}\affiliation\bnlaf
\collaboration{RBC and UKQCD Collaborations}\noaffiliation

\date{\today}

%=================================================
%The abstract
%=================================================
\begin{abstract}

We demonstrate a method for calculating the neutral $B$-meson decay constants and mixing matrix elements in unquenched lattice QCD with domain-wall light quarks and static $b$-quarks.  Our computation is performed on the ``2+1" flavor gauge configurations generated by the RBC and UKQCD Collaborations with a lattice spacing of $a \approx 0.11$~fm ($a^{-1} = 1.729$ GeV) and a lattice spatial volume of approximately $(1.8 \textrm{ fm})^3$.  We simulate at three different light sea quark masses with pion masses down to approximately 430~MeV, and extrapolate to the physical quark masses using a phenomenologically-motivated fit function based on next-to-leading order heavy-light meson $SU(2)$ chiral perturbation theory.  For the $b$-quarks, we use an improved formulation of the Eichten-Hill action with static link-smearing to increase the signal-to-noise ratio.  We also improve the heavy-light axial current used to compute the $B$-meson decay constant to $\CO(\alpha_s p a)$ using one-loop lattice perturbation theory.  We present initial results for the $SU(3)$-breaking ratios $f_{B_s}/f_{B_d}$ and $\xi = f_{B_s} \sqrt{B_{B_s}}/f_{B_d} \sqrt{B_{B_d}}$, thereby demonstrating the viability of the method.  For the ratio of decay constants, we find $f_{B_s}/f_{B_d} = 1.15(12)$ and for the ratio of mixing matrix elements, we find $\xi = 1.13(12)$, where in both cases the errors reflect the combined statistical and systematic uncertainties, including an estimate of the size of neglected $\CO(1/m_b)$ effects.
\end{abstract}

%\\
%\\
%\\
%\centerline{\normalsize In memory of Jan Wennekers.}
%}

\maketitle

%=================================================
\section{Introduction}
\label{sec:Intro}
%=================================================

Neutral $B$-meson mixing is a sensitive probe of quark flavor-changing interactions.  When combined with  experimental measurements of the $B^0_d$ and $B_s^0$ oscillation frequencies, precise QCD determinations of the $B^0_d$ and $B_s^0$-mixing hadronic matrix elements allow clean determinations of the Cabibbo-Kobayashi-Maskawa (CKM)~\cite{Cabibbo:1963yz,Kobayashi:1973fv} matrix elements $|V_{td}|$ and $|V_{ts}|$ with all sources of systematic uncertainty under control.  In the Standard Model, the mass-difference of the neutral $B$-meson mass eigenstates $\Delta m_q$ (often called the oscillation frequency) is given by~\cite{Buras:1990fn,Buchalla:1995vs}
\begin{equation}
\label{eq:ckm}
\Delta m_q = \frac{G^2_Fm^2_W}{16\pi^2m_{B_q}} \,
|V^*_{tq}V_{tb}|^2 S_0(x_t) \eta_B \mathcal M_q,
\end{equation}
where $q \in \{d,s\}$.  Both the Inami--Lim function $S_0(x_t)$ with $x_t = m_t^2/m_W^2$ \cite{Inami:1980fz} and the QCD coefficient $\eta_B$ can be computed in perturbation theory~\cite{Buras:1990fn}.  
The hadronic contribution to the $\Delta B = 2$ mixing matrix element, 
\begin{equation}
\label{eq:me}
\mathcal M_q = \Big\langle \bar B^0_q\Big\lvert
[\bar b\gamma^\mu(1-\gamma_5)q] [\bar b\gamma_\mu(1-\gamma_5)q]
\Big\rvert B^0_q \Big\rangle ,
\end{equation}
must be calculated nonperturbatively from first principles using lattice QCD.

The hadronic matrix element $\mathcal M_q$ is conventionally parametrized as
\begin{equation}
\label{eq:pme}
\mathcal M_q = \frac{8}{3} m^2_{B_q}f^2_{B_q}B_{B_q},
\end{equation}
where $m_{B_q}$ is the mass of the $B$-meson, $f_{B_q}$ is the $B$-meson decay constant and $B_{B_q}$ is the $B$-meson bag parameter.  Many statistical and systematic uncertainties from lattice calculations cancel in the $SU(3)$-breaking ratio $\mathcal M_s / \mathcal M_d$, which would be one in the limit $m_d \to m_s$.  It is therefore convenient and conventional to introduce the quantity
\begin{equation}
\xi \equiv \frac{f_{B_s}\sqrt{{B}_{B_s}}}{f_{B_d} \sqrt{{B}_{B_d}}} .
\end{equation}
Given improved lattice calculations of the nonperturbative factor $\xi$, recent experimental measurements of the oscillation frequencies $\Delta m_d$ and $\Delta m_s$ to about 1\% accuracy~\cite{Amsler:2008zzb,Abazov:2006dm,Abulencia:2006mq,Abulencia:2006ze} now allow the possibility of precisely determining the ratio of the CKM matrix elements $|V_{td}|/|V_{ts}|$ (see, for example, Ref.~\cite{Bernard:1998dg}).  

The ratio $|V_{td}|/|V_{ts}|$ constrains the apex of the CKM unitarity triangle~\cite{CKMfitter,UTfit}.
It is likely that new physics would give rise to new quark-flavor changing interactions and additional CP-violating phases;  these would manifest themselves as apparent inconsistencies among different measurements of quantities which should be identical within the standard CKM picture. Thus a precise determination of the ratio $\xi$ will help to constrain physics beyond the Standard Model.  Furthermore, possible indications of new physics in $B_d^0$-mixing at the $\sim$2.7-$\sigma$ level~\cite{Lunghi:2008aa} 
make the lattice calculation of $B$-meson mixing parameters timely.

Recently, the HPQCD Collaboration published the first unquenched determination of $\xi$ with an accuracy of  2.6\%~\cite{Gamiz:2009ku}, and the Fermilab Lattice and MILC Collaborations expect to have a result with similar errors soon~\cite{Evans:2009du}.   The HPQCD calculation employs a nonrelativistic QCD (NRQCD) action for the heavy $b$-quark~\cite{Lepage:1992tx}, while the Fermilab/MILC calculation uses the relativistic ``Fermilab" action for the $b$-quark~\cite{ElKhadra:1996mp}.  Both of these computations, however, rely on the same ``2+1" flavor asqtad-improved staggered ensembles generated by the MILC Collaboration~\cite{Bernard:2001av}, which include the effects of two degenerate light quarks and one heavier close-to-strange quark in the sea sector.

For such a phenomenologically important  quantity as $\xi$, it is valuable to have an independent crosscheck using different formulations of the lattice action for both the light and heavy quarks. Our calculation employs the 2+1 flavor dynamical domain-wall ensembles generated by the
RBC and UKQCD Collaborations with a lattice spacing of $a \approx 0.11$~fm ($a^{-1} = 1.729$~GeV)~\cite{Allton:2007hx}.  The use of domain-wall fermions~\cite{Kaplan:1992bt,Shamir:1993zy,Furman:1994ky} has the advantage over other light-quark formulations that the chiral perturbation theory expressions needed to extrapolate domain-wall lattice results to the physical $u$- and $d$-quark masses are closer to the continuum forms and have fewer parameters than in the Wilson or staggered cases~\cite{Sharpe:1995qp,Stag_ChPT}.   We compute the $b$-quarks in the static limit ($m_b \to \infty$), which leads to correlation functions that are noisier than those with propagating $b$-quarks such as in the Fermilab~\cite{ElKhadra:1996mp} or NRQCD actions~\cite{Lepage:1992tx}.  We therefore use the static-quark formulation of Refs.~\cite{DellaMorte:2005yc,DellaMorte:2003mn} with either APE~\cite{Albanese:1987ds,Falcioni:1984ei} or HYP~\cite{Hasenfratz:2001hp} smearing of the static-quark gauge links to increase the signal-to-noise ratio and reduce scaling violations (for some quantities) as compared to the Eichten-Hill action~\cite{Eichten:1989zv}.  Furthermore, the approximate chiral symmetry of the domain-wall action combined with the spin symmetry of  the static action simplifies the lattice-to-continuum operator matching as compared to the Wilson case by reducing the number of additional lattice operators which appear~\cite{Becirevic:2003hd,Loktik:2006kz}.   The results of this work extend an earlier study with two flavors of dynamical quarks and heavier light-quark masses in Ref.~\cite{Gadiyak:2005ea}.

The primary purpose of this paper is to demonstrate the viability of our method for computing the $B$-meson decay constants and $\Delta B = 2$ mixing matrix elements.  We therefore use the small-volume ($16^3$) ensembles with only one lattice spacing and have relatively heavy light-quark masses and limited statistics.  A novel feature of this work is the use of $SU(2)$ heavy-light meson chiral perturbation theory (HM$\chi$PT) to extrapolate $N_f = 2+1$ lattice QCD results for $B$-meson quantities to the physical quark masses.  This follows the approach taken by the RBC and UKQCD Collaborations in the light pseudoscalar meson sector in Ref.~\cite{Allton:2008pn}, and differs from the calculations of HPQCD and Fermilab/MILC, both of whom use $SU(3)$ HM$\chi$PT for their chiral and continuum extrapolations~\cite{Gamiz:2009ku,Evans:2009du}.  The use of $SU(2)$ $\chi$PT is based on on the fact that the strange quark is much heavier than the up and down quarks, and can therefore be integrated out of the chiral effective theory.  Because lattice QCD simulations at the physical strange-quark mass are possible via tuning, interpolation, or reweighting, $SU(3)$ $\chi$PT is generally not needed to extrapolate the strange quark to its physical value.  When the masses of the light valence and sea quarks are sufficiently small such that $SU(2)$ chiral perturbation theory is applicable, $SU(2)$ $\chi$PT  for light pseudoscalar meson masses and decay constants converges more rapidly than $SU(3)$ $\chi$PT~\cite{Allton:2008pn,Aoki:2008sm,Kadoh:2008sq,Bazavov:2009ir}.  Although we do not have enough data or sufficiently light quark masses to perform a thorough comparison of $SU(2)$ and $SU(3)$ HM$\chi$PT in this work, we believe that the use of $SU(2)$ HM$\chi$PT provides a promising alternative to $SU(3)$ and warrants further study when better data is available.  

In this work we compute both the ratio of decay constants, $f_{B_s}/f_{B_d}$, and the ratio of $\Delta B=2$ matrix elements, $\xi$. We focus on the $SU(3)$-breaking ratios because both the statistical and systematic errors are smaller and under better control than for the individual decay constants and mixing matrix elements.   Our results have large total uncertainties compared to those of HPQCD and Fermilab/MILC.  Within errors, however, our results for the $SU(3)$-breaking ratios are consistent with the values presented in the literature and we expect to improve upon them and present values for the individual decay constants and matrix elements in a future work.

%Because we are working in the static limit, our results neglect contributions of $\CO(1/m_b)$, which may be as large as $\sim 10\%$ for the decay constants and individual $B$-parameters.  Thus one must be cautious in using them to constrain new physics such as in the unitarity triangle analyses.  We believe, however, that our result for the phenomenologically-important parameter $\xi$ is more robust.  This is because relativistic corrections to the $SU(3)$-breaking ratio should be proportional to $(m_s - m_d)$, and thus of order $(m_s - m_d)/m_b \sim 2 \%$.  Furthermore, discretization errors due to the use of only one lattice spacing and truncation errors due to the use of one-loop lattice perturbation theory should also largely cancel in the ratio.  Thus our determination of $\xi$ provides a valuable independent crosscheck of other unquenched lattice results. (Possibly save this paragraph for the next static paper \ldots)

\bigskip

This paper is organized as follows.  First, in Section \ref{sec:setup}, we present the actions and parameters used in our lattice simulations.  Next, in Sec.~\ref{sec:PT}, we briefly discuss the perturbative matching of the heavy-light current and the four-fermion operators; the details of the lattice perturbation theory calculation will be presented in another publication~\cite{PT_Oa}.  We compute 2-point and 3-point lattice correlation functions and extract the decay constants and mixing matrix elements in Sec.~\ref{sec:BreakingRatios}, and extrapolate these results to the physical light-quark masses using a phenomenologically-motivated function based on next-to-leading order (NLO) $SU(2)$ heavy-meson chiral perturbation theory in Sec.~\ref{sec:ChExp}.  In Section \ref{sec:Err} we estimate the contributions of the various systematic uncertainties to $f_{B_s}/f_{B_d}$ and $\xi$, discussing each item in the error budget separately.  We present our final results and conclude in Section \ref{sec:Conc}. 

This paper also contains four appendices.  Appendix~\ref{sec:proj_details} 
specifies the $SU(3)$ projection methods that are used in our 
APE and HYP-smeared gauge links.  In App.~\ref{app:cont_match} some
details of the perturbative formulae used to match 
from continuum QCD to HQET as well as for the HQET running are presented.  
Appendix~\ref{app:Norman} discusses the large
ground state degeneracy present in HQET and how it can be 
exploited to compute $B$-meson mixing matrix elements using 
localized sources and sinks.
The $SU(3)$ and $SU(2)$ NLO HM$\chi$PT expressions 
for the $B$-meson decay constants and mixing matrix elements 
relevant for $N_f = 2+1$ domain-wall lattice simulations are provided in App.~\ref{app:ChPT}; some of these results have not been presented previously in the literature.  

%=================================================
\section{Lattice actions and parameters}
\label{sec:setup}
%=================================================

In this section we briefly describe our numerical lattice simulations.    We use the unquenched lattices generated by the RBC and UKQCD Collaborations which include the effects of 2+1 dynamical flavors of domain-wall quarks~\cite{Allton:2007hx}.  We calculate the decay constants and matrix elements on configurations with a lattice spacing of $a^{-1} = 1.729(28)$~GeV~\cite{Allton:2008pn} and an approximate spatial volume of $L^3 \approx (1.8 \textrm{ fm})^3$.  For each ensemble, the masses of the up and down sea quarks are degenerate and the mass of the strange sea quark is slightly larger than its physical value.  In order to distinguish the dynamical quark masses used in our simulations from the physical $u$, $d$, and $s$-quark masses, we denote the lighter sea quark mass by $m_l$ and the heavier sea quark mass by $m_h$.  Our lightest pion mass is approximately 430~MeV.  Table~\ref{tab:dwf_ens} summarizes the parameters of the dynamical domain-wall ensembles used in our analyses.

In Section \ref{subsec:DWF} we present the domain-wall fermion action used for both the valence and sea light $u$, $d$, and $s$ quarks.  Next,  in Sec.~\ref{subsec:Gauge}, we show the Iwasaki gauge action used for the gluon fields.  Finally, we discuss the static action used for the heavy $b$ quarks in Sec.~\ref{subsec:HQET}.

\begin{CTable}
\caption{Available 2+1 flavor domain-wall ensembles~\cite{Allton:2007hx}.  The columns from left to right are the approximate lattice spacing in fm, the bare light and strange quark masses in the sea sector, the dimensions of the lattice in lattice units, the pion mass in the sea sector, the dimensionless factor $m_\pi L$, and the residual quark mass in the chiral limit. \medskip}
\label{tab:dwf_ens}

\begin{tabular}{cccccccc}
 \hline\hline
 $a$(fm) & \ $am_l/ am_h$ & \ Volume & \ $m_\pi$(MeV) & \ $m_\pi L$ &  \ $a m_\textrm{res}$  \\
 \hline
  $\approx 0.11$ & 0.01/0.04 & \ $16^3 \times 32$  & 430 & \ 3.9  & \ 0.00315 \\
  $\approx 0.11$ & 0.02/0.04 & \ $16^3 \times 32$  & 560 & \ 5.2  & \ 0.00315 \\
  $\approx 0.11$ & 0.03/0.04 & \ $16^3 \times 32$  & 670 & \ 6.3  & \ 0.00315 \\
 \hline\hline

\end{tabular}\end{CTable}

%=================================================
\subsection{Light-quark action}
\label{subsec:DWF} 
%=================================================

We use the five-dimensional domain-wall fermion action~\cite{Shamir:1993zy,Furman:1994ky} for the light $u$, $d$, and $s$ quarks in both the valence and sea sectors:
\begin{eqnarray}
\label{eq:dwf_action}
S_{\rm DW} &= & a^4 \left( \sum_{s,s'=0}^{L_s-1}
 \sum_{x,y}\overline\psi_{s}(x) D^{\rm DW}_{ss'}(x,y)\psi_{s'}(y) 
 -\sum_{x}m_f\overline q(x)q(x) \right), \\
D^{\rm DW}_{ss'}(x,y) & = & D^4(x,y)\delta_{s,s'} + D^5(s,s')\delta_{x,y} + a^{-1}
 (M_5 - 5)\delta_{s,s'}\delta_{x,y}, \\
D^4(x,y) & = & \sum_\mu \frac{1}{2a}
 \left[(1-\gamma_\mu)U_\mu(x)\delta_{x+\hat{\mu},y} +
       (1+\gamma_\mu)U^{\dagger}_\mu(y)\delta_{x-\hat{\mu},y}\right], \\
D^5(s,s') & = & \begin{cases}
a^{-1} P_L\delta_{1,s'} & (s=0) \\
a^{-1} \left( P_L\delta_{s+1,s'} + P_R\delta_{s-1,s'} \right) & (0<s<L_s-1) \\
a^{-1}P_R\delta_{L_s-2,s'} & (s = L_s-1) \\
\end{cases}
\end{eqnarray}
where $\psi_{s}(x)$ is a 5-d Wilson-type fermion
field. The fifth dimension extends from $0$ to $L_s-1$ and is labeled by
$s$.   The domain-wall height is set to $M_5 = 1.8$ in our simulations. 
The projectors $P_L = (1-\gamma_{5})/2$ and $P_R = (1+\gamma_{5})/2$ select left- and right-handed spinor components, respectively.  The physical four-dimensional
quark field $q(x)$ is constructed from the five-dimensional field $\psi_s(x)$ at
$s=0$ and $L_s-1$:
\begin{equation}\label{eq:phys_quark}
q(x) = P_{L}\psi_{0}(x) + P_{R}\psi_{L_s-1}(x) ,
\end{equation}  
\begin{equation}\label{eq:phys_antiquark}
\overline{q}(x) = \overline{\psi}_{0}(x)P_R + \overline{\psi}_{L_s-1}(x)P_L.
\end{equation}  

In the limit $L_s \to \infty$, the left-handed and right-handed modes decouple and exact chiral symmetry is recovered.  In practice, however, $L_s$ is large but finite in numerical lattice simulations. This leads to a small amount of chiral symmetry breaking which can be parameterized in terms of an additive shift to the bare domain-wall quark mass.  At the value $L_s = 16$ used in our simulations we obtain a residual quark mass of $a m_\text{res} = 0.00315(2)$~\cite{Allton:2008pn}.  Because mixing between heavy-light four-fermion operators of different chiralities is proportional to the value of $a m_\text{res}$, this indicates that the size of errors from mixing with wrong-chirality operators is negligible.

%=================================================
\subsection{Gluon action}
\label{subsec:Gauge} 
%=================================================

We use the Iwasaki gauge action for the gluons~\cite{Iwasaki:1983ck}:
\begin{equation}\label{eq:gauge_action}
S_{\rm gauge} = -\frac{\beta}{3} \left( (1-8c_{1}) \sum_{P} 
\text{\,Re\,Tr\,}[U_{P}] + c_{1}\sum_{R} {\rm ReTr}[U_{R}] \right) ,
\end{equation}
where $\beta \equiv 6/g^{2}_{0}$ and $g_{0}$ is the bare lattice coupling.  $U_{P}$ is the path-ordered product of gauge links around the $1\times1$ plaquette $P$ and $U_R$ is the path-ordered product of gauge links around the $1\times2$ rectangle $R$.  The constant  $c_1$ is set to $-0.331$ in the Iwasaki action and we use $\beta = 2.13$ in our simulations.  As was shown in Refs.~\cite{AliKhan:2000iv,Aoki:2002vt} for the quenched approximation, the use of the Iwasaki action in combination with domain-wall valence quarks leads to improved chiral symmetry and a smaller residual quark mass than for the Wilson gauge action~\cite{Wilson:1974sk}.  In the case of $N_f = 2+1$ dynamical domain-wall simulations, the use of the Iwasaki action also allows frequent tunneling between topological sectors~\cite{Antonio:2008zz}.

%=================================================
\subsection{Heavy-quark action}
\label{subsec:HQET}
%=================================================

We use an improved static action for the $b$-quarks in the 2-point and 3-point correlation functions needed to compute the decay constants and matrix elements.  We build upon the original lattice formulation of the static effective action that was constructed by Eichten and Hill~\cite{Eichten:1989kb}:
\begin{equation}
	S_{\rm static} = a^3 \sum_{x,y} \left( \bar{h}(x) [\delta_{x,y} - U_{0}^{\dagger}(y)\delta_{x-\hat0,y} ] P_+h(y) - \bar{h}(x) [ \delta_{x,y} - U_{0}(x)\delta_{x+\hat0,y} ] P_-h(y) \right) ,
\label{eq:hqa}
\end{equation}
where $h(x)$ is the static quark field at site $x$, $U_{0}(x)$ is the gauge link in the temporal direction and $\hat0$ denotes the unit vector along the temporal direction.   The projectors $P_{\pm}=\frac{1}{2}(1\pm\gamma_0)$ select the parity even and odd components of $h(x)$, which we denote by $h^{(+)}(x)$ and $h^{(-)}(x)$. respectively. The components of the static-quark field $h^{(\pm)}$ in the Eichten-Hill action satisfy the relation $\gamma_0 h^{(\pm)} = \pm h^{(\pm)}$.  Furthermore, the static quark propagator $H(x,y) = \langle h(x) \bar{h}(y) \rangle$ can be expressed as the product of gauge links:
\begin{align}
H(x,y) & = H^{+}(x,y) + H^{-}(x,y), \\
H^{(+)}(x,y) &= \frac{1}{a^3}\theta(t_{x} - t_{y}) \delta_{{\vec x},{\vec y}} \left[U^{\dagger}_{0}(x-\hat0) \ldots U^{\dagger}_{0}(y + \hat0) U^{\dagger}_{0}(y)\right]P_+,\\
H^{(-)}(x,y) &=  -\frac{1}{a^3} \theta(t_{y} - t_{x}) \delta_{{\vec x},{\vec y}} \left[U_{0}(x) U_{0}(x+\hat0)\ldots U_{0}(y-\hat0)\right]P_- .
\end{align}

These properties make the Eichten-Hill formulation computationally simple.  In practice, however, numerical simulations with this action are quite noisy. Therefore we use instead smeared (or ``fat") link actions:
\begin{equation}
	S_{\rm fat} = a^3 \sum_{x,y} \left( \bar{h}(x) [\delta_{x,y} - \bar{V}_{0}^{\dagger}(y)\delta_{x-\hat0,y} ] P_+h(y) - \bar{h}(x) [ \delta_{x,y} - \bar{V}_{0}(x)\delta_{x+\hat0,y} ] P_-h(y) \right) ,
\label{eq:hqa_fat}	
\end{equation}
where the new gauge link $\bar{V}$ is obtained from the thin link $U$ by either APE blocking~\cite{Albanese:1987ds,Falcioni:1984ei} or hypercubic (HYP) blocking~\cite{Hasenfratz:2001hp}.  The heavy quark propagator for the improved action is given by
\begin{align}\label{eq:hq_imp_prop}
H^{(+)}_\textrm{fat}(x,y) &= \frac{1}{a^3}\theta(t_{x} - t_{y}) \delta_{{\vec x},{\vec y}} \left[\bar{V}^{\dagger}_{0}(x-\hat0) \ldots \bar{V}^{\dagger}_{0}(y + \hat0) \bar{V}^{\dagger}_{0}(y)\right]P_+,\\
H^{(-)}_\textrm{fat}(x,y) &=  -\frac{1}{a^3} \theta(t_{y} - t_{x}) \delta_{{\vec x},{\vec y}} \left[\bar{V}_{0}(x) \bar{V}_{0}(x+\hat0)\ldots \bar{V}_{0}(y-\hat0)\right]P_- .
\end{align}
The replacement of the simple gauge link $U_0$ by a smeared link significantly
improves the signal-to-noise ratio~\cite{DellaMorte:2003mn}.

We construct the fattened APE link by adding a weighted sum of the staples to the original thin link, and restrict the smearing to links in the temporal direction along which the heavy quark propagates.   We use the APE parameter $\alpha =1$, for which the smeared link is given by~\cite{Falcioni:1984ei}
\begin{eqnarray}
\label{eq:APE_smear}
\bar{V}_0(x) &=& {\rm\bf Proj}_{\text{SU(3)}}[V_0(x)], \\
V_0(x) &=& \frac{1}{6}\sum^3_{\nu=1} \left(U_\nu(x)U_0(x+\hat{\nu})U^{\dagger}_\nu(x+\hat{0}) + U^{\dagger}_\nu(x-\hat{\nu})U_0(x-\hat{\nu})U_\nu(x+\hat{0}-\hat{\nu})\right) .
\end{eqnarray}
We build the HYP link from three iterative steps of APE smearing which are
restricted to the hypercube around the original link.  For links in the temporal direction the construction is as follows:
\begin{eqnarray}
 \bar{V}_0(x) &=& {\rm\bf Proj}_{\text{SU(3)}}\left[(1-\alpha_1)U_0(x)
 +\frac{\alpha_1}{6}\sum_{\nu = \pm 1}^{\pm 3}\tilde{V}_{\nu;0}(x)\tilde{V}_{0;\nu}
 (x+\hat{\nu})\tilde{V}^{\dagger}_{\nu;0}(x+\hat{0})\right], \label{eq:HYP_smear_1}\\
 \tilde{V}_{\mu;\nu}(x) &=& {\rm\bf Proj}_{\text{SU(3)}}\left[(1-\alpha_2)U_\mu(x)
 +\frac{\alpha_2}{4}\sum_{\substack{\rho = \pm 0\\ \rho\neq \mu,\nu}}^{\pm 3}V_{\rho;\nu
 \mu}(x)V_{\mu;\rho\nu}(x+\hat{\rho})
 V^{\dagger}_{\rho;\nu\mu}(x+\hat{\mu})\right], \label{eq:HYP_smear_2}\\
 V_{\mu;\nu\rho}(x) &=& {\rm\bf Proj}_{\text{SU(3)}}\left[(1-\alpha_3)U_\mu(x)
 +\frac{\alpha_3}{2}\sum_{\substack{\eta = \pm 0 \\ \eta\neq\mu,\nu,\rho}}^{\pm 3}U_\eta(x)U_\mu(x+\hat{\eta})
 U^{\dagger}_\eta(x+\hat{\mu})\right]. \label{eq:HYP_smear_3}
\end{eqnarray}
We use the HYP smearing parameters $(\alpha_1,\alpha_2,\alpha_3)  = (1.0,1.0,0.5)$, sometimes referred to as HYP2. These were shown to approximately minimize the noise-to-signal ratio in Ref.~\cite{DellaMorte:2005yc}.

In Eqs.~(\ref{eq:APE_smear})--(\ref{eq:HYP_smear_3}), ${\rm\bf Proj}_{\text{SU(3)}}(V)$ indicates the projection of the link $V$ onto an $SU(3)$ matrix. This projection reduces the statistical noise and thus enhances the smearing effect without increasing the level of smearing.  In some cases the $SU(3)$ projection also suppresses quantum corrections to lattice operators in perturbation theory \cite{Lee:2002fj}.  The projection of the smeared link onto $SU(3)$ is not unique, and we use two different schemes: for the case of APE smearing, Eq. (\ref{eq:APE_smear}),
we project by the unit circle method based on polar decomposition \cite{Kamleh:2004xk}, while for the case of HYP smearing, Eqs.~(\ref{eq:HYP_smear_1})--(\ref{eq:HYP_smear_3}), we obtain the projected matrix by an iterative procedure seeking the $SU(3)$ matrix $U_\text{max}$ that maximizes $\text{Re}\ \text{Tr}\ (U_\text{max} V^\dagger)$, where $V$ is the HYP smeared link matrix~\cite{Bali:1992ab}. We describe the details of the two schemes and show their equivalence in the weak coupling limit in Appendix \ref{sec:proj_details}.

%=================================================
\section{Perturbative matching of heavy-light current and four-fermion operators}
\label{sec:PT}
%=================================================
In order to renormalize the heavy-light axial current and $\Delta B = 2$ four-fermion operator, we adopt a two-step matching procedure.  In the first step, we  match the QCD operators renormalized in the $\overline{\rm MS}$ scheme using naive dimensional regularization (NDR) at a scale $\mu_b$ to continuum static effective theory operators renormalized at a scale $\mu$.  This step is described in Sec.~\ref{sec:cont_match}.  In this paper, we choose $\mu_b$ to be the $b$-quark mass $m_b$ and $\mu$
to be the inverse lattice spacing $a^{-1}$. In the second step, we match the continuum static effective theory operators to the lattice ones.  This step is described in Sec.~\ref{sec:stat_match}.  We combine the results of the two steps and present  the results for the complete matching coefficients in Table~\ref{tab:PT}.

%=================================================
\subsection{Continuum matching}
\label{sec:cont_match}
%=================================================

The QCD operators considered in this paper are the axial vector current
\begin{eqnarray}
A_0^\textrm{QCD}=\bar{b}\gamma_0\gamma_5q,
\end{eqnarray}
and the $\Delta B=2$ four-quark operator
\begin{eqnarray}
O_L^\textrm{QCD}(\mu_b)=
[\bar{b}\gamma_{\mu}(1-\gamma_5)q][\bar{b}\gamma_{\mu}(1-\gamma_5)q].
\end{eqnarray}
These are related to the continuum HQET operators by
\begin{eqnarray}
A_0^\textrm{QCD}&=&
C_A(\mu){A}^\textrm{HQET}_0(\mu)+O(\Lambda_{\rm QCD}/m_b), \label{eq:A0_QCD_to_HQET} \\
O_L^\textrm{QCD}(\mu_b)&=&
Z_1(\mu_b,\mu){O}^\textrm{HQET}_L(\mu)+Z_2(\mu_b,\mu){O}^\textrm{HQET}_S(\mu)
+O(\Lambda_{\rm QCD}/m_b), \label{eq:OL_QCD_to_HQET}
\end{eqnarray}
where
\begin{eqnarray}
{A}^\textrm{HQET}_0&=&\bar{h}\gamma_0\gamma_5q,\\
{O}^\textrm{HQET}_L&=&
[\bar{h}\gamma_{\mu}(1-\gamma_5)q][\bar{h}\gamma_{\mu}(1-\gamma_5)q],\\
{O}^\textrm{HQET}_S&=&
[\bar{h}(1-\gamma_5)q][\bar{h}(1-\gamma_5)q] .
\end{eqnarray}
Note that, because $A_0^\textrm{QCD}$ is a conserved current, it does not depend upon the renormalization scale $\mu_b$. The coefficients $C_A$ and $\vec Z = (Z_1, Z_2)$ in Eqs.~(\ref{eq:A0_QCD_to_HQET}) and~(\ref{eq:OL_QCD_to_HQET}) are products of three factors:
\begin{eqnarray}
C_A(\mu) & = & \tilde C_A(m_b)\cdot U_A^{(4)}(m_b,m_c) \cdot
U_A^{(3)}(m_c,\mu),\label{eq:C_A}\\
\vec{Z}(\mu_b,\mu) & = & \vec{\tilde{Z}}(\mu_b,m_b) \cdot U_L^{(4)} 
(m_b,m_c) \cdot U_L^{(3)}(m_c,\mu) , \label{eq:ZDB2}
\end{eqnarray}
where $\tilde C_A$ and $\vec{\tilde{Z}} \equiv (\tilde Z_1, \tilde Z_2)$ are the matching coefficients from HQET to QCD.  The factors $U_A^{(N_f)}(\mu',\mu)$ and the $2 \times 2$ matrix $U_L^{(N_f)}(\mu',\mu)$ account for the renormalization group running between scales:  We first match the continuum QCD operators onto continuum HQET operators at one-loop;  this occurs at a scale $\mu_b = m_b$ in the $\overline{\mbox{MS}}$(NDR) scheme.  We then run the matching coefficients in four-flavor continuum HQET from $m_b$ to $m_c$ at two-loops.  Finally, we run the coefficients at two-loops in three-flavor continuum HQET from $m_c$ to the scale $\mu = 1/a$, where the matching to three-flavor lattice HQET is done. 

The matching factors and anomalous dimensions needed to compute these coefficients are given in Refs.~\cite{Eichten:1989zv,Ji:1991pr,Broadhurst:1991fz} for the heavy-light current and in Refs.~\cite{Flynn:1990qz,Gimenez:1992is,Ciuchini:1996sr,Buchalla:1996ys} for the four-quark operator.  For completeness, we present them in Appendix~\ref{app:cont_match}. Here we simply quote the results for the matching coefficients, which already contain some terms of $O(\alpha_s^2)$:
\begin{align}
&C_A(\mu)=\left[\frac{\alpha_s(m_b)}{\alpha_s(m_c)}\right]^{-\frac{6}{25}} \cdot \left[\frac{\alpha_s(m_c)}{\alpha^{(3)}_s(\mu)}\right]^{-\frac{2}{9}} \cdot \left(1-\frac{8}{3}\frac{\alpha_s(m_b)}{4\pi}\right)\nonumber \\
&\qquad \cdot \left(1 + J_A^{(4)}
\frac{\alpha_s(m_b)-\alpha_s(m_c)}{4\pi}\right) \cdot\left(1 + J_A^{(3)}
\frac{\alpha_s(m_c)-\alpha_s^{(3)}(\mu)}{4\pi}\right)+ \CO(\alpha_s^2), \\
&Z_1(m_b,\mu)= \left[\frac{\alpha_s(m_b)}{\alpha_s(m_c)}\right]^{-\frac{12}{25}} \cdot \left[\frac{\alpha_s(m_c)}{\alpha^{(3)}_s(\mu)}\right]^{-\frac{4}{9}} \nonumber \\
&\qquad
\cdot \Bigg\{ \left(1-14\frac{\alpha_s(m_b)}{4\pi}\right)
\cdot \left( 1+J_{11}^{(4)}\frac{\alpha_s(m_b)-\alpha_s(m_c)}{4\pi} \right) \cdot \left( 1+J_{11}^{(3)}\frac{\alpha_s(m_c)-\alpha_s^{(3)}(\mu)}{4\pi} \right) \nonumber \\
&\qquad +2\frac{\alpha_s(m_b)}{4\pi} \Bigg\{ \left(1-\left[\frac{\alpha_s(m_b)}{\alpha_s(m_c)}\right]^{\frac{8}{25}}  \right) \left( 1+J_{11}^{(3)}\frac{\alpha_s(m_c)-\alpha^{(3)}_s(\mu)}{4\pi} \right) \nonumber \\
&\qquad\quad+\left[\frac{\alpha_s(m_b)}{\alpha_s(m_c)}\right]^{\frac{8}{25}} \cdot \left( 1-\left[\frac{\alpha_s(m_c)}{\alpha^{(3)}_s(\mu)}\right]^{\frac{8}{27}} \right) \Bigg\}
\Bigg\} + \CO(\alpha_s^2),\\
&Z_2(m_b,\mu)= -8\frac{\alpha_s(m_b)}{4\pi}
\Bigg[\frac{\alpha_s(m_b)}{\alpha_s(m_c)}\Bigg]^{-\frac{4}{25}} \Bigg[\frac{\alpha_s(m_c)}{\alpha_s^{(3)}(\mu)}\Bigg]^{-\frac{4}{27}}  + \CO(\alpha_s^2),
\end{align}
with the parameters $J_A$ and $J_{11}$ given by
\begin{align}
J_{A}^{(3)} = -0.7545,\qquad J_{A}^{(4)} = -0.9098,\\
J_{11}^{(3)} = -1.6980, \qquad J_{11}^{(4)} = -1.8637.
\end{align}
Note that the leading-order mixing between $O_L^{QCD}$ and $O_S^{HQET}$ is of $\CO(\alpha_s)$. To determine the coupling constant $\alpha_s$ at different scales,
which is required to obtain these results, we fix the value of $\alpha_s$ at the $Z$-boson mass to the PDG value
$\alpha_s(m_Z=91.1876~{\rm GeV})=0.1176$~\cite{Amsler:2008zzb}.  Using four-loop running~\cite{Chetyrkin:1997dh,Vermaseren:1997fq}, we obtain $\alpha_s(m_b=4.20~{\rm GeV}~\mbox{\cite{Amsler:2008zzb}})=0.2228$,
$\alpha_s(m_c=1.27~{\rm GeV}~\mbox{\cite{Amsler:2008zzb}})=0.3819$, and $\alpha_s^{(3)}(a^{-1}=1.729~{\rm GeV}~\mbox{\cite{Allton:2008pn}})=0.3141$.  Hence we find
\begin{eqnarray}
\label{eq:QCD-HQET-numbers}
C_A(a^{-1})=1.0459,\;\;\;
Z_1(m_b,a^{-1})=0.9100,\;\;\;
Z_2(m_b,a^{-1})=-0.1502
\end{eqnarray}
for the matching coefficients in Eqs.~(\ref{eq:A0_QCD_to_HQET})
and~(\ref{eq:OL_QCD_to_HQET}) that relate the operators in continuum QCD
to those in the continuum static effective theory.

%=================================================
\subsection{Static effective theory matching}
\label{sec:stat_match}
%=================================================

We next discuss the lattice-to-continuum operator matching in the static effective theory.   Although both the domain-wall and static quark actions are on-shell ${\cal O}(a)$ improved, the domain-wall-static vertices still receive ${\cal O}(a)$ corrections.  This was shown in Ref.~\cite{Ishikawa:1998rv} for the case of clover light quarks and static heavy quarks even when the parameter in the clover action $r \to 0$ and chiral symmetry is restored in the light-quark sector.   In this work, we remove $\CO(a)$ lattice discretization errors from the axial vector current in the light-quark chiral limit at one-loop in lattice perturbation theory by adding dimension-four operators containing derivatives; we refer to these errors as $\CO(pa)$ to distinguish them from $\CO(m_q a)$ errors that vanish in the light-quark chiral limit.  The improved operators are determined by requiring that the on-shell quark scattering amplitudes in the $B$-meson rest frame agree in the lattice and continuum theories through ${\cal O}(\alpha_s p a)$, where $p$ is the momenta of the light quarks in the $B$-meson and is typically of $\CO(\Lambda_\textrm{QCD})$.  The details of this calculation are presented in Refs.~\cite{PT_Oa}. 
We neglect $\CO(m_q a)$ errors because these are estimated to be small~\cite{PT_Oa}, but account for them when estimating the systematic errors in Sec.~\ref{sec:Err}.  In this work we also neglect $\CO(\alpha_s pa)$ corrections to the four-quark operator since we estimate the size of the resulting contribution to the $SU(3)$ breaking ratio $\xi$ would be much smaller than our current statistical precision, but we again account for them in the systematic error budget.  The one-loop $\CO(\alpha_s pa)$ corrections for the four-quark operator are available in Ref.~\cite{PT_Oa}, however, for use in future simulations.  

For domain-wall light quarks, the one-loop lattice perturbation theory calculations are presented in Refs.~\cite{Loktik:2006kz,PT_Oa,Christ:2007cn}.  The lattice-to-continuum matching has the form:
\begin{align}
{A}^\textrm{HQET}_0(\mu)&=
\frac{\sqrt{u_0}}
{\sqrt{(1-(w_0^{\rm MF})^2)Z_w^{\rm MF}}}Z_A^{\rm MF}(\mu,a^{-1})
\left[{A}_0^{\rm lat}(a^{-1})
+c_A^{\rm MF}(\mu,a^{-1})a {A}_{\partial,0}^{\rm lat}(a^{-1})
\right]\nonumber\\
&\equiv {\cal Z}_A^{\rm MF}(\mu,a^{-1}) \left[{A}_0^{\rm lat}(a^{-1}) + c_A^{\rm MF}(\mu,a^{-1})a {A}_{\partial,0}^{\rm lat}(a^{-1}) \right].
\label{EQ:AMU_lat}\\
{O}^\textrm{HQET}_L(\mu)&=
\frac{u_0}{(1-(w_0^{\rm MF})^2)Z_w^{\rm MF}}Z_L^{\rm MF}(\mu,a^{-1})
{O}_L^{\rm lat}(a^{-1}) \equiv{\cal Z}_L^{\rm MF}(\mu,a^{-1}){O}_L^{\rm lat}(a^{-1}),
\label{EQ:OL_lat}\\
{O}^\text{HQET}_S(\mu)&=
\frac{u_0}{(1-(w_0^{\rm MF})^2)}
{O}_S^{\rm lat}(a^{-1}) \equiv{\cal Z}_S^{\rm MF}(\mu,a^{-1}) {O}_S^{\rm lat}(a^{-1}),
\label{EQ:OS_lat}
\end{align}
with the constant
\begin{equation}
	w_0^{\rm MF} = 1 - M_5 + 4(1-u_0) ,
\end{equation}
where the mean-field link $u_0 = P^{1/4}$ is obtained from the expectation value of the plaquette $P$.   In Eqs.~(\ref{EQ:AMU_lat})-(\ref{EQ:OS_lat}) we match ${A}^\textrm{HQET}$ and ${O}^\textrm{HQET}_L$ at one-loop, but match ${O}^\textrm{HQET}_S$ at tree-level.  This is sufficient because the leading-order mixing between the continuum QCD operator $O_L^{QCD}$ and the continuum HQET operator $O_S^{HQET}$ is already of $\CO(\alpha_s)$, and has no tree-level component.  The lattice operators have the same form as in the continuum static effective theory
\begin{eqnarray}
{A}^\textrm{lat}_0&=&\bar{h}\gamma_0\gamma_5q, \label{eq:A_lat} \\
{O}^\textrm{lat}_L&=&
[\bar{h}\gamma_{\mu}(1-\gamma_5)q][\bar{h}\gamma_{\mu}(1-\gamma_5)q], \label{eq:OL_lat} \\
{O}^\textrm{lat}_S&=&
[\bar{h}(1-\gamma_5)q][\bar{h}(1-\gamma_5)q] . \label{eq:OS_lat}
\end{eqnarray}
The $O(p a)$ derivative operator in the equation for the axial current is given by
\begin{eqnarray}
{A}^\textrm{lat}_{\partial,0}=
\partial_0 \left(\bar{h}\gamma_5 q\right) ,
\label{EQ:AMU_Opa}
\end{eqnarray}
where we have simplified the expression using the equations-of-motion.  The domain-wall specific renormalization factor $Z_w^{\rm MF}$ that enters the above equations was calculated perturbatively in Ref.~\cite{Aoki:2002iq}. 

The superscript ``MF" denotes mean-field improvement~\cite{Lepage:1992xa}, in which we modify the bare lattice coupling using the value of the mean-field link. 
Use of this ``boosted" coupling as the new expansion parameter improves the convergence of lattice perturbation theory.  The mean-field improved coupling $\alpha^\textrm{MF}$ has several definitions which differ only at higher-order in perturbation theory than we consider.  These differences enter our estimate of the systematic error in Section \ref{sec:Err}. Our choice for obtaining the mean-field improved coupling from the bare lattice coupling $g_0^2$ is
\begin{eqnarray}
\frac{1}{(g^{\rm MF})^2}=\frac{P}{g_0^2}+d_g+c_p+N_fd_f,
\end{eqnarray}
where $N_f=3$ is the number of dynamical flavors.  We use the plaquette value averaged over different light sea quark mass,  $P=0.5881$, because the difference in the value of $P$ for different light sea quark masses is less than $0.05\%$ for the ensembles used in this work.   The constants $d_g = 0.1053$ and $c_p = 0.1401$ were calculated for the Iwasaki gauge action in Ref.~\cite{Aoki:2002iq} and $d_f =  -0.001465$ was obtained for $M^\textrm{MF}_5 = 1.303$ from a linear interpolation of the results in Table II of Ref.~\cite{Aoki:2003uf}.  On our ensembles, the mean-field improved coupling is $\alpha^{\rm MF}=0.1769$.

Given the value of the plaquette in our simulations, $u_0=0.8757$ and $w_0^{\rm MF}=-0.3029$.  After setting $\mu=a^{-1}$, the coefficients appearing in Eqs.~(\ref{EQ:AMU_lat}), (\ref{EQ:OL_lat}) and (\ref{EQ:OS_lat}) are:
\begin{eqnarray}
Z_w^{\rm MF}&=&1+\frac{\alpha^{\rm MF}}{4\pi} \frac{4}{3} \times 5.250,\\
Z_A^{\rm MF}&=&1+\frac{\alpha^{\rm MF}}{4\pi}  \frac{4}{3} \times
\begin{cases}
-1.584\;\;\mbox{APE} \\
~~0.077\;\;\mbox{HYP}
\end{cases}\\
c_A^{\rm MF}&=&\frac{\alpha^{\rm MF}}{4\pi}  \frac{4}{3} \times
\begin{cases}
3.480\;\;\mbox{APE} \\
6.412\;\;\mbox{HYP}
\end{cases}\\
Z_L^{\rm MF}&=&1+\frac{\alpha^{\rm MF}}{4\pi}\times
\begin{cases}
-4.462\;\;\mbox{APE} \\
~~1.076\;\;\mbox{HYP}
\end{cases}
\end{eqnarray}
where the values are given for both the APE and HYP link-smearings used in this work.  These results can be combined to determine the overall multiplicative renormalization factors denoted by $\mathcal{Z}$ in Eqs.~(\ref{EQ:AMU_lat})--(\ref{EQ:OS_lat}), which we present for completeness in Table~\ref{TAB:factor}.

%%%%%%%%%%%%%%%%%%%%
\begin{CTable}
\caption{Lattice-to-continuum operator matching factors in the static effective theory for the choice of coupling $\alpha^\textrm{MF}_s$.} 
\label{TAB:factor}
\begin{tabular}{ccccc}
\hline\hline
smearing & ${\cal Z}_A^{\rm MF}$ &$c_A^{\rm MF}$ &${\cal Z}_L^{\rm MF}$ &${\cal Z}_S^{\rm MF}$ \\\hline
APE      & 0.9090 & 0.0653 & 0.8225 & 0.9642 \\
HYP      & 0.9382 & 0.1204 & 0.8909 & 0.9642 \\
\hline\hline
\end{tabular}
\end{CTable}

%=================================================
\subsection{Complete matching coefficients}
\label{sec:total_match}
%=================================================

In order to match the lattice HQET operators at scale $\mu = a^{-1}$ directly onto the desired continuum QCD operators at $\mu_b = m_b$, we must combine the coefficients obtained in the two steps.  We define the complete lattice HQET-to-continuum QCD matching coefficients as  
\begin{align}
	Z_\Phi(a^{-1})  &=  C_A(a^{-1}) \cdot \mathcal{Z}_A^{\rm MF}(a^{-1}, a^{-1}), \\
	Z_{VA}(\mu_b,a^{-1})  &=  Z_1(\mu_b,a^{-1}) \cdot \mathcal{Z}_L^{\rm MF}(a^{-1}, a^{-1}), \\
	Z_{SP}(\mu_b,a^{-1})  &=  Z_2(\mu_b,a^{-1}) \cdot \mathcal{Z}_S^{\rm MF}(a^{-1}, a^{-1}),
\end{align}
and present their values for our choice of simulation parameters in Table~\ref{tab:PT}.  These will be used in the following section to extract the physical decay constants and $\Delta B=2$ four-fermion matrix elements via the relations
\begin{align}
A^\text{QCD}_0 &= Z_\Phi(a^{-1}) \left( A^\text{lat}_0(a^{-1}) + c_A^{\rm MF}a A^\text{lat}_{\partial,0}(a^{-1})\right),\\
O_L^\text{QCD}(\mu_b) &= Z_{VA}(\mu_b,a^{-1})\, O_L^\text{lat}(a^{-1}) + Z_\text{SP}(\mu_b,a^{-1})\, O_S^\text{lat}(a^{-1}).
\end{align}

In practice, the renormalization factor $Z_\Phi$ cancels in the ratio $f_{B_s}/f_{B_d}$, and only the quantity $Z_{SP}/Z_{VA}$ enters the ratio $\xi$.  Therefore we do not need $Z_\Phi$ (or $Z_{VA}$ and $Z_{SP}$ by themselves) for our current analysis of the $SU(3)$-breaking ratios.  We present all three matching coefficients for completeness, however, because they will be necessary for calculating the individual decay constants and four-fermion operator-mixing matrix elements in future work.

\begin{CTable}
\caption{Perturbative matching factors for the decay constants and mixing matrix elements evaluated
for APE smeared and HYP smeared static-quark gauge links for the choices of the strong coupling constant $\alpha^\textrm{MF}_s$.  \medskip }
\label{tab:PT}

\begin{tabular}{ccccc}
\hline\hline
 smearing &  $Z_\Phi$  & \ $Z_{VA}$ & \ $Z_{SP}$ \\ \hline
APE & 0.9507 & 0.7485 & -0.1448 \\
HYP & 0.9813 & 0.8108 & -0.1448 \\
\hline\hline
\end{tabular}\end{CTable}

%=================================================
\section{Lattice calculation of $SU(3)$ breaking ratios}
\label{sec:BreakingRatios}
%=================================================

In this section we calculate the ratios of the $B$-meson decay constants and mixing matrix elements at unphysical values of the light and strange quark masses.  On each sea quark ensemble, we compute the necessary 2-point and 3-point correlation functions at two values of the valence quark mass:  the unitary point $m_x = m_l$ and a point tuned to the physical strange quark mass $a m_{s} = 0.0359$~\cite{Allton:2007hx}.  We also use two different link smearings (APE and HYP) to improve the static heavy quark action in order to help estimate discretization effects.  Table \ref{tab:SimData} presents the parameters chosen for our matrix element computations.  In the first subsection we calculate the ratio of $B$-meson decay constants and in the second we calculate the ratio of $\Delta B=2$ mixing matrix elements.

\begin{CTable}
\caption{Parameters used in our simulations.  The columns from left to right are  the light and (approximately) strange sea quark masses, the light and strange valence quark masses, and the number of configurations analyzed for both of our setups using APE or HYP link smearing in the static quark gauge links. \medskip}
\label{tab:SimData}

\begin{tabular}{lccr}
 \hline\hline
  & & \multicolumn{2}{c}{$\#$ configs.}\\[-1.7mm] 
  \ $am_l/am_h$ & \ $am_x$ & APE & HYP   \\
  \hline
 \ 0.01/0.04 & \ 0.01, 0.0359 & 298 & 300\\
 \ 0.02/0.04 & \ 0.02, 0.0359 & 298 & 300\\
 \ 0.03/0.04 & \ 0.03, 0.0359 & 298 & 300\\ 
 \hline\hline
\end{tabular}\end{CTable}

%=================================================
\subsection{Calculation of the ratio of $B$-meson decay constants}
%=================================================

In QCD the decay constant $f_{B_q}$ for the $B_q$-meson is defined by the vacuum-to-meson matrix element
\beq 
\label{eq:fB_def}
	\langle 0 | \bar{b} \gamma_\mu \gamma_5 q | B_q(p) \rangle = i f_{B_q} p_\mu.
\eeq
Because the decay constant $f_{B_q}$ behaves as $1/\sqrt{m_{B_q}}$ in the limit of large $B_q$-meson mass, we calculate the combined decay amplitude
\beq
	\Phi_{B_q} = f_{B_q} \sqrt{m_{B_q}},
\eeq
where $m_{B_q}$ is the physical mass of the $B_q$-meson.  We determine  the quantity $\Phi_{B_q}$ by computing two-point correlation functions of the static-light axial current $A_\mu^{(\pm)\textrm{stat}} = \bar h^{(\pm)} \gamma_\mu\gamma_5 q$.\footnote{For the case of static $b$-quarks, we can relate the $B$-meson interpolating operator $\bar h^{(-)} \gamma_5 q$ to the axial current operator in the temporal direction using the relation $h^{(\pm)}(x) \gamma_0 = \pm h^{(\pm)}(x)$, and thereby express all correlation functions entirely in terms of the axial current.}  

In practice, we use Coulomb gauge-fixed wall sources for the $b$-quark to calculate the local-wall ($LW$) and wall-wall ($WW$) correlation functions
\begin{align}
\label{eq:two_lw}
	\mathcal C^{LW}(t, t_0) & = a^3 \sum_{\vec x \in V} \langle0|A^{L}_0(\vec x,t)A^{W}_0(t_0)^\dagger|0\rangle , \\
\label{eq:two_ww}
	\mathcal C^{WW}(t, t_0) & = \langle0|A^{W}_0(t)A^{W}_0(t_0)^{\dagger}|0\rangle , 
\end{align}
with the local ($L$) and wall-source ($W$) axial currents given by
\begin{align}
	A^{L}_{0}(\vec x, t) &= \bar{h}^{(+)}(\vec x, t)\gamma_{0}\gamma_{5}q(\vec x, t) + \bar{h}^{(-)}(\vec x, t)\gamma_{0}\gamma_{5}q(\vec x, t) , \\
	A^{W}_{0}(t) &= a^6 \sum_{\vec y \in V} \sum_{\vec z \in V} \left( \bar{h}^{(+)}(\vec y,t)\gamma_0 \gamma_5 q(\vec z,t)+ \bar{h}^{(-)}(\vec y,t)\gamma_0 \gamma_5 q(\vec z,t)\right). 
\end{align}
From the ratio of ${\mathcal C}^{LW}$ to ${\mathcal C}^{WW}$ we obtain the combined decay amplitude
\begin{equation}
	\Phi_{B_q}^{\rm lat} = \lim_{ t \gg t_0} \sqrt{\frac{2}{L^3}} \frac{| {\cal C}^{LW}(t, t_0) |}{\sqrt{{\cal C}^{WW}(t, t_0)\text{e}^{-m_{B_q}^*(t-t_0)}}}, \label{eq:PhiUnRen}
\end{equation}
where we determine the unphysical $B$-meson rest mass $m_{B_q}^*$ via
\beq
	a m_{B_q}^* = \lim_{ t \gg t_0} \textrm{log} \left( \frac{\calC^{LW}(t, t_0)}{\calC^{LW}(t+a, t_0)} \right).
\eeq
A derivation of Eq.~(\ref{eq:PhiUnRen}) is presented in Appendix~\ref{app:Norman}.  Finally we compute the renormalized decay amplitude 
\beq
	\Phi_{B_q}^\mathrm{ren} = Z_{\Phi} \Big[ 1 + c_A^{\rm{MF}} \sinh\big(  a m_{B_q}^* \big) \Big] \Phi^{\rm lat}_{B_q}, \label{eq:PhiRen}
\eeq
using the perturbative matching factors given in Tables~\ref{TAB:factor} and~\ref{tab:PT}.  The contribution proportional to $c_A^{\rm{MF}}$ improves the heavy-light axial current operator  through $\CO(\alpha_s p a)$, where the $\sinh$ arises from the symmetric derivative in the $\CO(pa)$ operator.   The overall multiplicative factor $Z_{\Phi}$ is needed to obtain the combined decay amplitude in the continuum.

The computation of the statistical errors throughout this paper follows the prescription for numerically computing the autocorrelation function as proposed in reference \cite{Wolff:2003sm}. The autocorrelation function quantifies the degree of correlation between two measurements made at different trajectories, and depends upon the observable of interest.  By summing the autocorrelation function over the separation between measurements, one obtains the integrated autocorrelation time. We obtain a better estimate of the true statistical error by inflating the variance of the measured Monte Carlo data using the integrated autocorrelation time.  In many instances we must compute the errors in a quantity which itself depends on several lattice correlators;  we refer to this as a derived quantity and refer to the lattice correlators as primary observables.  When calculating the errors in a derived quantity, we account for the correlations between primary observables by using the functional dependence of the derived quantity on the primary observables.  As a cross-check of the statistical error estimate, we compared the results obtained with this approach with those obtained using a single-elimination jackknife procedure; we find that both the central values and statistical errors are consistent between the two error estimation methods.

\begin{figure}[hpt]
\begin{picture}(160,140)
\put(-3,70){\includegraphics[clip,scale=0.44]{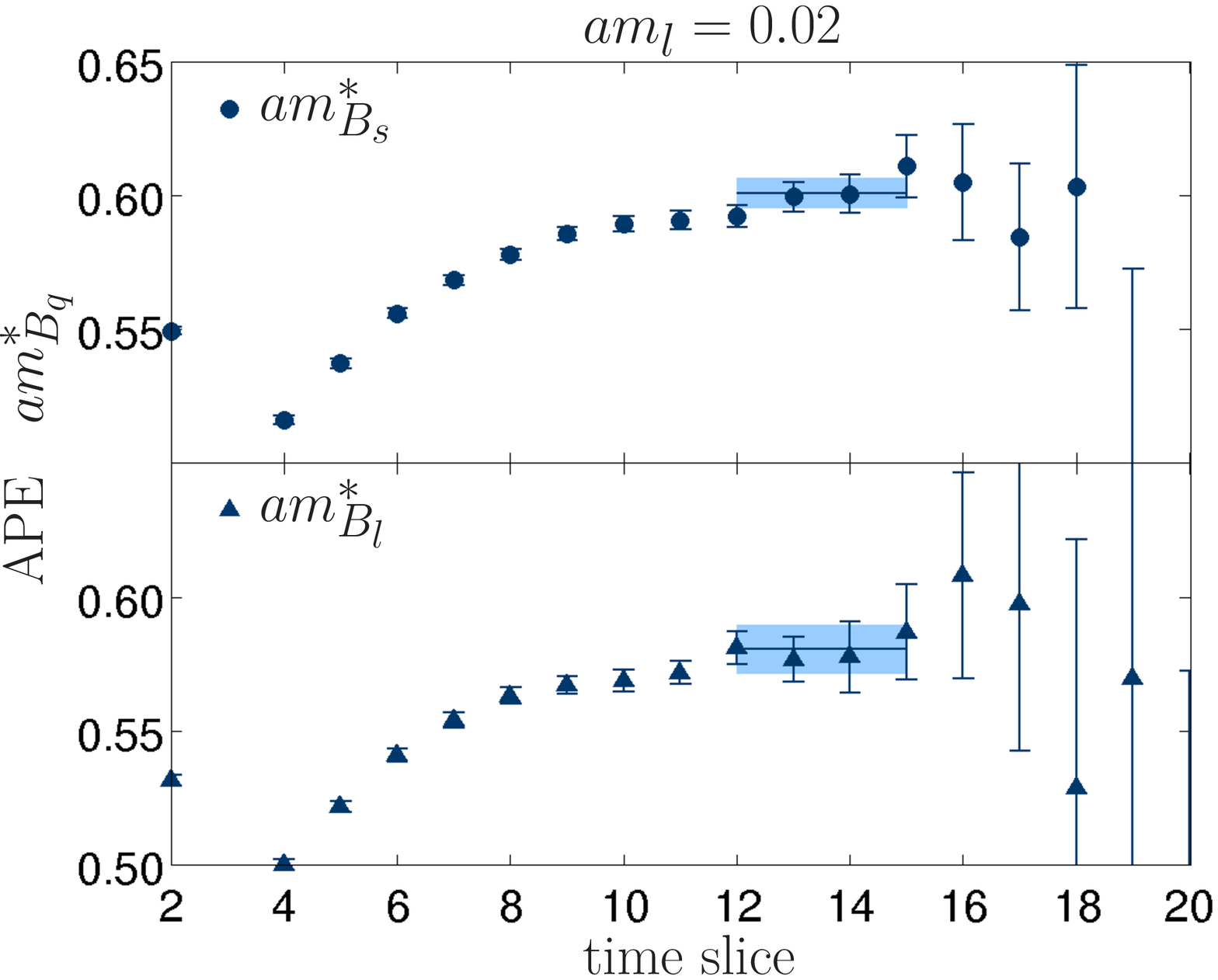}}
\put(81,70){\includegraphics[clip,scale=0.44]{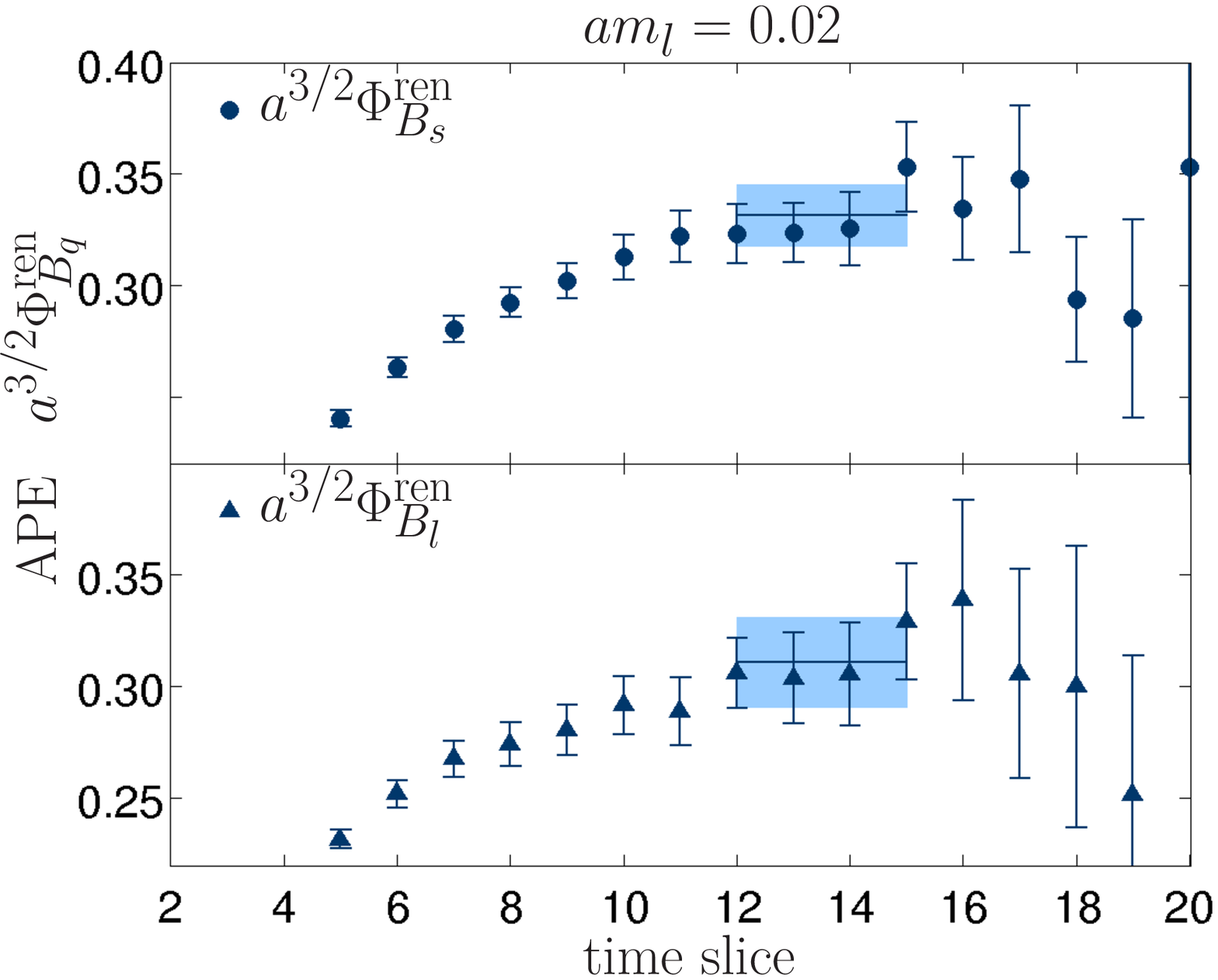}}
\put(-3,0){\includegraphics[clip,scale=0.44]{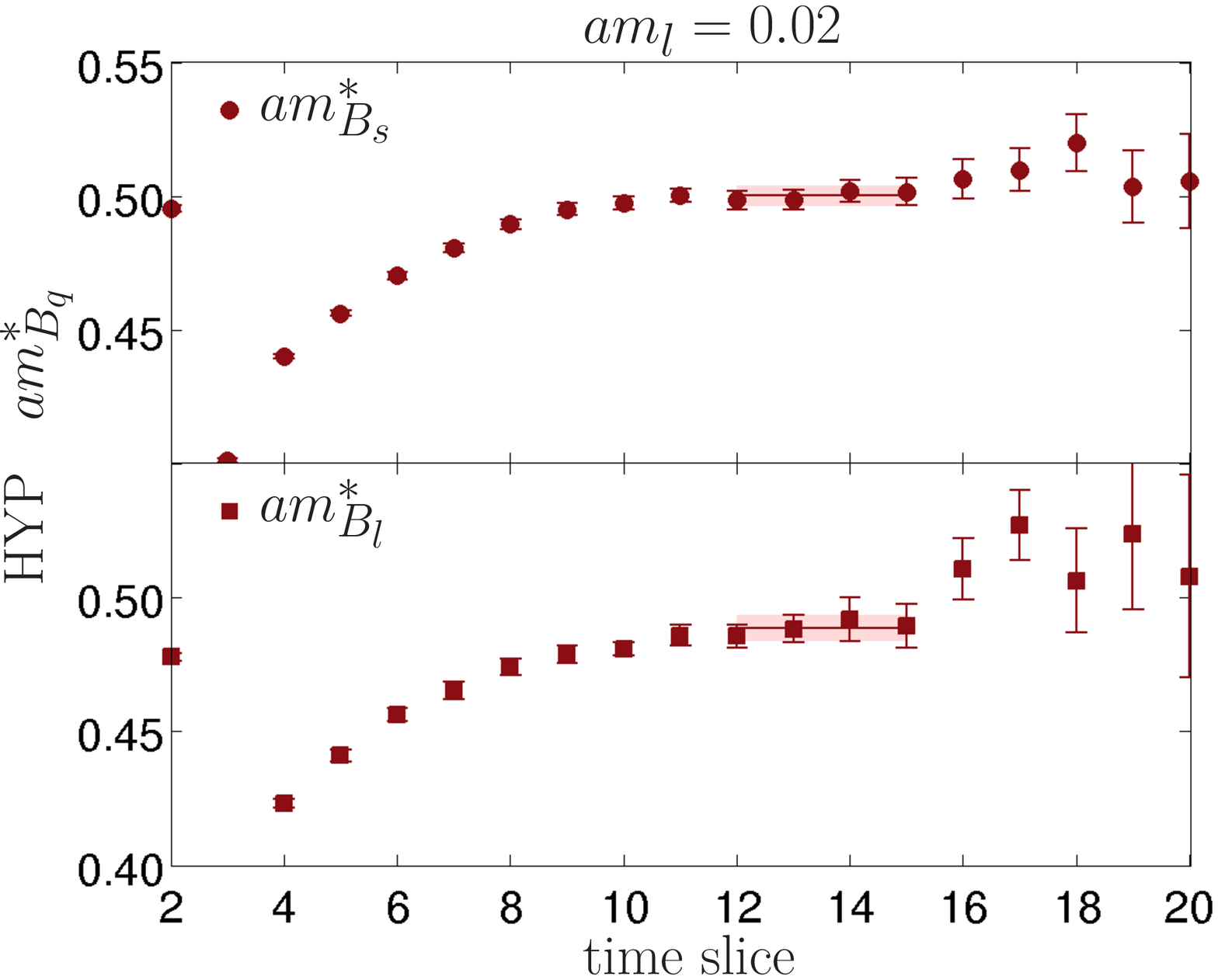}}
\put(81,0){\includegraphics[clip,scale=0.44]{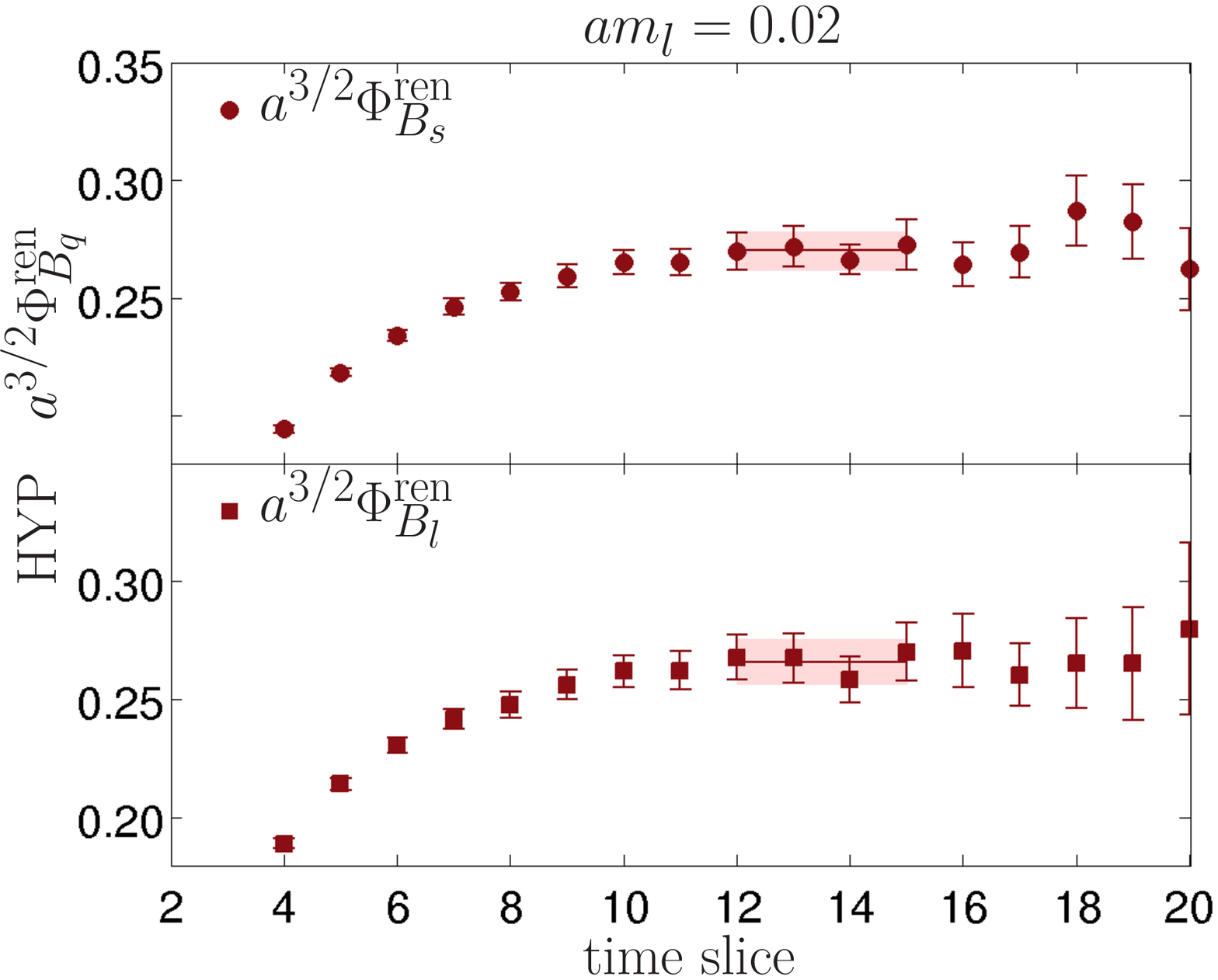}}
\end{picture}
\caption{Determination of $m_{B_q}^*$ (left) and $\Phi_{B_{q}}^\textrm{ren}$ (right) using the average of two time sources on the $am_l = 0.02$ ensemble.  The upper plots show the APE data, while the lower plots show the HYP-smeared data.  For each panel, the shaded band corresponds to the plateau extracted from averaging the data over four consecutive time slices.  Errors shown are statistical only.}
\label{fig:mBstar_Phi_plots}
\end{figure}

For example, Fig.~\ref{fig:mBstar_Phi_plots} shows the determinations of $m_{B_q}^*$ and $\Phi_{B_q}^\mathrm{ren}$ on the $am_l = 0.02$ ensemble for the APE (upper plots) and HYP data sets (lower plots).  The central value and statistical error of each data point in Fig.~\ref{fig:mBstar_Phi_plots} are computed as functions of the primary observables ${\cal C}^{LW}$ and ${\cal C}^{WW}$.  Then the value of the plateau and its error are computed using a function which averages the values of $m_{B_q}^*$ (or $\Phi_{B_q}^\mathrm{ren}$) on time-slices 12, 13, 14 and 15 because we do not observe excited-state contamination in this region.  In order to reduce the size of the statistical errors, we average the correlators beginning at two time sources.  We achieve the averaging of our two sources by replacing e.g.~${\cal C}^{LW}(t,0)$ with $\left[{\cal C}^{LW}(t,0) +{\cal C}^{LW}(20a-t,20a)\right]/2$ in Eqs.~(\ref{eq:PhiUnRen})-(\ref{eq:PhiRen}).  For the case of the HYP-smeared data, these are located at $t/a=0$ and $t/a=20$, whereas for the APE-smeared data the second source is located either at $t/a = 20$, 21, or 24. 

\begin{figure}[hpt]
\begin{picture}(160,140)
\put(-3,70){\includegraphics[clip,scale=0.45]{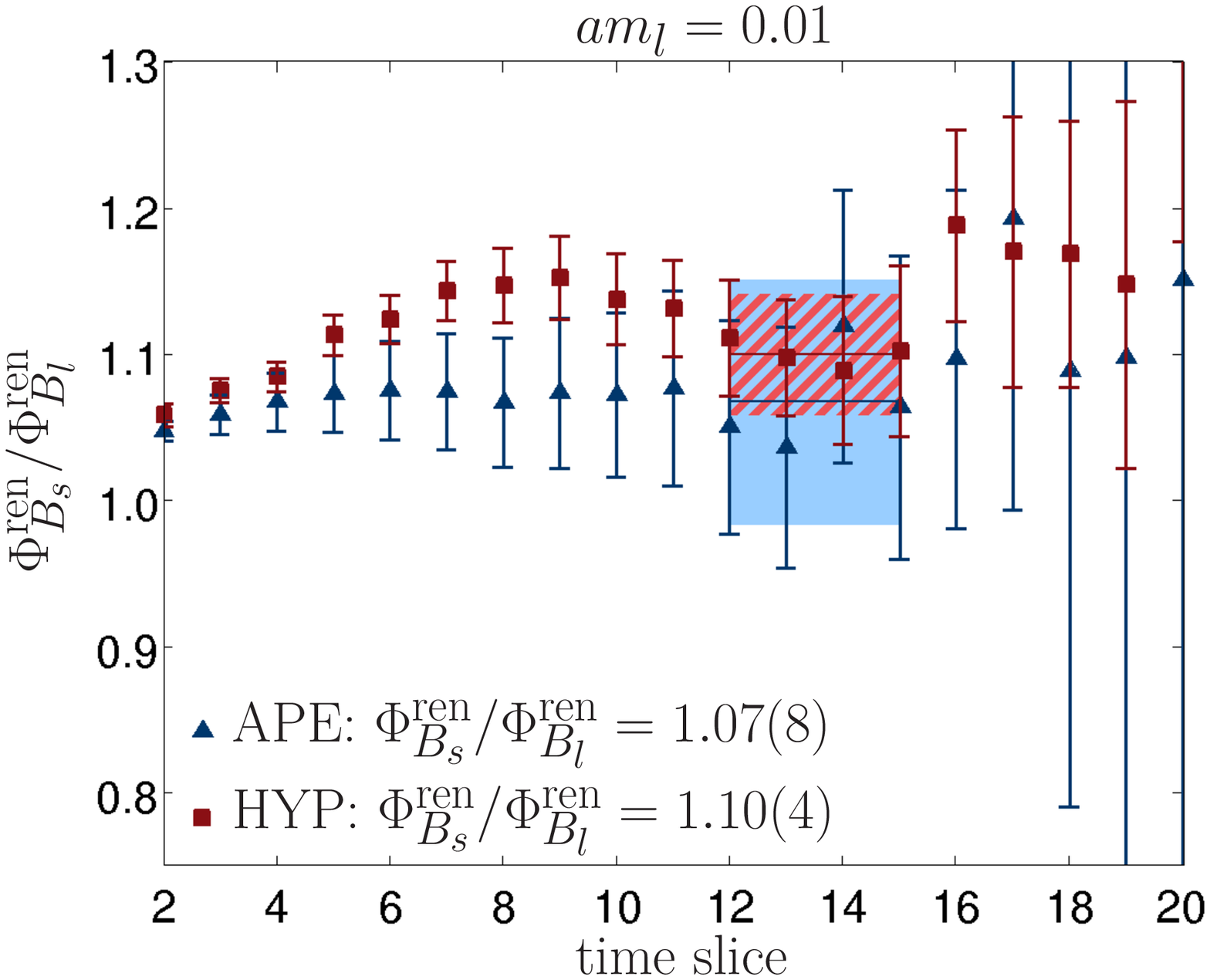}}
\put(83,70){\includegraphics[clip,scale=0.45]{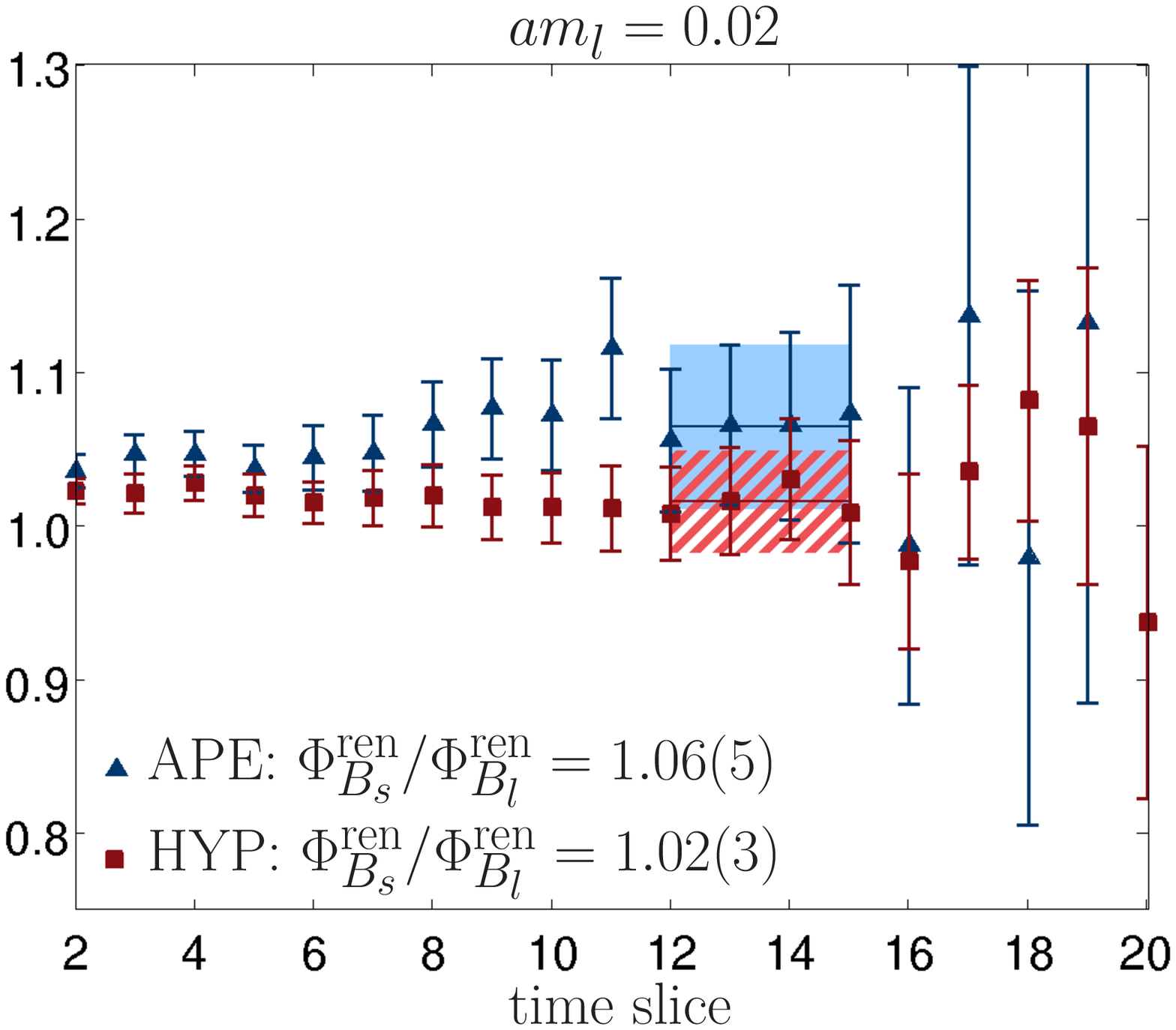}}
\put(43,0){\includegraphics[clip,scale=0.45]{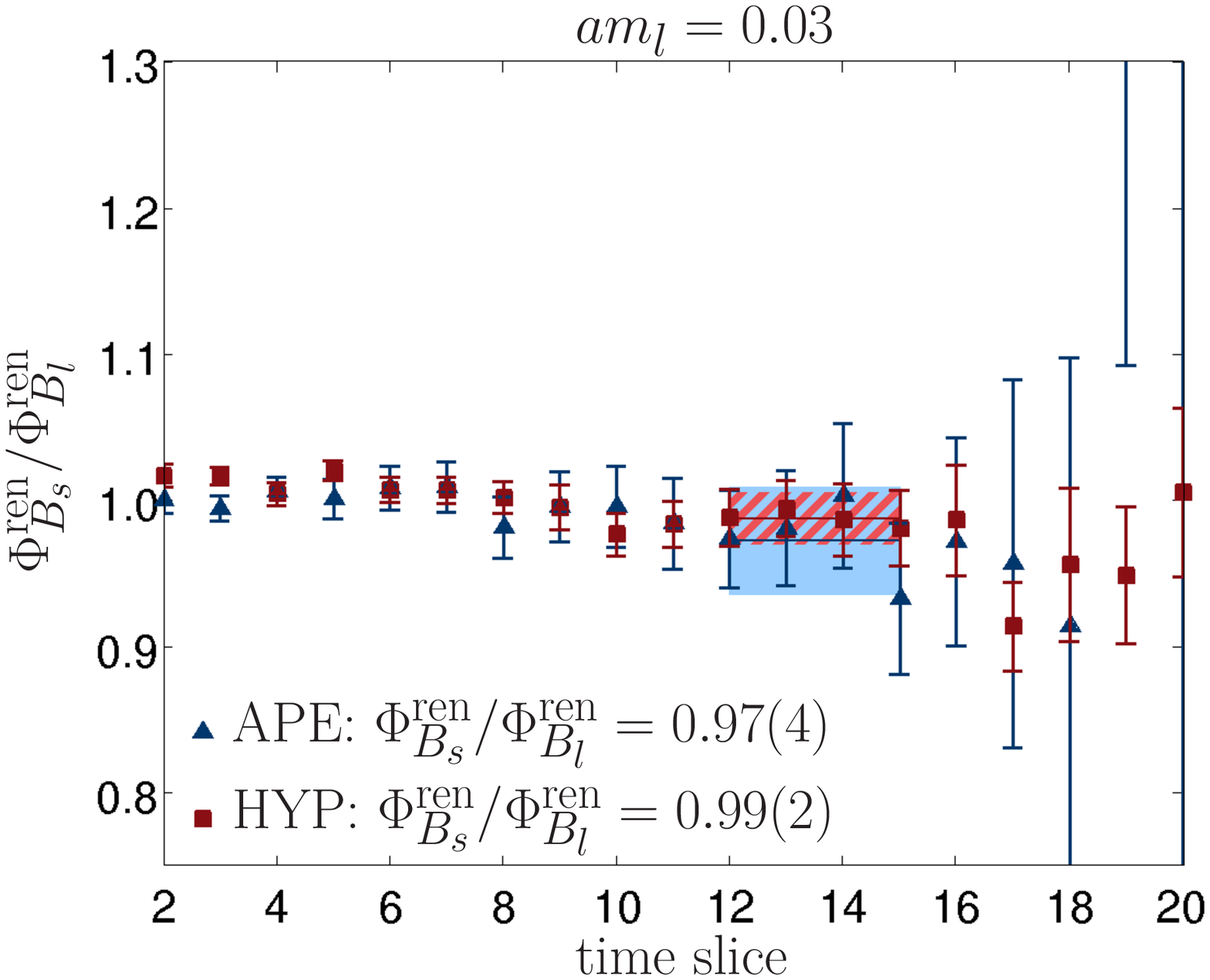}}
\end{picture}
\caption{Determination of $\Phi_{B_{s}}^\textrm{ren}/\Phi_{B_l}^\textrm{ren}$ using the average of two time sources on the three sea quark ensembles.  The blue (triangle) points denote the APE data, while the red (square) points denote the HYP-smeared data.  The shaded (hatched) band corresponds to the plateau extracted from averaging the APE (HYP) data over four consecutive time slices.  Errors shown are statistical only.}
\label{fig:fDs_fD_plots}
\end{figure}

Finally, we compute the ratio $\Phi_{B_{s}}^\textrm{ren} / \Phi_{B_{l}}^\textrm{ren}$ on each ensemble as a function of $\Phi_{B_{s}}^\textrm{ren}$ and $\Phi_{B_{l}}^\textrm{ren}$, which themselves depend on the corresponding primary observables ${\cal C}^{LW}$ and ${\cal C}^{WW}$; this is shown in Fig.~\ref{fig:fDs_fD_plots}.  We obtain the plateau for the ratio from time-slices 12--15, where we do not observe excited-state contamination in the numerator $\Phi_{B_{s}}^\textrm{ren}$ or the denominator $\Phi_{B_{l}}^\textrm{ren}$.  Table~\ref{tab:fDs_fD_results} presents the values of $\Phi_{B_{s}}^\mathrm{ren}/\Phi_{B_l}^\mathrm{ren}$ on the three sea-quark ensembles.   Despite the use of two time sources, the statistical errors in the ratio are as large as $7.5\%$ in case of the data using APE smearing.

\begin{CTable}
\caption{The renormalized decay amplitude ratio $\Phi_{B_{s}}^\textrm{ren}/\Phi_{B_{l}}^\textrm{ren}$ for both the APE- and HYP-smeared data sets.  Errors shown are statistical only. \medskip}
\label{tab:fDs_fD_results}

\begin{tabular}{ccc} \hline\hline
 & \multicolumn{2}{c}{\ \ $\Phi_{B_{s}}^\textrm{ren}/\Phi_{B_{l}}^\textrm{ren}$} \\[-1.5mm] 
 $am_l/am_h$ & \ APE & \ HYP \\
 \hline
  0.01/0.04 & \ 1.07(8) & \ 1.10(4)  \\	
  0.02/0.04 & \ 1.06(5) & \ 1.02(3)  \\	
  0.03/0.04 & \ 0.97(4) & \ 0.99(2)  \\	
 \hline\hline  
\end{tabular}\end{CTable}

%=================================================
\subsection{Calculation of the ratio of $B$-meson mixing matrix elements}
%=================================================
%
The $B_q-\bar B_q$ mixing parameter in continuum QCD is defined in terms of the matrix element of the $\Delta B=2$ four-fermion operator via Eqs.~(\ref{eq:me}) and (\ref{eq:pme}).  Because the matrix element $\cM_q$ behaves as $m_{B_q}$ in the limit of large $B_q$-meson mass, we calculate the desired matrix element divided by the $B_q$-meson mass:
\beq
	M_{B_q} = \cM_q / m_{B_q} .
\eeq

For the determination of $M_{B_q}$, we use different spatial sources for the two choices of link smearing:  in the case of APE-smeared links we use box sources of size $8^3$, while for HYP smearing we use wall sources.  The local-wall two-point functions are already defined in Eq.~(\ref{eq:two_lw}), while the box-box two-point functions are given by 
\begin{align}
\label{eq:two_bb}
	\mathcal C^{BB}(t, t_0) &= \langle 0|A_0^{B}(t) A_0^{B}(t_0)^\dagger |0\rangle,
\end{align}
with
\begin{align}
A_0^{B}(t) &= a^6 \sum_{\vec x, \vec y \in \Delta V}\left( \bar h^{(+)}(\vec x, t) \gamma_0 \gamma_5 q(\vec y, t)+ \bar h^{(-)}(\vec x, t) \gamma_0 \gamma_5 q(\vec y, t)\right),
\end{align}
where the superscript $B$ denotes a box source in the region $\Delta V$.
We also compute the three-point correlation functions 
\begin{equation}\label{eq:three_bb}
	\mathcal C^I_{O_{i}}(t_f,t,t_0) = a^3 \sum_{\vec x \in V}\langle  0|{A}^{(+) I}_{0}(t_f)^\dagger O^{\rm lat}_{i}(\vec x,t)A^{(-) I}_{0}(t_0)^{\dagger}|0\rangle ,
\end{equation}
where $A^{(\pm) I}_0(t)$ is either the box-source axial current $A_0^{(\pm) B}(t)$ or  the wall-source axial current $A_0^{(\pm) W}(t)$.  The operator $O_i$ can be either the sum of the squared vector plus squared axial vector current or the squared scalar plus squared pseudoscalar current, respectively,\footnote{We neglect the odd-parity parts of $O_L^{\rm lat}$ and $O_S^{\rm lat}$, Eqs.~(\ref{eq:OL_lat}) and~(\ref{eq:OS_lat}), when computing the lattice three-point correlation functions because only the parity-conserving components contribute to the desired matrix element.}
\begin{align}
O^\textrm{lat}_{VV+AA} &= 2\left(\bar h^{(+)} \gamma^\mu q \right)\left(\bar h^{(-)} \gamma_\mu q\right)+ 2\left(\bar h^{(+)} \gamma^\mu \gamma_5 q\right)\left(\bar h^{(-)} \gamma^\mu \gamma_5 q\right),\\
O^\textrm{lat}_{SS+PP} &= 2\left(\bar h^{(+)} q\right)\left(\bar h^{(-)}q\right)+ 2\left(\bar h^{(+)} \gamma_5 q\right)\left(\bar h^{(-)} \gamma_5 q\right).
\end{align}
Although the $SS+PP$ operator does not contribute to $B_q-\bar B_q$ mixing in continuum QCD,  its counterpart in
HQET is introduced through the QCD $\to$ HQET matching as shown in Eq.~(\ref{eq:OL_QCD_to_HQET}).

Because of the different spatial wavefunctions used for the APE- and HYP-smeared data, we extract the $B_q-\bar B_q$ matrix element in different ways for the two data sets.  With the APE-smeared box source data, we can compute the matrix element directly from the ratio of correlators \cite{Christ:2007cn}
\begin{align}
\label{eq:box_Mq}
M_{\CO_{i}} &= \lim_{ t_f \gg t \gg t_0} 2 \frac{C^B_{O_{i}}(t_f,t,t_0) e^{m_{B_q}^*( t_f - t_0) /2} }{ \sqrt{C^{BB}(t, t_f) C^{BB}(t, t_0)} }. 
\intertext{For the HYP-smeared wall-source data, however, we must first extract the bag parameter $B_{B_q}$}
\label{eq:wall_Bq}
B_{\CO_{i}} &= \lim_{ t_f \gg t \gg t_0} \frac{3}{8} {L^3}  \frac{C^W_{O_{i}}(t_f,t,t_0)}{ C^{LW}(t, t_f) C^{LW}(t, t_0) },
\end{align}
then we obtain the matrix element via
\begin{equation}
M_{\CO_{i}} = \frac{8}{3} B_{\CO_{i}} {(\Phi_{B_q}^\textrm{ren})}^{2} .
\end{equation}
The derivation Eq.~(\ref{eq:box_Mq}) is more complex than that 
of either Eq.~(\ref{eq:PhiUnRen}) or Eq.~(\ref{eq:wall_Bq}) because the box sources in the
amplitudes appearing in both the numerator and denominator
are not translationally invariant and will create $B$-meson
states carrying a variety of spatial momenta.  As explained
in greater detail in App.~\ref{app:Norman}, these states are degenerate
with the lowest energy $B$-meson state and so cannot be suppressed
by simply going to large Euclidean time separations.  Instead,
Eq.~(\ref{eq:box_Mq}) is derived in App.~\ref{app:Norman} using the local conservation
of heavy-quark number, a property special to the static
approximation.  Finally, for both the APE and HYP-smeared data, we compute the renormalized matrix element through $\CO(\alpha_s)$,
\beq
M_{B_q}^{\textrm{ren}} = Z_{VA} M_{VV+AA}^{\textrm{lat}} + Z_{SP} M_{SS+PP}^{\textrm{lat}},
\label{eq:Mb}
\eeq
using the perturbative matching factors given in Table~\ref{tab:PT}.

\begin{figure}[hpt]
\begin{picture}(160,70)
\put(-3,0){\includegraphics[clip,scale=0.44]{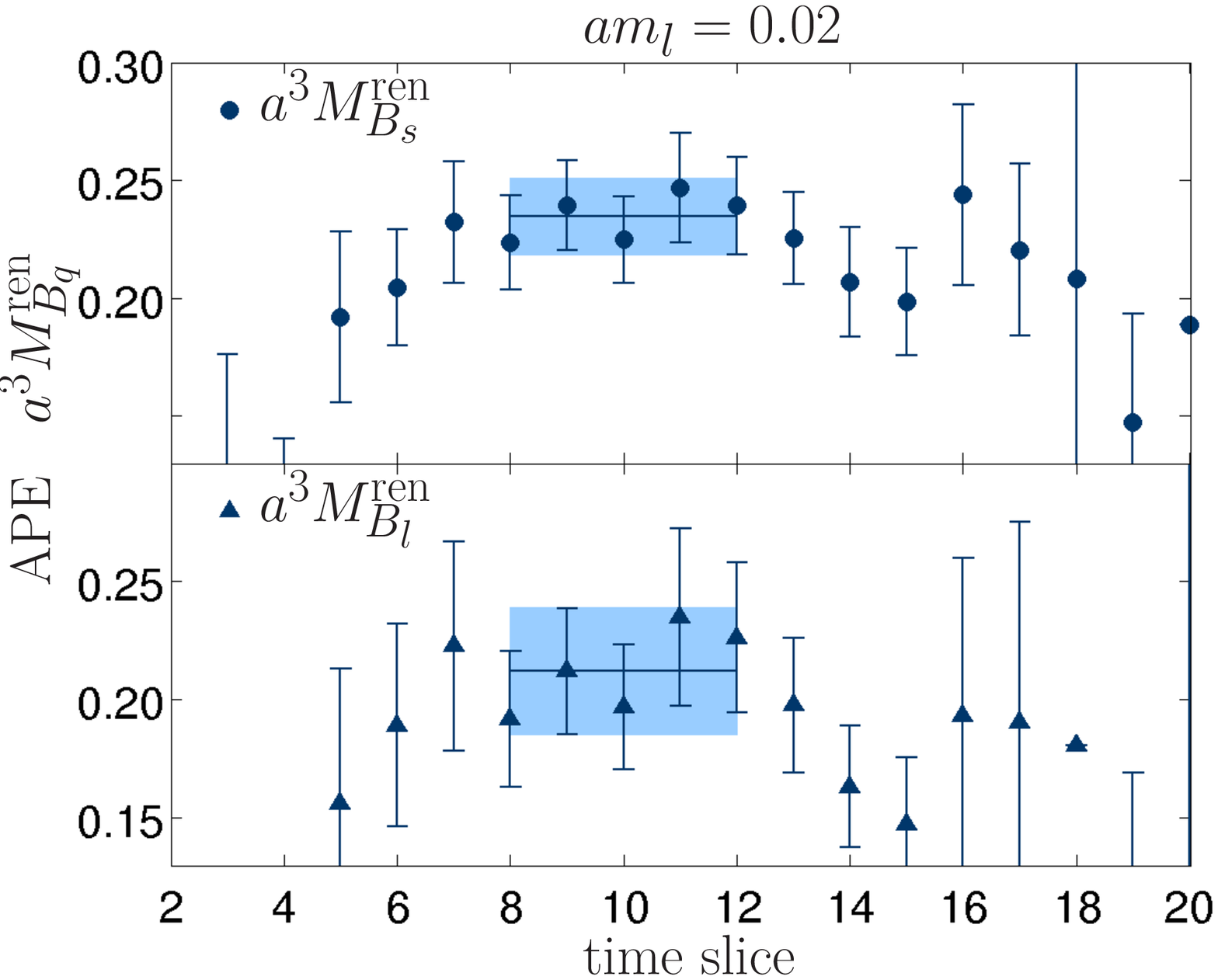}}
\put(81,0){\includegraphics[clip,scale=0.44]{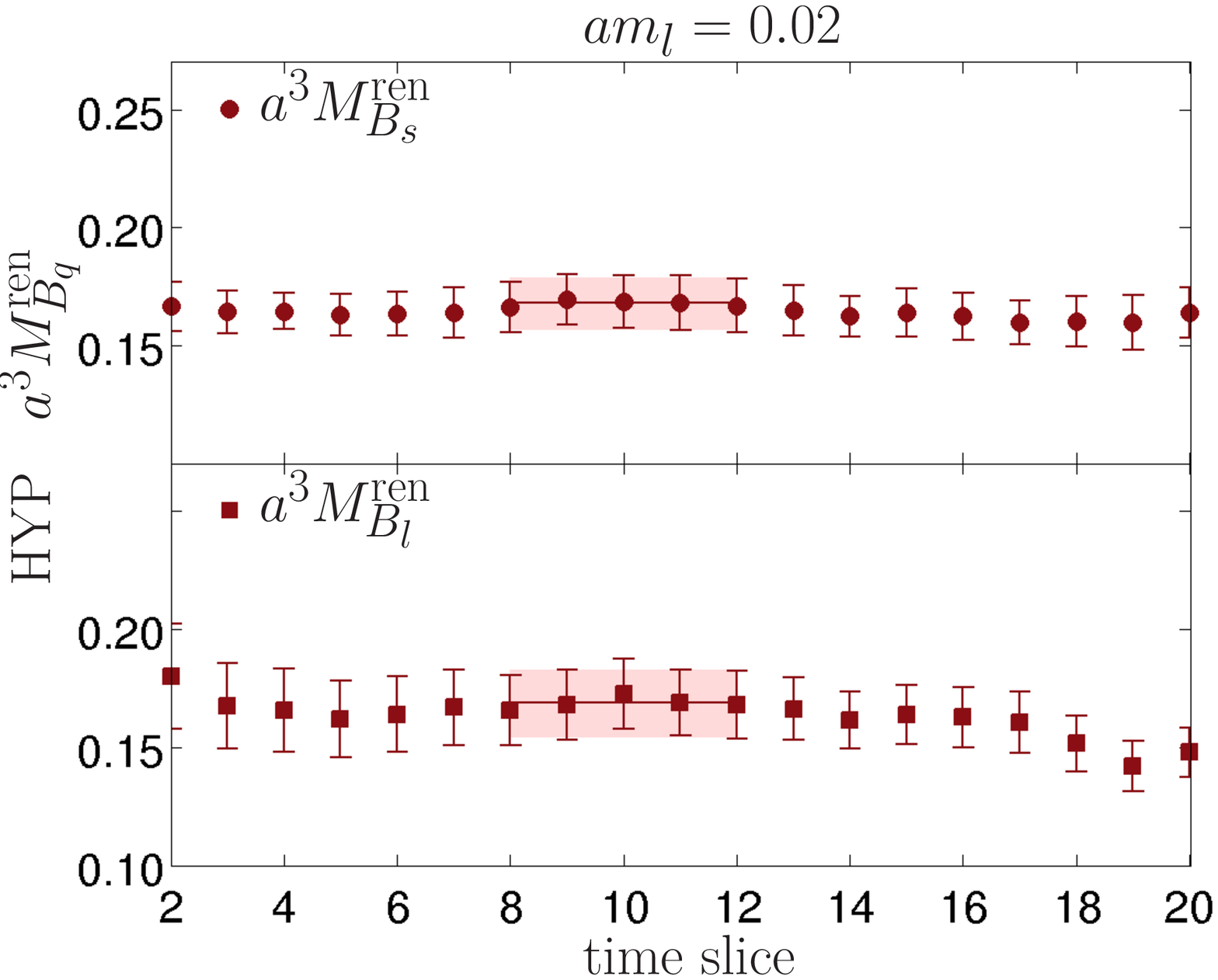}}
\end{picture}
\caption{Determination of $M_{B_{s}}^\textrm{ren}$ and $M_{B_{l}}^\textrm{ren}$ on the $am_l = 0.02$ ensemble.  The left-hand plot shows the APE data, while the right-hand plot shows the HYP-smeared data.  For each panel, the shaded band corresponds to the plateau extracted from averaging the data over four consecutive time slices.  Errors shown are statistical only.}
\label{fig:Mren_plots}
\end{figure}

\begin{figure}[hpt]
\begin{picture}(160,140)
\put(-3,70){\includegraphics[clip,scale=0.45]{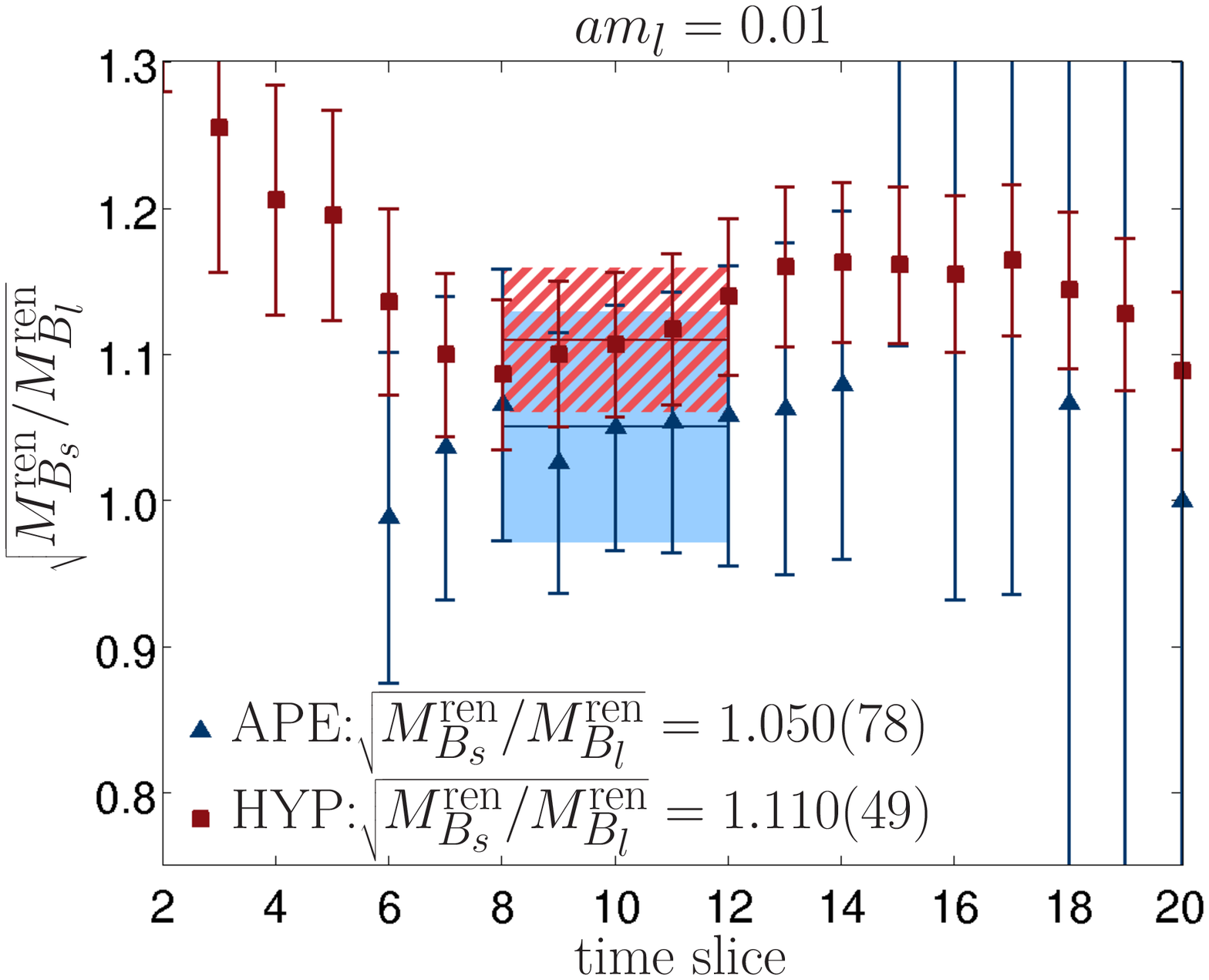}}
\put(83,70){\includegraphics[clip,scale=0.45]{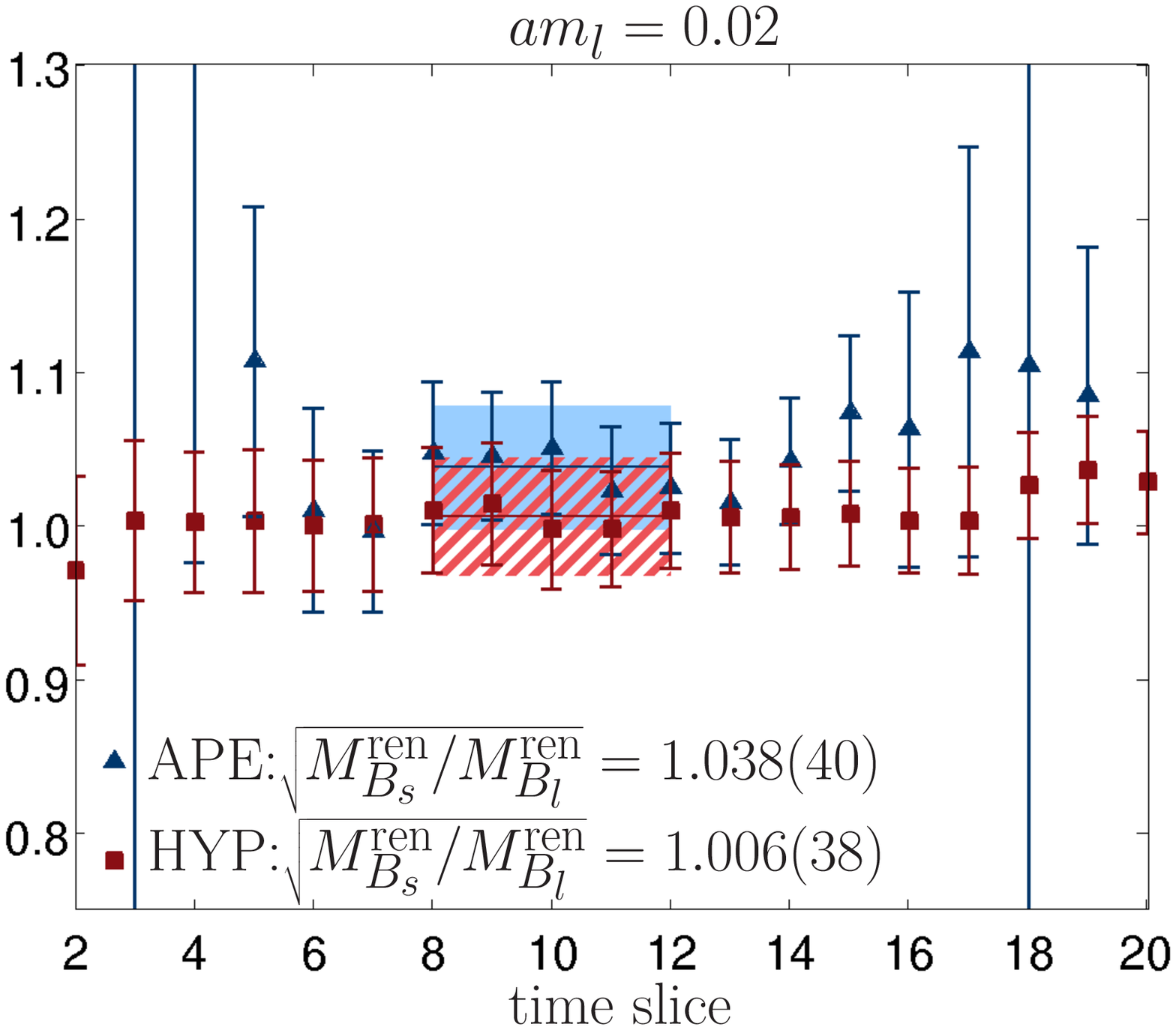}}
\put(43,0){\includegraphics[clip,scale=0.45]{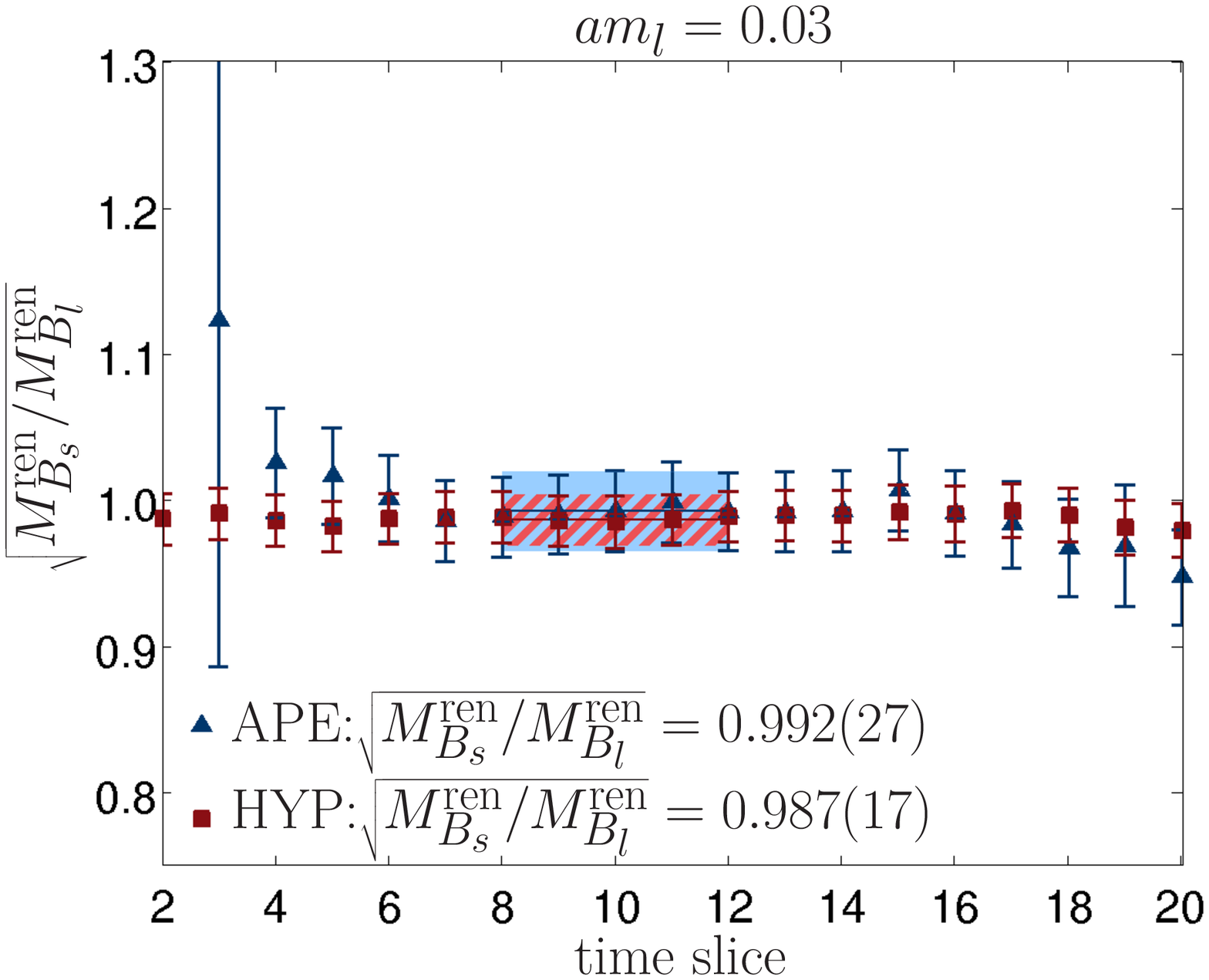}}
\end{picture}
\caption{Determination of $\sqrt{M_{B_{s}}^\textrm{ren}/M_{B_{l}}^\textrm{ren}}$ using the average of two time sources on the three sea quark ensembles.  The blue (triangle) points denote the APE data, while the red (square) points denote the HYP-smeared data.  The shaded (hatched) band corresponds to the plateau extracted from averaging the APE (HYP) data over four consecutive time slices.  Errors shown are statistical only.}
\label{fig:xi_plots}
\end{figure}

We compute the ratio of mixing matrix elements in the same manner as we compute the ratio of decay constants in the previous subsection.  For example, Fig.~\ref{fig:Mren_plots} shows the determination of the numerator $M_{B_{s}}^{\textrm{ren}}$ and the denominator $M_{B_l}^{\textrm{ren}}$ on the $am_l = 0.02$ ensemble for the APE (left-hand plot) and HYP data sets (right-hand plot).  We compute the values of the plateaux using time-slices 8--12 because we do not observe excited-state contamination in this region.  Table~\ref{tab:xi_results} shows the results for the $SU(3)$-breaking ratios $\sqrt{M_{B_{s}}^\textrm{ren}/M_{B_{l}}^\textrm{ren}}$ on the three sea quark ensembles; the corresponding plateau plots are shown in Fig.~\ref{fig:xi_plots}.

\begin{CTable}
\caption{The renormalized $SU(3)$-breaking ratio $\sqrt{M_{B_{s}}^\textrm{ren}/M_{B_{l}}^\textrm{ren}}$ for both the APE and HYP-smeared data sets.  Errors shown are statistical only. \medskip}
\label{tab:xi_results}

\begin{tabular}{ccc}
 \hline\hline
 & \multicolumn{2}{c}{\ \ $\sqrt{M_{B_{s}}^\textrm{ren}/M_{B_{l}}^\textrm{ren}}$} \\[-1.5mm] 
 $am_l/am_h$ & \ APE & \ HYP \\
 \hline
  0.01/0.04 & \ 1.050(78) & \ 1.110(49)  \\	
  0.02/0.04 & \ 1.038(40) & \ 1.006(38)  \\	
  0.03/0.04 & \ 0.992(27) & \ 0.987(17)  \\	
 \hline\hline
\end{tabular}\end{CTable}

%=================================================
\section{Chiral Extrapolation}
\label{sec:ChExp}
%=================================================

We extrapolate our results for the ratio of the decay constants and for the ratio of mixing matrix elements to the physical point using a phenomenologically-motivated function based on next-to-leading order partially quenched $SU(2)$ heavy-light meson chiral perturbation theory.  In $SU(2)$ $\chi$PT, the pesudoscalar mesons containing strange quarks (i.e. kaons and $\eta$'s) are integrated out of the theory.  Thus $SU(2)$ $\chi$PT does not require an expansion in the strange quark mass about the chiral limit, and the $SU(2)$ $\chi$PT expansion parameter in isospin-symmetric simulations is $m_l / \Lambda_\chi$, where $m_l$ is the light up-down sea quark mass and $\Lambda_\chi$ is a typical hadronic scale.  This improves the convergence of the chiral expansion relative to $SU(3)$ $\chi$PT, as long as $m_l$ is sufficiently light that corrections of $\CO(m_l/m_s)$ are small.  Studies by the RBC and UKQCD Collaborations, the PACS-CS Collaboration, and the MILC Collaboration confirm this picture and show that, for light pseudoscalar meson masses and decay constants, $SU(2)$ $\chi$PT within its applicable region converges more quickly than $SU(3)$ $\chi$PT~\cite{Allton:2008pn,Aoki:2008sm,Kadoh:2008sq,Bazavov:2009ir}.  For the case of $SU(3)$-breaking ratios such as $f_{B_s}/f_{B_d}$ and $\xi$, however, $SU(3)$ HM$\chi$PT has the advantage that the chiral extrapolation formulae manifestly preserve the fact that the ratios must be equal to one in the limit $m_l \to m_s$.  Within the framework of $SU(2)$ HM$\chi$PT, this fact must be introduced in a more {\it ad hoc} manner such as by matching the $SU(2)$ HM$\chi$PT expression at small quark masses onto an analytic form at large quark masses that becomes one when $m_l \to m_s$.  We therefore plan to perform a study comparing the use of $SU(2)$ versus $SU(3)$ HM$\chi$PT for the extrapolation of $B$-meson decay constants and mixing matrix elements to the physical quark masses in a future analysis when we have lighter data with smaller statistical errors in order to see which procedure leads to a more accurate determination of these quantities.

Although we know that HM$\chi$PT is the correct low-energy effective description of the lattice theory when the simulated quark masses are sufficiently light, we do not know {\it a priori} at what mass the range of validity of HM$\chi$PT ends.  Studies of the light pseudoscalar meson sector on the RBC/UKQCD domain-wall ensembles show that NLO $\chi$PT does not describe the numerical data for the masses and decay constants when the pion masses are above about 420~MeV~\cite{Allton:2008pn}.  Since the lightest pion mass in our analysis is approximately 430~MeV, this suggests that most of our data may be too heavy for NLO HM$\chi$PT to apply and that the inclusion of NNLO terms may be necessary.  Unfortunately, because we only have three data points for each of the $SU(3)$-breaking ratios, we do not have enough data points to reliably constrain the values of the higher-order terms (there are two free parameters at NNLO in the $SU(3)$ HM$\chi$PT expressions, and even more at NNLO in SU(2) HM$\chi$PT).  When the masses of the light pseudoscalar mesons in the chiral logarithms are sufficiently heavy, however, the logarithms can be well-approximated by polynomials.  We therefore choose to fit the data to a linear fit function that is constrained to be equal to one in the $SU(3)$ limit ($m_l \to m_{s}$) and then match this result onto the NLO $SU(2)$ HM$\chi$PT expression at the location of our lightest data point ($am_l = 0.01$), which we believe is sufficiently light that NLO $SU(2)$ HM$\chi$PT should apply.  Although the behavior of the data is unlikely to be strictly linear in the heavy-mass region, the statistical errors in our data points are sufficiently large (as great as $\sim 8\%$) that we can successfully perform a linear fit and obtain a good $\chi^2/\rm{dof}$ without the addition of higher-order polynomial terms.\footnote{Because the data points for the $SU(3)$-breaking ratios in Tables~\ref{tab:fDs_fD_results} and~\ref{tab:xi_results} were generated on three different sea-quark ensembles, they are statistically independent;  thus the uncorrelated $\chi^2/\rm{dof}$ correctly reflects the goodness-of-fit.}  Furthermore, we cannot reliably determine the size of a quadratic term if we include one in the fit.  Given our poor statistical errors, however, we cannot exclude the possibility of other fit functions, and we use alternate fit forms as one way to estimate the chiral extrapolation error.  We also vary the location of the $SU(2)$ HM$\chi$PT matching point and the parameters that enter the $SU(2)$ HM$\chi$PT expressions, and consider matching onto $SU(3)$ HM$\chi$PT.  All of these variations are discussed in greater detail when we estimate the systematic uncertainty due to the chiral extrapolation in Sec.~\ref{sec:ChPT_err}.

In the case of the decay constants, we extrapolate the ratio 
\beq
	\frac{\Phi_{B_{s}}^\textrm{ren}}{\Phi_{B_l}^\textrm{ren}} = \sqrt{\frac{m_{B_{s}}}{m_{B_l}}}\, \frac{f_{B_{s}}}{f_{B_l}},
\eeq
while for the mixing matrix elements we consider
\beq
	\sqrt{\frac{M^\textrm{ren}_{B_{s}}}{M^\textrm{ren}_{B_l}}} = \sqrt{\frac{m_{B_{s}}}{m_{B_l}}}\, \left( \frac{f_{B_{s}}\sqrt{B_{B_{s}}}}{f_{B_l}\sqrt{B_{B_l}}}\right).
\eeq
From here on, we drop the superscript ``ren'' for simplicity because we only refer to the renormalized quantities.  The expressions for the heavy-light meson decay constant and mixing matrix element to NLO in the light-quark expansion, but zeroth order in $1/m_b$, are given in Appendix~\ref{app:ChPT}.  For completeness, we present formulae for both $SU(3)$ and $SU(2)$ HM$\chi$PT.  The $SU(3)$ fit functions depend on the valence and sea light-quark masses ($m_{s}, m_l, m_h$) and the lattice spacing $a$.  The $SU(2)$ functions are obtained from the $SU(3)$ expressions by taking the limit $\left( \frac{m_l}{m_{s}}, \frac{m_l}{m_h}\right) \ll 1$.  Thus they only apply in the region in which the value of the average up-down quark mass is much smaller than the valence and sea strange quark masses.  Furthermore, because the strange quark has been integrated out of the $SU(2)$ theory, the $SU(2)$ fit functions only depend upon the light-quark mass $m_l$ and the lattice spacing $a$.  Therefore the expressions for the $SU(3)$-breaking ratios at NLO in $SU(2)$ HM$\chi$PT are particularly simple:
\begin{eqnarray}
	\frac{\Phi_{B_{s}}}{\Phi_{B_l}} & =  & R_\Phi \Bigg\{ 1 + \frac{ 1+3 g_{B^*B\pi}^2 }{\left(4\pi f \right)^2}  \bigg( \frac{ 3 }{4} \bigg) m_L^2 \ln\left(\frac{m_L^2}{\Lambda_\chi^2}\right)  + C_l \frac{2 \textcolor{black}{B} m_l}{\left(4\pi f \right)^2}  \Bigg\}\,,  \label{eq:RPhi_SU2} \\
	\sqrt{\frac{M_{B_{s}}}{M_{B_l}}} & = & R_M \Bigg\{ 1 + \frac{2 + 3 g_{B^*B\pi}^2  }{\left(4\pi f \right)^2} \ \bigg( \frac{1}{2} \bigg) m_L^2 \ln\left(\frac{m_L^2}{\Lambda_\chi^2}\right) + D_l \frac{\textcolor{black}{B} m_l}{\left(4\pi f \right)^2}  \Bigg\}  \,, \label{eq:RM_SU2} \quad
\end{eqnarray}
where $m^2_L = 2\textcolor{black}{B} (m_l + m_\textrm{res})$ is the tree-level mass-squared of a pseudoscalar meson composed of two quarks with mass $m_l$, $\textcolor{black}{B,}$ and $f$ are the leading-order low-energy constants of $\chi$PT, and the quark masses in the analytic terms are expressed in terms of dimensionless ratios to make the coefficients $C_l$ and $D_l$ of ${\cal O}(1)$.  These functional forms are derived from Eqs.~(\ref{eq:fBx_SU2_ChPT})-- (\ref{eq:MBs_SU2_ChPT}) by taking the ratio of the expressions for the valence quark $y = s$ over the expressions for $x = l$.  Because we are working at a single lattice spacing, the analytic terms proportional to $a^2$ are absorbed into the values of the leading-order coefficients $R_\Phi$ and $R_M$.  Note that in the limit $m_l \to m_{s}$ the $SU(3)$-breaking ratios are not constrained to unity, as would be the case in $SU(3)$ HM$\chi$PT.  (In fact, the point $m_l = m_{s}$ does not even lie within the range of validity of SU(2) HM$\chi$PT and hence of Eqs.~(\ref{eq:RPhi_SU2}) and (\ref{eq:RM_SU2}).)  This is because, once the strange quark has been integrated out of the $SU(2)$ theory, the expressions no longer contain explicit strange-quark mass dependence.  All of the effects of the strange quark are encoded in the values of the low-energy constants, which differ in the $SU(2)$ and $SU(3)$ theories.  

Although the coefficients of the chiral logarithms depend on the low-energy constants $g_{B^*B\pi}$, $f$, and $\textcolor{black}{B}$, once these are fixed as we now describe, there are only two free parameters each in Eqs.~(\ref{eq:RPhi_SU2}) and~(\ref{eq:RM_SU2}):  the overall normalization and the coefficient of the analytic term proportional to $m_l$.  This allows us to smoothly match the $SU(2)$ expressions onto the linear fit of the heavy data without ambiguity.  In the chiral extrapolation we obtain our central value using $g_{B^*B\pi} = 0.516$ for the $B^*$-$B$-$\pi$ coupling, which comes from a two-flavor lattice determination in the static heavy quark limit by Ohki, Matsufuru, and Onogi~\cite{Ohki:2008py}.  We then vary the value of $g_{B^*B\pi}$ over a reasonable spread of values based on both lattice calculations and phenomenological fits to experimental data in order to estimate the systematic uncertainty, as described in further detail in Sec.~\ref{sec:gpi}.  Moreover we set the leading-order pseudoscalar meson decay constant $f$ to the experimental value of $f_\pi = 130.4 \pm 0.04 \pm 0.2$~MeV~\cite{Amsler:2008zzb}.  This is consistent to the order in $\chi$PT at which we are working since it only modifies higher-order NNLO terms.  Studies by both the MILC and JLQCD Collaborations suggest that the use of a physical parameter in the chiral coupling ($f \to f_\pi$) leads to improved convergence of $\chi$PT~\cite{Aubin:2004fs,Noaki:2008iy}. The scale in the chiral logarithms is fixed by setting $\Lambda_{\chi} = 1$ GeV. For the low-energy constant $\textcolor{black}{B}$ we use the value $a \textcolor{black}{B} = 2.414(61)$ obtained from a NLO fit of the pseudoscalar meson masses~\cite{Allton:2008pn}.  Finally, whenever the residual quark mass appears, we use its value in the chiral limit  $am_\textrm{res} = 0.00315$.   
%This effectively accounts for some higher-order chiral corrections and hence gives a better approximation of the pion mass squared at a given light-quark mass than use of $\textcolor{black}{B}$ in the chiral limit.  

\begin{figure}[hpt]
\includegraphics[clip,scale=0.6]{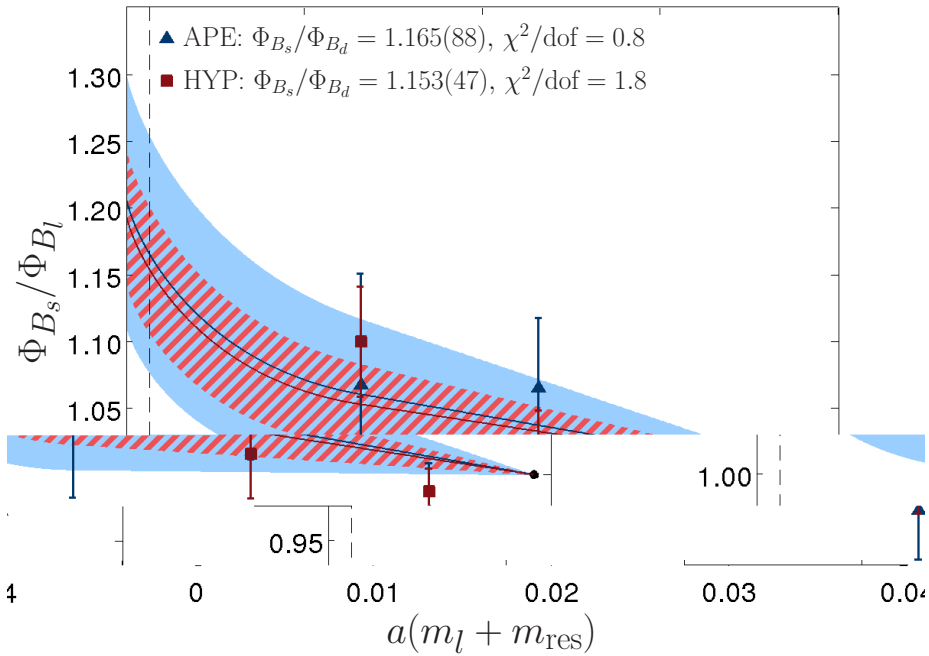}
\caption{Chiral extrapolation of $\Phi_{B_{s}}/\Phi_{B_l}=\sqrt{m_{B_{s}}/m_{B_l}}\cdot f_{B_{s}}/f_{B_l}$. The blue (triangle) points denote the APE data, while the red (square) points denote the HYP-smeared data.   The color of the shaded (hatched) error band corresponds to those of the APE (HYP) data points. The dashed vertical line denotes the physical average $u$-$d$ quark mass and the black dot denotes the physical strange quark mass, at which the $SU(3)$-breaking ratio must be one.  The $SU(2)$ HM$\chi$PT coefficients obtained from the fit are $R_\Phi = 1.21(9)$ and $C_l =1.1(6)$ in the case of APE smearing and $R_\Phi = 1.19(5)$ and $C_l = 1.2(3)$ in the case of HYP smearing.  Errors shown are statistical only.}
\label{fig:fDs_fD_fit}
\end{figure}

\begin{figure}[hpt]
\includegraphics[clip,scale=0.6]{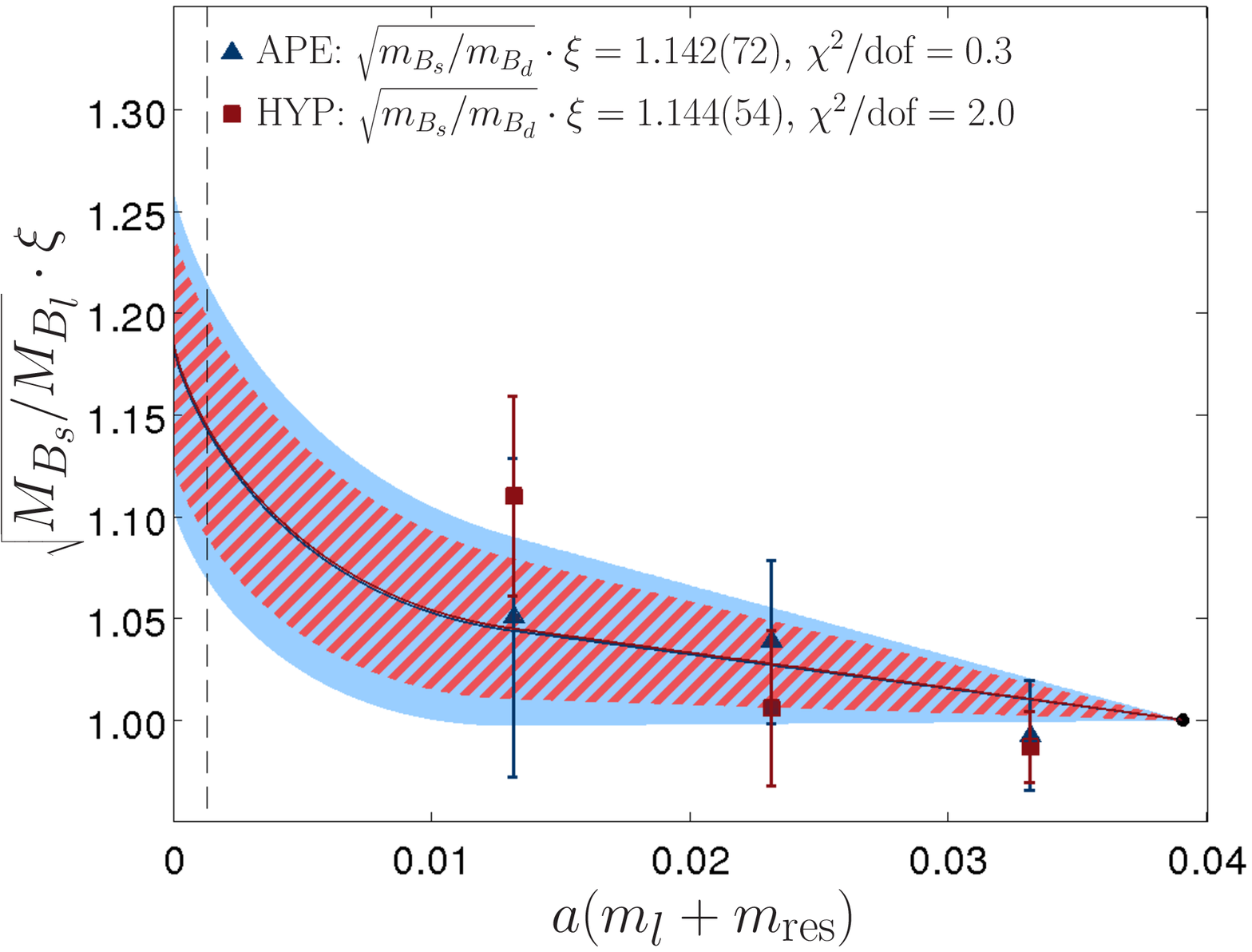}
\caption{Chiral extrapolation of $\sqrt{M_{B_{s}}/M_{B_l}} = \sqrt{m_{B_{s}}/m_{B_l}}\,\left( {f_{B_{s}}\sqrt{B_{B_{s}}}}/{f_{B_l}\sqrt{B_{B_l}}}\right) $. The blue (triangle) points denote the APE data, while the red (square) points denote the HYP-smeared data.    The color of the shaded (hatched) error band corresponds to those of the APE (HYP) data points. The dashed vertical line denotes the physical average $u$-$d$ quark mass and the black dot denotes the physical strange quark mass, at which the $SU(3)$-breaking ratio must be one.  The $SU(2)$ HM$\chi$PT coefficients obtained from the fit are $R_M = 1.18(8)$ and $D_l = 2.6(1.1)$ in the case of APE smearing and $R_M = 1.18(6)$ and $D_l =  2.6(8)$ in the case of HYP smearing.  Errors shown are statistical only.}
\label{fig:Xi_fit}
\end{figure}

The results of the chiral extrapolation are shown in Figs.~\ref{fig:fDs_fD_fit} and \ref{fig:Xi_fit}. The blue triangles (red squares) show the data obtained using APE (HYP) link smearing and are plotted versus the light sea quark mass. We indicate the location of the physical strange quark mass $m_s$ by the black dot. The dashed vertical line marks the physical average $u$-$d$ quark mass, which is the point at which we extract the physical values for $\Phi_{B_s}/\Phi_{B_d}$ and $\sqrt{m_{B_s}/m_{B_d}}\,\xi$. The agreement between the two smearings is good.  For the case of the APE data, the $\chi^2/\textrm{dof}$ for the fit of both $SU(3)$-breaking ratios is below one, indicating that the data are well-described by the linear fit function.  For the HYP data, the $\chi^2/\textrm{dof's}$ are $1.8$ for $\Phi_{B_s}/\Phi_{B_d}$ and $2.0$ for $\sqrt{m_{B_s}/m_{B_d}}\,\xi$, respectively.  These still correspond to confidence levels of greater than 10\%, however, and are therefore consistent with the data.  The error bands in Figs.~\ref{fig:fDs_fD_fit} and \ref{fig:Xi_fit} are fairly broad and hence the statistical uncertainty at the physical point is large, $\sim$ 4.1--7.6\% for $\Phi_{B_s}/\Phi_{B_d}$ and $\sim$ 4.7--6.3\% for $\sqrt{m_{B_s}/m_{B_d}}\,\xi$.   This is due to the large statistical errors in our data points, as well as the fact that our lightest mass is still quite heavy, which forces us to extrapolate over a large mass range.  We summarize the results of our preferred chiral extrapolation in Tab.~\ref{tab:ChiralFits} and discuss the estimation of our systematic errors in the following section.

\begin{CTable}
\caption{Results for $\Phi_{B_s}/\Phi_{B_d}$ and $\sqrt{m_{B_s}/m_{B_d}}\cdot \xi$ at the physical point. Only statistical errors are shown. }
\label{tab:ChiralFits}
\begin{tabular}{ccc} \hline\hline
& \multicolumn{2}{c}{link smearing} \\[-1.7mm]
 & APE & HYP \\ \hline
$\Phi_{B_s}/\Phi_{B_d}$ & 1.165(88) & 1.153(47) \\
$\sqrt{m_{B_s}/m_{B_d}}\cdot \xi$ \qquad & 1.142(72) & 1.144(54) \\ \hline\hline
\end{tabular}
\end{CTable}

%=================================================
\section{Estimation of systematic errors}
\label{sec:Err}
%=================================================

In this section we estimate the systematic uncertainties in the $SU(3)$-breaking ratios $f_{B_s}/f_{B_d}$ and $\xi$.  For clarity, we present each source of error in a separate subsection.  The total error budgets for both quantities are given at the end of the section in Table~\ref{tab:total_err}.

%=================================================
\subsection{Chiral extrapolation fit ansatz}
\label{sec:ChPT_err}
%=================================================

As described in the previous section, we extrapolate our data to the physical quark masses using a linear fit matched onto NLO $SU(2)$ HM$\chi$PT at the value of our lightest data point.  The $SU(2)$ expressions are derived using the symmetries of the lattice theory; therefore they contain the correct dependence of the $B$-meson decay constants and mixing matrix elements on the quark mass and lattice spacing through NLO when the quark masses are sufficiently light.  Given the large light-quark masses used in our simulations, however, it may be that the $SU(2)$ chiral logarithms do not become important until the pion mass is even lighter than the range of our data.  Thus we vary the location of the matching point as one way to estimate the systematic uncertainty due to the chiral extrapolation.  In addition, because we use NLO $SU(2)$ HM$\chi$PT to obtain our central value, we must estimate the systematic uncertainty due to the truncation of higher-orders in HM$\chi$PT.  We do this in several ways:  (i) by explicitly adding higher-order terms to the linear plus $SU(2)$ HM$\chi$PT fit function, (ii) by matching the linear fit onto the NLO $SU(3)$ HM$\chi$PT expressions, and (iii) by varying the value of the low-energy constant $f$ in the coefficient of the NLO chiral logarithms.

\bigskip

\begin{figure}[hpt]
\includegraphics[clip,scale=0.405]{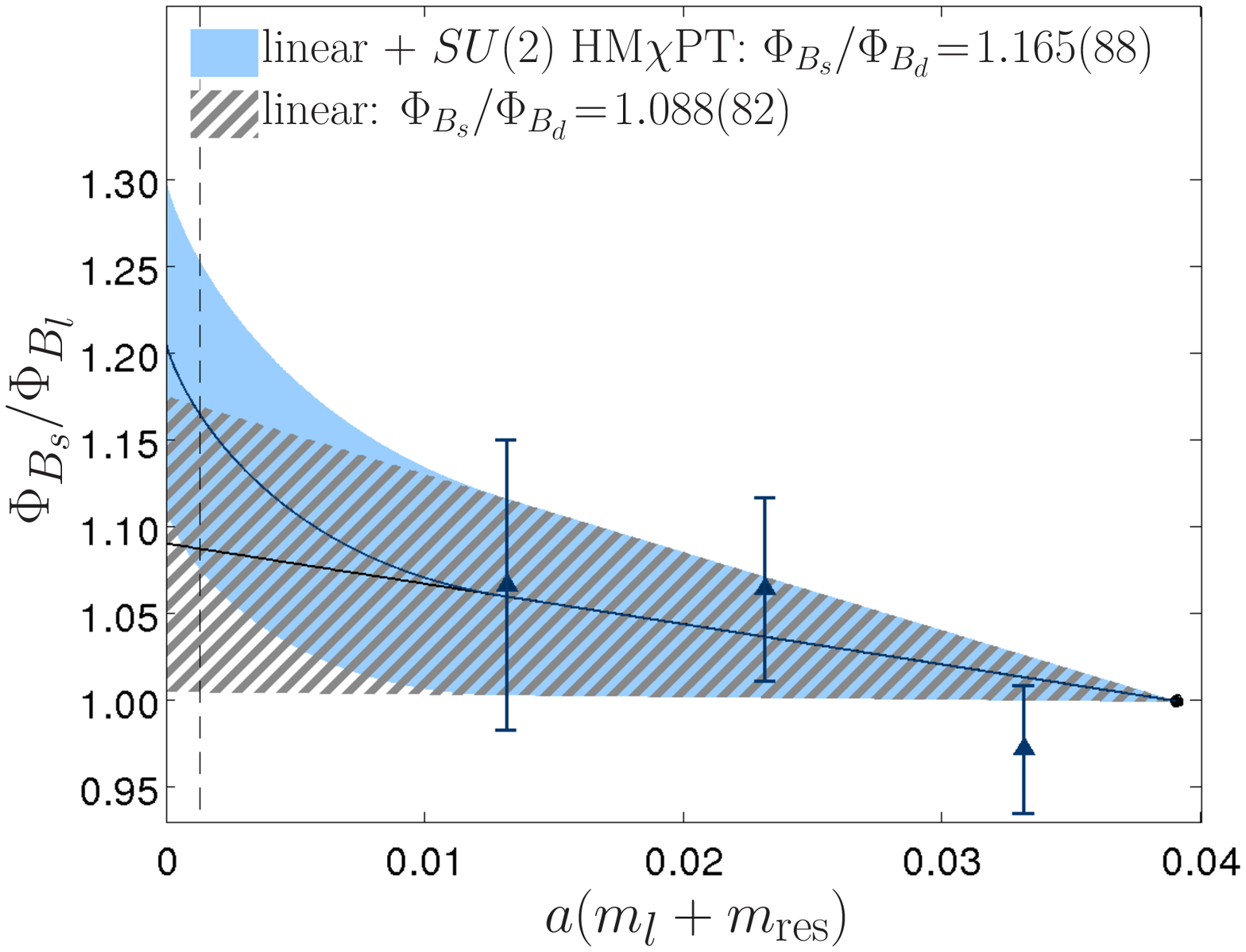}
\includegraphics[clip,scale=0.405]{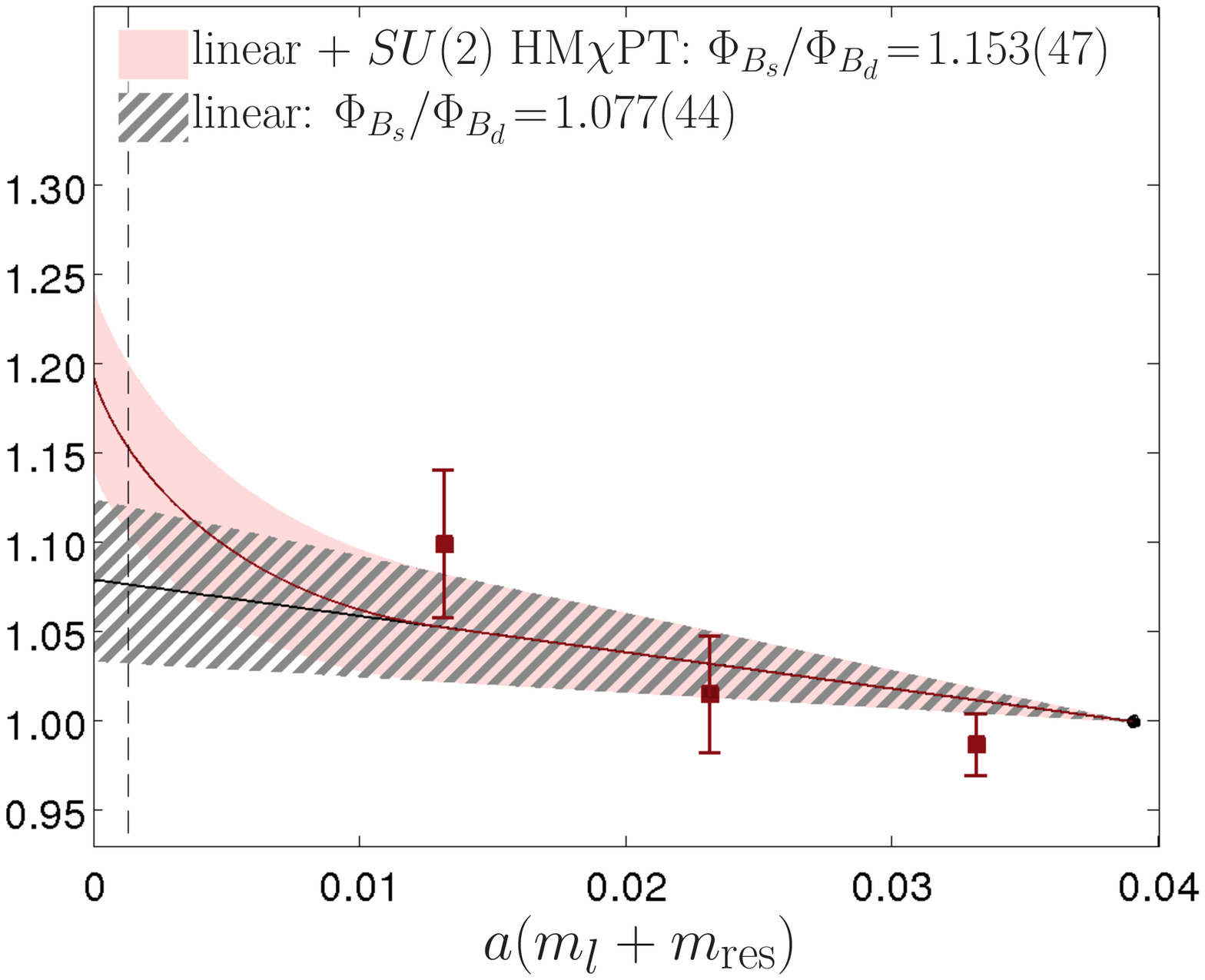}
\caption{Chiral extrapolation of $\Phi_{B_{s}}/\Phi_{B_l}=\sqrt{m_{B_{s}}/m_{B_l}}\cdot f_{B_{s}}/f_{B_l}$ using a linear fit matched onto NLO $SU(2)$ HM$\chi$PT at the lightest data point (solid band) and a linear extrapolation all the way to the chiral limit (hatched band).  Errors shown are statistical only.  The left plot shows the comparison for the APE data, while the right plot shows the same comparison for they HYP-smeared data.}
\label{fig:fB_lin}
\end{figure}

\begin{figure}[hpt]
\includegraphics[clip,scale=0.405]{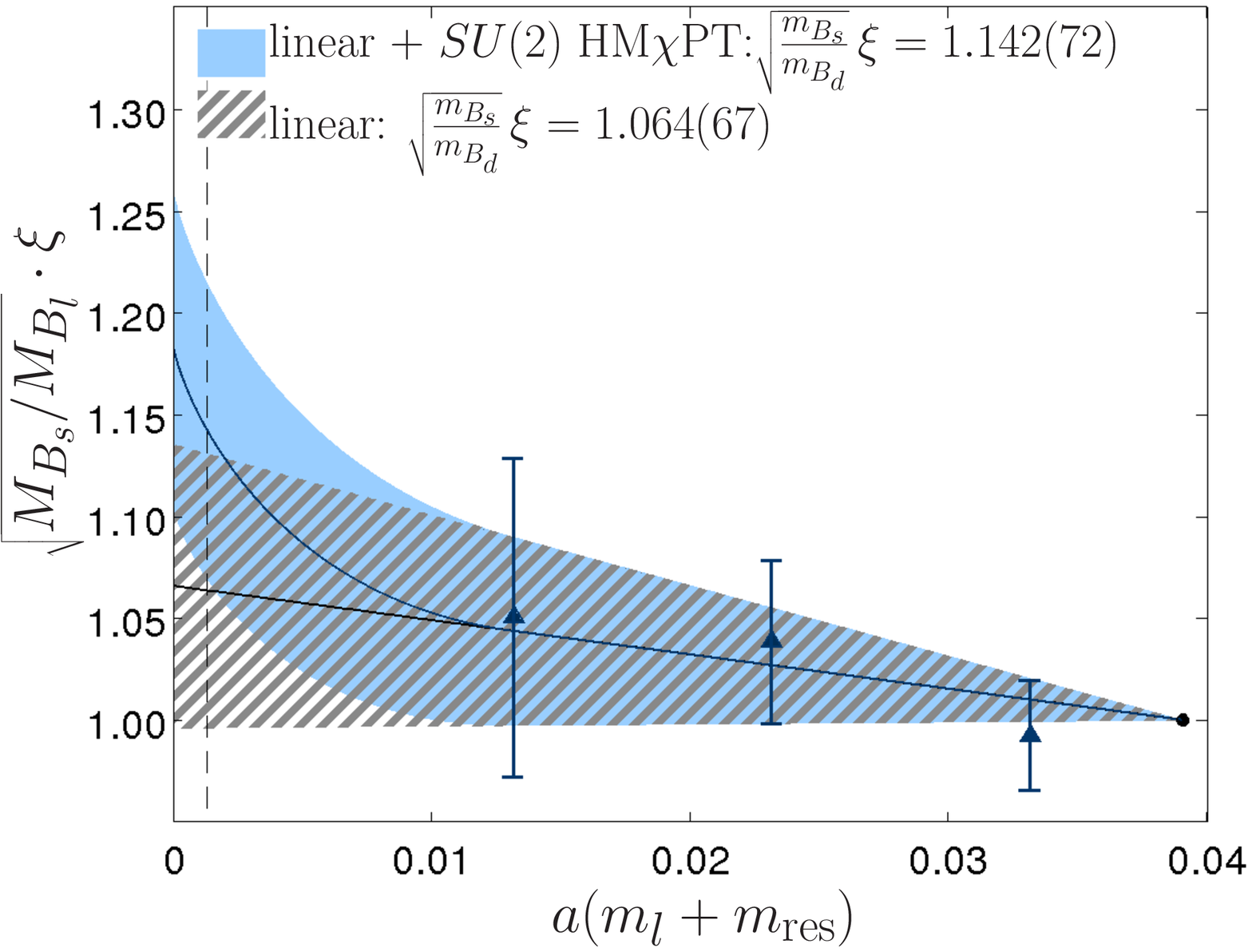}
\includegraphics[clip,scale=0.405]{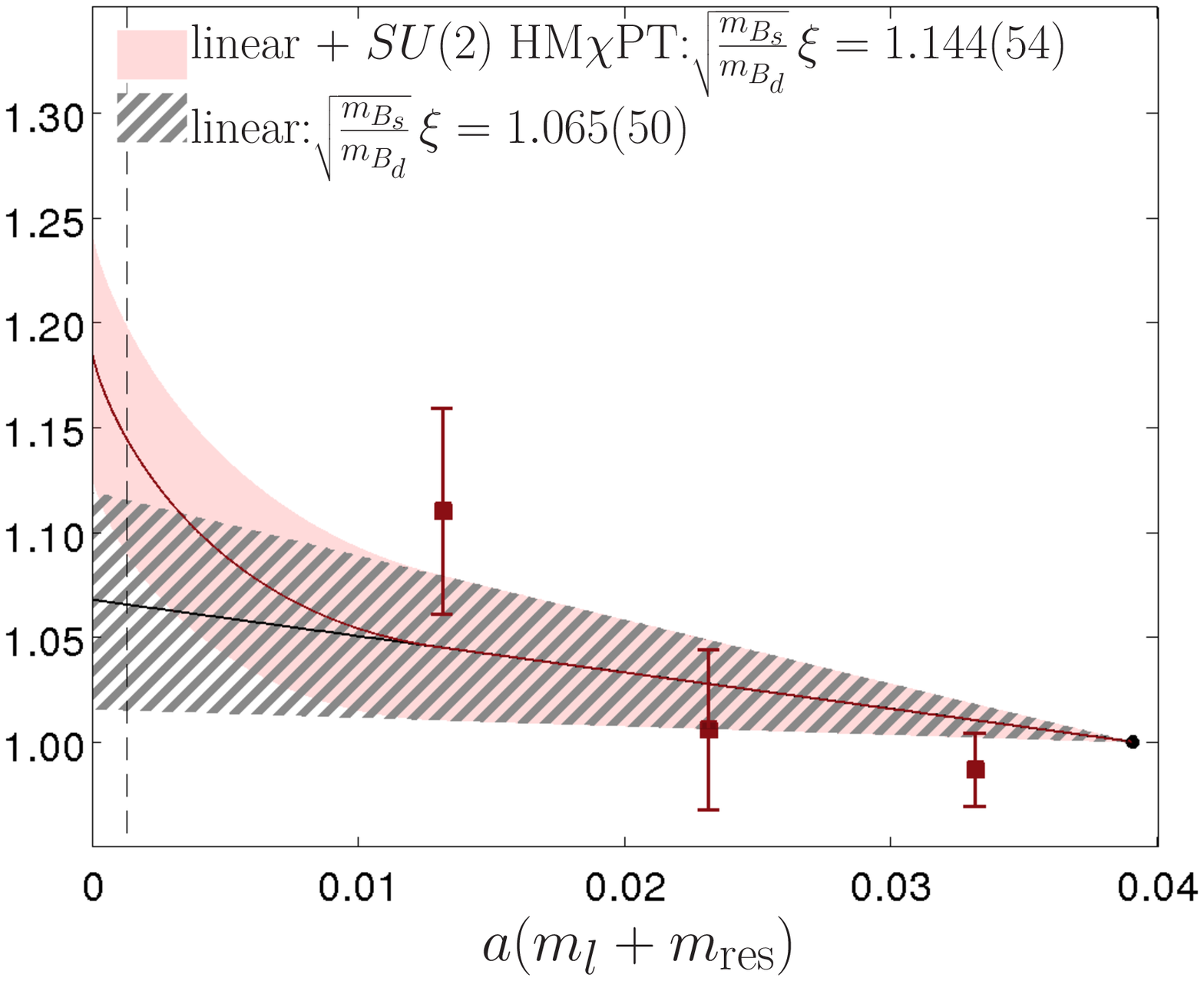}
\caption{Chiral extrapolation of $\sqrt{M_{B_{s}}/M_{B_l}} = \sqrt{m_{B_{s}}/m_{B_l}}\,\left( {f_{B_{s}}\sqrt{B_{B_{s}}}}/{f_{B_l}\sqrt{B_{B_l}}}\right) $ using a linear fit matched onto NLO $SU(2)$ HM$\chi$PT at the lightest data point (solid band) and a linear extrapolation all the way to the chiral limit (hatched band).  Errors shown are statistical only.  The left plot shows the comparison for the APE data, while the right plot shows the same comparison for they HYP-smeared data.}
\label{fig:xi_lin}
\end{figure}

Because we do not know {\it a priori} at what mass the $SU(2)$ chiral logarithms become important, we vary the point at which we match the linear fit onto NLO $SU(2)$ HM$\chi$PT in order to estimate the systematic uncertainty due to the choice of matching point.  At the matching point used in the preferred fit, the ratio of the light up-down quark mass to the strange valence-quark mass is $am_l / am_{s} \approx 0.28 $ and to the strange sea quark mass is $am_l / am_h = 0.25$.  Since both of these quantities are small, we expect the strange quark can be integrated out of the chiral effective theory and that $SU(2)$ HM$\chi$PT is applicable in this region.  As the light-quark mass decreases, $SU(2)$ becomes an even better approximation.  At light-quark masses above this point, however, the ratios $am_l / am_{s}$ and $am_l / am_h$ are no longer small expansion parameters, and $SU(2)$ will eventually cease to apply.  Therefore, when we estimate the systematic error due to the choice of matching point, we only consider extrapolations in which the matching point is closer to the chiral limit than in the preferred fit.  The limiting case is where the matching point is at the chiral limit, which corresponds to a purely linear extrapolation.  Although the choice of a linear fit is not based on effective field theory, it cannot be ruled out by the data.  We therefore take the difference between the extrapolated values for $f_{B_s}/f_{B_d}$ and $\xi$ obtained with the preferred fit and with a purely linear fit to estimate the systematic error due to the choice of matching point; this leads to an error in $f_{B_s}/f_{B_d}$ of 6.6\% for both the APE and HYP data and in $\xi$ of 6.8\% (6.9\%) for the APE (HYP) data.  We compare the results of the preferred fit with those of the purely linear extrapolation in Fig.~\ref{fig:fB_lin} for $\Phi_{B_{s}}/\Phi_{B_l}$ and in Fig.~\ref{fig:xi_lin} for $\sqrt{m_{B_{s}}/m_{B_l}}\,\xi$.  Comparison with the linear extrapolation leads to a conservative systematic error estimate since we know that, when the pion mass is sufficiently light, the chiral logarithms will become important and the extrapolation function will begin to curve upward.   However, the linear fit provides a clear lower-bound on the possible extrapolated value.  

In order to estimate the error due to the omission of higher-order terms in the chiral expansion, we compare the result of the linear plus NLO $SU(2)$ HM$\chi$PT fit to a fit supplemented by NNLO analytic terms.  Once the valence and sea strange quarks have been integrated out of the chiral effective theory, the $SU(2)$ HM$\chi$PT expressions can only depend upon the light up-down quark mass $m_l$ and the lattice spacing $a$.  Thus the only possible NNLO analytic terms are those proportional to $m_l^2$, $m_l a^2$, and $a^4$.  Because we are working at a single lattice spacing, the contribution proportional to $a^4$ is indistinguishable from the LO normalization factors $R_\Phi$ and $R_M$ in Eqs.~(\ref{eq:RPhi_SU2}) and~(\ref{eq:RM_SU2}) and the contribution proportional to $m_l a^2$ is indistinguishable from the NLO terms $C_l m_l$ and $D_l m_l$.  Thus there is only one additional free parameter at NNLO.  In practice, we implement the ``NNLO fit" by performing a quadratic fit in $m_l$ to our data and then matching to the NLO $SU(2)$ HM$\chi$PT expressions supplemented by the NNLO analytic term proportional to $m_l^2$ at the value of our lightest data point.  When we include the term quadratic in $m_l$, however, the extrapolated values for $f_{B_s}/f_{B_d}$ and $\xi$ have significantly larger statistical errors (about $20 - 40\%$) than those obtained with the preferred fit (about $4 - 8\%$).   Therefore we cannot draw any meaningful conclusion about the change in the central values of $f_{B_s}/f_{B_d}$ and $\xi$, since the outcome of the NNLO extrapolation is consistent with the NLO extrapolation within the large statistical errors.  This increase in statistical errors is to be expected since we have introduced an extra free parameter, and it is difficult to constrain two parameters with only three data points.  Thus we do not use the ``NNLO fit" as a way to estimate the chiral extrapolation error in this work, but leave it as a way to estimate the uncertainty due to the omission of higher-order terms in future analyses when we have more data points and smaller statistical errors.

We can also estimate the systematic uncertainty in the chiral extrapolation by comparing the values of $f_{B_s}/f_{B_d}$ and $\xi$ obtained by matching onto $SU(2)$ HM$\chi$PT with those obtained by matching onto $SU(3)$ HM$\chi$PT.  Although the $SU(2)$ chiral logarithms are a subset of the $SU(3)$ chiral logarithms, the $SU(2)$ and $SU(3)$ theories have different series expansions and different degrees-of-convergence within their ranges of applicability.  Therefore the comparison with NLO $SU(3)$ HM$\chi$PT provides another means of estimating the error due to the use of NLO $SU(2)$ HM$\chi$PT.  The expressions for $SU(3)$-breaking ratios at NLO in $SU(3)$ HM$\chi$PT are, schematically,
\begin{eqnarray}
	\frac{\Phi_{B_{s}}}{\Phi_{B_l}}  & = & 1 + \textrm{``chiral logs"} + \frac{2\textcolor{black}{B}}{(4\pi f)^2}\tilde{c}_\textrm{val} (m_{s} - m_l),\label{eq:RPhi_SU3} \\
	\sqrt{\frac{M_{B_{s}}}{M_{B_l}}}  & = & 1 + \textrm{``chiral logs"} + \frac{\textcolor{black}{B}}{(4\pi f)^2} \tilde{d}_\textrm{val} (m_{s} - m_l), \label{eq:RM_SU3}
\end{eqnarray}
where ``chiral logs" indicate non-analytic functions of the pseudo-Goldstone meson masses, e.g. $m_L^2 \log (m_L^2 / \Lambda_\chi^2)$.  These are derived from Eqs.~(\ref{eq:fB_PQChPT}) and (\ref{eq:MB_PQChPT}) by taking the ratio of the expressions for the valence quark $x = s$ over the expressions for $x = l$.  Because the strange quark is treated in the same manner as the up and down quarks in the $SU(3)$ chiral effective theory, the low-energy constants are the same in the numerator and denominator.  Thus the overall normalizations cancel in the expressions in Eqs.~(\ref{eq:RPhi_SU3}) and~(\ref{eq:RM_SU3}) and the ratios are constrained to unity in the limit $m_l \to m_{s}$.  Therefore the expressions for the $SU(3)$-breaking ratios have one free parameter each, instead of two as in the $SU(2)$ HM$\chi$PT case.  This means that the $SU(3)$ HM$\chi$PT expressions cannot be matched smoothly onto the linear fit of the heavy data.  We choose to make the value of the extrapolation function continuous, while leaving a discontinuity in the slope at the matching point.  The difference between the linear plus NLO $SU(2)$ HM$\chi$PT fit and the linear plus NLO $SU(3)$ HM$\chi$PT fit leads to a difference in $f_{B_s}/f_{B_d}$ of 2.3\% (2.4\%) for the APE (HYP) data and in $\xi$ of 2.6\% (2.5\%) for APE (HYP).

In the preferred linear plus NLO $SU(2)$ HM$\chi$PT fit, we set the leading-order pseudoscalar meson decay constant $f$ equal to the experimentally-measured value of $f_\pi$.  This fixes the coefficient of the chiral logarithms and improves the convergence of the chiral expansion~\cite{Aubin:2004fs,Noaki:2008iy}.  At NLO in $\chi$PT, however, it is equally consistent to use the pseudoscalar decay constant in the $SU(2)$ chiral limit $f_0$ or the kaon decay constant $f_K$ because the different choices only affect NNLO terms that are of higher-order than we consider.  We therefore vary $f$ between $f_0 = 115$~MeV~\cite{Allton:2008pn} and $f_K = 155.5$~MeV~\cite{Amsler:2008zzb} in order to estimate the systematic uncertainty due to the omission of higher-order terms in the chiral expansion.  The use of $f_0$ in the $SU(2)$ chiral limit, which the RBC and UKQCD Collaborations find to be about $10\%$ lower than $f_\pi$, leads to a difference from the central value for $f_{B_s}/f_{B_d}$ of 2.1\% (2.2\%) for APE (HYP) and from the central value for $\xi$ of 2.3\% for both APE and HYP.  The use of $f_K$ leads to similar changes in $f_{B_s}/f_{B_d}$ of 2.1\% and in $\xi$ of 2.2\%.

\bigskip

We take the largest of the uncertainties enumerated above, which is obtained from the difference between the preferred linear plus NLO $SU(2)$ HM$\chi$PT fit and the purely linear fit, for the final estimate of the chiral extrapolation error;  this leads to the values in the row labeled ``chiral extrapolation" in Tab.~\ref{tab:total_err}.  

%=================================================
\subsection{Uncertainty due to $g_{B^*B\pi}$}
\label{sec:gpi}
%=================================================

Although we fix the value of the $B^*$-$B$-$\pi$ coupling (and hence the coefficient of the one-loop chiral logarithms) in the chiral extrapolation of our lattice data, $g_{B^*B\pi}$ is in fact poorly known from phenomenology.  Hence we must vary the value of $g_{B^*B\pi}$ over a sensible range based on lattice QCD calculations and phenomenology in order to estimate the systematic errors in $f_{B_s}/f_{B_d}$ and $\xi$ due to the uncertainty in the coefficient of the one-loop chiral logarithms.

There has only been one unquenched lattice QCD calculation of the $B^*$-$B$-$\pi$ coupling with a complete associated error budget, which gives $g_{B^*B\pi} = 0.516(5)(50)$, where the first uncertainty is statistical and the second is systematic~\cite{Ohki:2008py}.  Although this value was computed in the static heavy-quark limit and neglects the effects of the dynamical strange quark, we take this as our central value because of its small statistical errors and full systematic error budget, as well as because it lies in the middle of the range of values presented in the literature.  Another recent determination of  $g_{B^*B\pi} = 0.44 \pm 0.03^{+0.07}_{-0.00}$ in the static limit in two-flavor lattice QCD, where the errors are due to statistics and the chiral extrapolation uncertainty, is consistent with this result~\cite{Becirevic:2009yb}.  QCD sum rules and the relativistic quark model give a slightly lower value of $g_{B^*B\pi} \approx 0.38 \pm 0.08$~\cite{Casalbuoni:1996pg}.   The value of the $B^*$-$B$-$\pi$ coupling is expected to be close to the $D^*$-$D$-$\pi$ coupling because of heavy-quark symmetry.  We can therefore also use lattice QCD calculations and phenomenological extractions of $g_{D^*D\pi}$ as estimates.  Be\'cirevi\'c and Haas recently computed $g_{D^*D\pi} = 0.71(7)$ in 2-flavor lattice QCD, but this result is from only a single lattice spacing and presents no estimate of the systematic error~\cite{Becirevic:2009xp}.  Stewart computed the value of $g_{D^*D\pi}$ in 1998 by fitting to experimental data~\cite{Stewart:1998ke}, and recently updated his result to include the experimental measurement of the $D^*$ decay width~\cite{Anastassov:2001cw}.  His latest determination is $g_{D^*D\pi} = 0.51$, but with no error quoted~\cite{Arnesen:2005ez}.  The most sophisticated extraction of the $D^*$-$D$-$\pi$ coupling was performed by Kamenik and Fajfer, and gives $g_{D^*D\pi} = 0.66^{+0.08}_{-0.06}$, where the uncertainty only reflects the error due to counterterms~\cite{Fajfer:2006hi}.  Finally, we note that the chiral-continuum extrapolations of $B$-meson quantities by the Fermilab/MILC Collaboration~\cite{Bailey:2008wp} and the HPQCD Collaboration~\cite{Gamiz:2009ku} tend to prefer even smaller values of the $B^*$-$B$-$\pi$ coupling than those in the literature.

Most of the results presented above do not have complete error budgets, and are inconsistent within the quoted errors, so for this work we take $g_{B^*B\pi} = 0.516 \pm 0.2$ to account for the spread of values.  We then vary $g_{B^*B\pi}$ within this range to determine how much it changes the central values for $f_{B_s}/f_{B_d}$ and $\xi$. This leads to an uncertainty in $f_{B_s}/f_{B_d}$ of $3.2\%$ for both APE and HYP and in $\xi$ of $2.1\%$.

%=================================================
\subsection{Discretization errors}
%=================================================

In this work we only analyze data at a single, relatively coarse lattice spacing of $a \approx 0.11$~fm, so we estimate the size of discretization errors with power-counting.  As a consistency-check of our estimation procedure, however, we can compare the estimated errors in the individual decay constants and matrix elements with the observed differences in those quantities for the APE and HYP-smeared data.  This is because, aside from statistical errors, the differences in the values obtained from the two smearings are due to discretization effects and higher-order corrections to the renormalization factors.  We find about 15--20\% differences in the decay constants and matrix elements obtained from the APE and HYP-smeared data, whereas we estimate by power-counting that discretization errors in $f_{B_q}$ should be about 15\% and in $\cM_q$ should be about 20\%.  Thus our observations are consistent with the scaling behavior expected from power-counting.    We can also compare our power-counting estimates with the findings of the ALPHA Collaboration, who performed a study of $f_{B_s}$ at several lattice spacings using static $b$-quarks with similar link-smearings in the quenched approximation~\cite{DellaMorte:2007ij}.  Although ALPHA observes a violation of $\CO(a^2)$ scaling behavior at inverse lattice spacings below $a^{-1} \approx 2.5$ GeV, the difference between the predicted value for $f_{B_s}$ at $a \approx 0.11$~fm given $\CO(a^2)$ scaling and the value of $f_{B_s}$ that they obtain in the continuum limit using data within the scaling region is only about 25\%, which is again close to our power-counting estimate.  Thus we expect that naive power-counting should lead to a reasonable estimate of the discretization error in the $SU(3)$-breaking ratios.  

For the error estimates in this subsection, we evaluate the strong coupling constant at the lattice scale, $\alpha_s^{\bar{\textrm{MS}}} (1/a) \sim 1/3$.  We choose $\Lambda_\textrm{QCD} = 500$~MeV because the typical QCD scale that enters heavy-light quantities tends to be larger than for light-light quantities, as indicated by fits to moments of inclusive $B$-decays using the heavy-quark expansion~\cite{Buchmuller:2005zv}.  Fortunately, some of the finite lattice-spacing effects cancel in the ratios $f_{B_s}/f_{B_d}$ and $\xi$.  This can be seen by the fact that, although the APE and HYP data differ by about 15--20\% for the individual decay constants and matrix elements, they agree within statistical errors for the ratios.  In $SU(3)$-breaking quantities, errors must be proportional to the difference in quark masses $(m_s - m_d)$.  Dimensional analysis therefore suggests that contributions to the total discretization error are suppressed by the factor $(\tilde{m}_s - \tilde{m}_d)/\Lambda_\textrm{QCD} \sim 1/5$, where we use $\tilde{m}_s$ and $\tilde{m}_d$ to denote the renormalized quark masses in the $\bar{\textrm{MS}}$ scheme~\cite{Amsler:2008zzb} (as opposed to the bare lattice quark masses) in this subsection and the next.  The observed size of $SU(3)$-breaking effects in the $B$-meson decay constants ($f_{B_s}/f_{B_d} - 1$) and in the $B-$mixing matrix elements ($\xi - 1$) are consistent with this expectation.  Discretization errors in $f_{B_s}/f_{B_d}$ and $\xi$ can arise from both the actions and the operators.  We estimate each source of error separately, and add them in quadrature to obtain the total discretization error.  

None of the actions that we are using are $\CO(a^2)$-improved.  Therefore the leading discretization errors from the domain-wall fermion action and Iwasaki gauge action are of $\CO(a^2 \Lambda_\textrm{QCD}^2)$.  When combined with the $SU(3)$-breaking suppression factor, this leads to discretization errors in the ratios $f_{B_s}/f_{B_d}$ and $\xi$ of $\CO(a^2 \Lambda_\textrm{QCD}^2 \times (\tilde{m}_s - \tilde{m}_d)/\Lambda_\textrm{QCD}) \sim 1.7\%$.  The leading heavy-quark discretization errors from the static action are also of $\CO(a^2 \Lambda_\textrm{QCD}^2)$.  Hence heavy-quark discretization errors also contribute $\sim 1.7\%$ to the total error in the ratios.  Because we improve the heavy-light axial current used to compute the decay constant through $\CO(\alpha_s a p)$, the leading discretization errors from the heavy-light current are of $\CO(\alpha_s a m)$, $\CO(\alpha_s^2 a \Lambda_\textrm{QCD})$, and $\CO(a^2 \Lambda_\textrm{QCD}^2)$.\footnote{There are also discretization errors from mixing with operators of other chiralities that are proportional to $a m_\text{res}$.  These effects, however, are expected to be sub-percent level in the matrix elements~\cite{Aoki:2007zz}, and therefore negligible in the $SU(3)$-breaking ratios.}   When combined with the $SU(3)$-breaking suppression factor, this leads to discretization errors in the ratio $f_{B_s}/f_{B_d}$ of $\CO(\alpha_s \times (am_s - am_d)) \sim 1.2\%$ plus $\CO(\alpha_s^2 a \Lambda_\textrm{QCD} \times (\tilde{m}_s - \tilde{m}_d)/\Lambda_\textrm{QCD}) \sim 0.6\%$ plus $\CO(a^2 \Lambda_\textrm{QCD}^2 \times (\tilde{m}_s - \tilde{m}_d)/\Lambda_\textrm{QCD}) \sim 1.7\%$.  Although we do not improve the heavy-light four fermion operator used to compute the $B$-mixing matrix element, the operator does not have any tree-level $\CO(a)$ errors~\cite{PT_Oa}.  Thus the leading discretization errors in the ratio $\xi$  from the four-fermion operator are of $\CO(\alpha_s \times (am_s - am_d)) \sim 1.2\%$ plus $\CO(\alpha_s a \Lambda_\textrm{QCD} \times (\tilde{m}_s - \tilde{m}_d)/\Lambda_\textrm{QCD}) \sim 1.9\%$ plus $\CO(a^2 \Lambda_\textrm{QCD}^2 \times (\tilde{m}_s - \tilde{m}_d)/\Lambda_\textrm{QCD}) \sim 1.7\%$. 

Adding the contributions from light-quark and gluon discretization errors, heavy-quark discretization errors, and discretization errors in the heavy-light current or four-fermion operator in quadrature, we estimate the error in $f_{B_s}/f_{B_d}$ to be $\sim$ 3.2\% and the error in $\xi$ to be $\sim$ 3.7\%.  %Because these estimates are derived using a conservative choice for the hadronic scale $\Lambda_\textrm{QCD}$ = 500~MeV, the discretization errors may in fact be smaller than this.  This can only be confirmed, however, with the inclusion of data at a second lattice spacing.

%=================================================
\subsection{Heavy-light current and four-fermion operator renormalization}
%=================================================

We compute the renormalization factors needed to match the lattice axial current and four-fermion operator to the continuum using one-loop lattice perturbation theory. This leaves a residual error due to the omission of higher-order terms.  Based on power-counting, we estimate the truncation error in the coefficients to be of $\CO(\alpha_s^2)$, which is the size of the first neglected term in the series.  As we noted earlier in Sec.~\ref{sec:total_match}, however, the matching coefficient $Z_\Phi$ cancels in the ratio of decay constants $f_{B_s}/f_{B_d}$; thus its contribution to the error in $f_{B_s}/f_{B_d}$ is zero.  Although such an exact cancellation does not occur for the ratio of mixing matrix elements $\xi$, the error in $\xi$ due to the uncertainty in the ratio of matching coefficients $Z_{SP}/Z_{VA}$ is suppressed by the $SU(3)$-breaking factor $(\tilde{m}_s - \tilde{m}_d)/\Lambda_\textrm{QCD}$.  This is because, in the $SU(3)$ limit, the four-fermion operator matrix elements would be equal in the numerator and denominator, so the error in $\xi$ from the renormalization factor uncertainty would be zero.  We therefore expect the error in $f_{B_s}/f_{B_d}$ to be $0\%$ and the error in $\xi$ to be of $\CO(\alpha_s^2 \times (\tilde{m}_s - \tilde{m}_d)/\Lambda_\textrm{QCD} ) \sim 2.2\%$.  This error will decrease with the inclusion of data at a finer lattice spacing because the smaller coupling constant will improve the convergence of the series.

%=================================================
\subsection{Scale uncertainty}
%=================================================

Because in this paper we only compute dimensionless ratios, the uncertainty in the determination of the lattice spacing only enters implicitly through the uncertainty in the light-quark masses and in the renormalization factors.  We estimate the systematic error due to the truncation of lattice perturbation theory in the previous subsection and due to the light-quark mass determinations in the following subsection.

%=================================================
\subsection{Light- and strange-quark mass uncertainties} 
%=================================================

We obtain the physical decay constants and mixing matrix elements by setting the light-quark masses to their physical values in the linear plus $SU(2)$ HM$\chi$PT chiral extrapolation formulae, once the low-energy constants have been determined from fits to numerical lattice data.  We use the bare-quark mass value determined from fits to the light pseudoscalar meson masses~\cite{Allton:2008pn}:
\bea
	a m_{ud} + a m_\textrm{res} &=& 0.001300(85) , 
\eea
where $m_{ud}$ is the average of the up and down quark masses.  The quoted error includes both statistics and the systematic uncertainties from the chiral extrapolation, finite-volume effects, discretization effects, and the unphysical strange sea quark mass.  In order to estimate the systematic uncertainty in the ratios $f_{B_s}/f_{B_d}$ and $\xi$, we vary the bare light-quark masses within their stated uncertainties.  We then take the maximal difference from the central value to be the systematic error.  From this method, we find that the systematic error in $f_{B_s}/f_{B_d}$ due to the uncertainty in the light-quark mass determination is 0.2\% for both smearings and the systematic error in $\xi$ is 0.2\% regardless of the smearing used for the heavy quark.

Because the strange-quark mass does not explicitly appear in the $SU(2)$ HM$\chi$PT extrapolation formulae, we cannot estimate the errors in $f_{B_s}/f_{B_d}$ and $\xi$ due to the simulated valence and sea strange-quark masses with the simple method used above for the light quarks.  We must instead address the errors due to the uncertainty in the strange valence-quark mass (which is set to the physical value of $m_s$) separately from those due to the choice of strange sea-quark mass (which is not tuned to the physical $m_s$).  

We calculate the decay constants and mixing matrix elements with the strange valence-quark mass tuned to the physical value obtained from fits of the light pseudoscalar meson masses~\cite{Allton:2007hx}:
\bea
\label{eq:ms}
	 a m_{s} + a m_\textrm{res} &=& 0.0390(21), 
\eea
where the error includes those due to statistics and to uncertainties in the pseudoscalar meson masses, residual mass, and lattice scale.  Since the time at which our numerical computation was performed, however, we have improved the strange-quark mass determination using data on the larger-volume $24^3$ ensemble.  This analysis yields $a m_s + a m_\textrm{res} = 0.0375(17)$~\cite{Allton:2008pn}, where the error includes statistics and the systematic uncertainties from the chiral extrapolation, finite-volume effects, discretization effects, and the unphysical strange sea quark mass.  Fortunately, the two determinations are consistent within their stated uncertainties and the small change in valence strange-quark mass leads to a negligible difference in the the decay constants and mixing matrix elements.  A linear interpolation of $\Phi_{B_s}$ and $M_{B_s}$ from $am_s = 0.0359 \to 0.0343$ leads to at most a percent-level change in these quantities, which is too small to resolve within our large statistical errors.  We therefore conclude that it is sufficient to forgo the interpolation to $am_s = 0.0343$ in our current analysis.  

Although the strange-quark mass is integrated out in $SU(2)$ HM$\chi$PT, the value of the strange valence-quark mass enters the chiral extrapolation described in Sec.~\ref{sec:ChExp} via the linear fit in the heavy mass region that is constrained to unity in the limit $m_l \to m_s$.  Once the slope in the linear region has been determined from fits to numerical lattice data, we estimate the errors in $f_{B_s}/f_{B_d}$ and $\xi$ due to the uncertainty in $m_s$ by fixing the slope to this value and varying $m_s$ within its stated uncertainty $\sigma_{m_s}$ given in Eq.~(\ref{eq:ms}).  This shifts the location of the $SU(3$) limit and hence the linear portion of the extrapolation in the horizontal direction by an amount $\pm \sigma_{m_s}$;  consequently it shifts the location of the point at which we match onto $SU(2)$ HM$\chi$PT in the vertical direction by the slope times $\pm \sigma_{m_s}$.    From this method, we find that the systematic error in $f_{B_s}/f_{B_d}$ due to the uncertainty in the valence strange-quark mass determination is 0.4\% regardless of the smearing used for the heavy quark and the systematic error in $\xi$ is 0.4\% (0.3\%) for the APE (HYP) data.

The value of the strange sea-quark mass does not explicitly enter the chiral extrapolation formulae.  Nevertheless, it implicitly affects the determinations of $f_{B_s}/f_{B_d}$ and $\xi$ through the values of the $SU(2)$ HM$\chi$PT low-energy constants.  The latest results of the RBC and UKQCD Collaborations for $f_\pi, f_K$, and $B_K$~\cite{Kelly:2009fp} rely on reweighting~\cite{Hasenfratz:2008fg} to self-consistently tune the strange sea-quark mass to its physical value during the chiral extrapolation of the pseudoscalar meson masses and decay constants.  We can therefore estimate the effect of the incorrect strange sea-quark mass on $f_{B_s}/f_{B_d}$ and $\xi$ based on observations in the light pseudoscalar meson sector.  Preliminary studies by RBC and UKQCD find that reweighting to the physical strange sea-quark mass leads to no statistically significant change in $m_\pi$ and a 2\% change in $f_\pi$~\cite{Mawhinney:2009jy}.  If we assume that the unphysical strange sea-quark mass leads to a 2\% error in the decay constants and matrix elements, then it will lead to a $2\% \times (\tilde{m}_s - \tilde{m}_d)/\Lambda_\textrm{QCD} \sim 0.4\%$ error in the $SU(3)$-breaking ratios $f_{B_s}/f_{B_d}$ and $\xi$.

In order to obtain the total systematic error in $f_{B_s}/f_{B_d}$ and $\xi$ due to the light- and strange-quark masses, we add the three estimates for the light-quark error, the strange valence-quark error, and the strange sea-quark error in quadrature.  This leads to an error in $f_{B_s}/f_{B_d}$ of  0.6\% for both the APE and HYP data and an error in $\xi$ of 0.6\% (0.5\%) for the APE (HYP) data.

%=================================================
\subsection{Finite-volume errors}
%=================================================

We estimate the uncertainties in the ratios $f_{B_s}/f_{B_d}$ and $\xi$ due to finite volume effects using one-loop finite-volume HM$\chi$PT.  The effect of the finite spatial lattice volume is simply to turn the one-loop integrals that appear in the HM$\chi$PT expressions into sums, which we show in Eqs.~(\ref{eq:FV_logs})-(\ref{eq:FV_logs_2}).  It is therefore straightforward to compute the corrections to our data given the parameters of our simulations.  We find that the one-loop finite volume corrections are about $1\%$ at our lightest quark mass and even less at heavier masses.  We therefore take $1\%$ to be the systematic uncertainty in $f_{B_s}/f_{B_d}$ and $\xi$ due to the finite spatial lattice volume.

%=================================================
\subsection{$1/m_b$ corrections}
%=================================================

Because we work in the static heavy-quark limit, our results for $f_{B_s}/f_{B_d}$ and $\xi$ neglect relativistic effects due to the finite $b$-quark mass.  We therefore estimate the size of the omitted relativistic corrections using power-counting.  The leading $1/m_b$ corrections to the decay constants and matrix elements are of $\CO(\Lambda_\textrm{QCD}/m_b)$.  Because we are only computing the $SU(3)$-breaking ratios $f_{B_s}/f_{B_d}$ and $\xi$, however, relativistic effects in these quantities are further suppressed by a factor of $(m_s - m_d)/\Lambda_\textrm{QCD}$.  Therefore we expect the relativistic corrections to these quantities to be approximately
\beq
	\frac{\Lambda_\textrm{QCD}}{m_b} \times \frac{m_s - m_d}{\Lambda_\textrm{QCD}} \sim 2\% ,
\eeq
where the scale $\Lambda_\textrm{QCD}$ cancels in the ratio, and we use the average $\bar{\textrm{MS}}$ quark masses listed in the PDG~\cite{Amsler:2008zzb}.

\begin{table}
\caption{Total error budget for the $SU(3)$-breaking ratios $f_{B_s}/f_{B_d}$ and $\xi$.  Each source of uncertainty is discussed in Sec.~\ref{sec:Err}, and is rounded to the nearest percentage.}
\label{tab:total_err}
\begin{tabular}{lcccccc} \\ \hline\hline

& &\multicolumn{2}{c}{$f_{B_s}/f_{B_d}$} & &\multicolumn{2}{c}{$\xi$} \\[-1.7mm]
uncertainty                  	 		& \quad\quad &APE & HYP & \quad\quad & APE & HYP \\ \hline
statistics                 				&& 8\% & 4\% && 6\%& 5\% \\ \hline
chiral extrapolation       			&& 7\% & 7\% && 7\%& 7\% \\
uncertainty in $g_{B^*B\pi}$ 		&& 3\% & 3\%&& 2\%&  2\% \\
discretization error       			&& 3\% & 3\%&& 4\% & 4\%\\
renormalization factors    			&& 0\% & 0\%&& 2\% & 2\% \\
scale and quark mass uncertainties && 1\%& 1\% && 1\% & 1\% \\
finite volume error       			&& 1\% & 1\%&& 1\% & 1\%\\
$1/m_b$ corrections        			&& 2\% & 2\%&& 2\% & 2\%\\
\hline
total systematics       				&& 9\% & 9\%&& 9\%& 9\%\\
\hline\hline
\end{tabular}\end{table}

%=================================================
\section{Results and Conclusions}
\label{sec:Conc}
%=================================================

\begin{figure}[hpt]
\includegraphics[clip,scale=0.6]{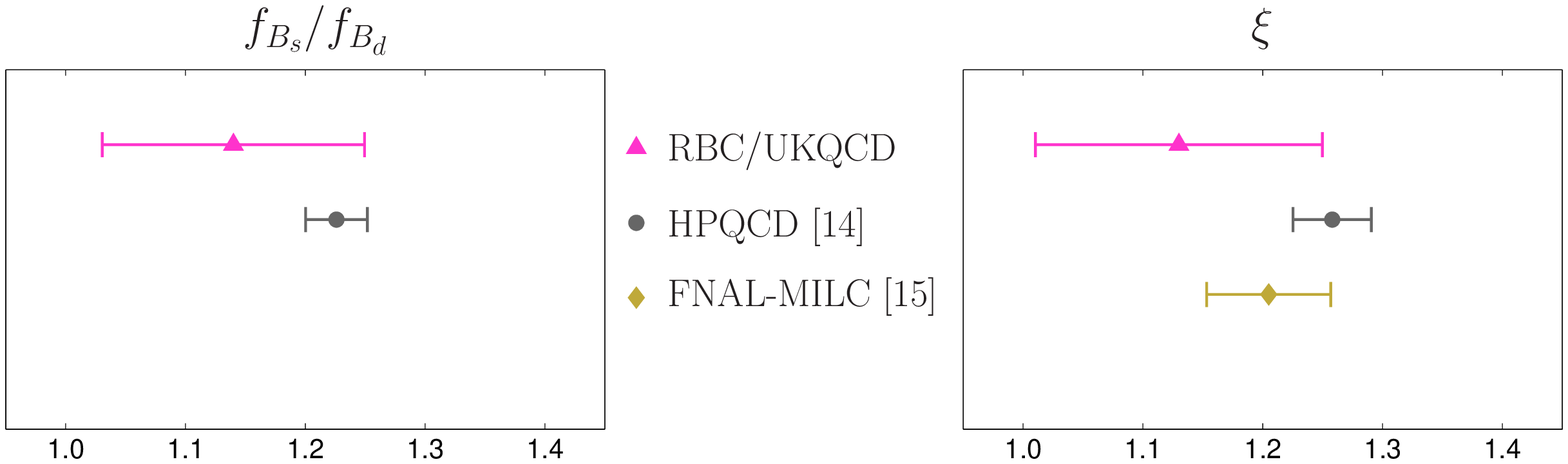}
\caption{Comparison of lattice QCD results for $f_{B_s}/f_{B_d}$ and $\xi$. The magenta (triangle) denotes our new values obtained by averaging the APE and HYP results, the gray (circle) the values published by the HPQCD Collaboration~\cite{Gamiz:2009ku} and the beige (diamond) the preliminary value presented by the FNAL-MILC Collaboration at Lattice 2009~\cite{Evans:2009du}.}
\label{fig:LatticeResults}
\end{figure}

Using the experimentally-measured ratio of masses $m_{B_s^0}/m_{B_d^0} = 5366.6 / 5279.5 = 1.0165$~\cite{Amsler:2008zzb}, we obtain the following values for the $SU(3)$-breaking ratios of $B$-meson decay constants and mixing matrix elements:
\begin{equation}
\frac{f_{B_s}}{f_{B_d}} = 
\left\{ \begin{aligned}
& 1.16(09)(10) \quad\textrm{APE}\\
& 1.14(05)(10) \quad\textrm{HYP}
\end{aligned}\right. \, 
\qquad \textrm{and} \qquad
\xi =
\left\{ \begin{aligned}
& 1.13(07)(10) \quad\textrm{APE}\\
& 1.13(05)(10) \quad\textrm{HYP}
\end{aligned}\right. \, ,
\end{equation}
where the first errors are statistical and the second are the sum of the individual systematic errors added in quadrature.  We find good agreement between the different link-smearings, indicating that, despite the use of a single lattice spacing, discretization errors are small in the ratios.  We therefore average the APE and HYP determinations to obtain our final results.  After adding the statistical and systematic errors for each link smearing in quadrature, we compute the average assuming that the two determinations are 100\% correlated using the method of Ref.~\cite{Schmelling:1994pz}:
\begin{eqnarray}
\frac{f_{B_s}}{f_{B_d}} &  = & 1.15(12) \label{eq:final_fB} \\
\xi & = & 1.13(12) ,  \label{eq:final_MB} 
\end{eqnarray}
where the errors reflect the combined statistical and systematic uncertainties.   Although we computed these quantities in the static $b$-quark limit, the 
inclusion of the neglected $1/m_b$ corrections (which we estimate to be about 2\%) produces a negligible change in the total errors in Eqs.~(\ref{eq:final_fB}) and~(\ref{eq:final_MB})  given the size of our other uncertaintites;  thus our results can be directly compared to phenomenological determinations and other lattice QCD results using relativistic $b$-quarks.  As shown in Fig.~\ref{fig:LatticeResults}, our results agree with the published results of the HPQCD Collaboration ($\xi = 1.258 \pm 0.025_\textrm{stat.} \pm 0.021_\textrm{sys.}$)~\cite{Gamiz:2009ku} and the preliminary results of the Fermilab Lattice and MILC Collaborations ($\xi = 1.205 \pm 0.037_\textrm{stat.} \pm 0.034_\textrm{sys}$)~\cite{Evans:2009du}.  

Although our results have significantly larger errors than the other $N_f = 2+1$ flavor determinations, in this work we have demonstrated the viability of our lattice computation method.  In particular, we have introduced the new approach of using $SU(2)$ heavy-light meson chiral perturbation theory to extrapolate $N_f = 2+1$ lattice QCD results for $B$-meson quantities to the physical quark masses. The largest sources of error in our calculation are from statistics and the chiral extrapolation, and we expect to reduce the sizes of both in a future work that analyzes the $24^3$ domain-wall ensembles with the same lattice spacing~\cite{Allton:2008pn}.  Some of the $24^3$ ensembles contain almost three times as many configurations as we have analyzed in this work.  Furthermore, the use of a larger spatial volume will allow us to simulate at lighter valence and sea quark masses.  Once we have made these improvements, our results will provide a valuable cross-check of these important inputs to the CKM unitarity triangle analysis and determination of the ratio of CKM matrix elements $|V_{td}|/|V_{ts}|$.

%=================================================
\section*{Acknowledgments}
%=================================================

We thank S.~Aoki, S.D.~Cohen, C.~Dawson, E.~Gamiz, C.~Jung, A.~Lenz, M.~Lin, and N.~Yamada for useful
discussions.   

Computations for this work were carried out in part on facilities of the USQCD Collaboration, which are funded by the Office of Science of the U.S. Department of Energy.  We thank BNL, PPARC, RIKEN, University of Edinburgh, and the U.S. DOE for providing the facilities essential for the completion of this work.  The Edinburgh QCDOC system was funded by PPARC JIF grant PPA/J/S/1998/00756 and operated through support from the Universities of Edinburgh, Southampton, and Wales, Swansea and from STFC grant PP/E006965/1.

This work was supported in part by the U.S. Department of Energy under grant No. DE-FG02-92ER40699, by
Columbia University's I.I.~Rabi grant, by PPARC grants PP/D000238/1 and PP/C503154/1, and by the Grant-in-Aid of the Ministry of Education, Culture, Sports, Science and Technology, Japan (MEXT Grant), No.~17740138, No.~19740134, and No.21540289.   PAB is supported by RCUK.   JMF and CTS acknowledge partial support from STFC grant ST/G000557/1 and EU contract MRTN-CT-2006-035482 (Flavianet).  This manuscript has been authored by employees of Brookhaven Science Associates, LLC under Contract No. DE-AC02-98CH10886 with the U.S. Department of Energy.  RV acknowledges support from BNL via the Goldhaber Distinguished Fellowship.  

\appendix

%=================================================
\section{$SU(3)$ projection}
\label{sec:proj_details}
%=================================================
In this appendix we describe the two $SU(3)$ projection schemes used in this paper
and show their equivalence in the weak coupling limit. Necessary properties of the projector mapping an arbitrary, complex $3\times 3$ matrix on the $SU(3)$ subgroup are idempotence and gauge-covariance. These properties are not sufficient to specify a unique projector and hence several choices exist. For APE smearing we use the unit circle projection~\cite{Kamleh:2004xk} which is based on polar decomposition, while for HYP smearing we seek iteratively the matrix $U_\text{max} \in SU(3)$ which maximizes $\text{Re} \, \text{Tr}( U_\text{max} V^\dagger )$~\cite{Bali:1992ab}.\medskip

First we describe the unit circle projection~\cite{Kamleh:2004xk}. For a complex $3\times 3$ matrix $V$ with $\det(V)\neq 0$, we calculate the matrix
\begin{equation}
W = V [V^\dagger V]^{-1/2} ,
\label{eq:PolarDecomposition1}
\end{equation}
which is unitary by construction and has a spectrum lying on the unit circle. The square root is obtained by Jacobi matrix diagonalization. From $W$ we obtain a special unitary matrix by computing
\begin{equation}
\overline{V} =  [(\det(W)]^{-1/3} W.
\label{eq:PolarDecomposition2}
\end{equation}
This projection is idempotent since an element of $SU(3)$ is projected by Eqs.~(\ref{eq:PolarDecomposition1}) and (\ref{eq:PolarDecomposition2}) back onto itself. The projection is also gauge covariant, as we now show. The matrices $V$ and $V^\dagger$ transform as 
\begin{align}
V \to G_L V G_R^\dagger \qquad \text{and}\qquad V^\dagger \to G_R V^\dagger G_L^\dagger,
\end{align}
and hence
\begin{align}
V^\dagger V \to G_R V^\dagger V G_R^\dagger.
\end{align}
Since $[V^\dagger V]^{-1/2}$ has the same transformation properties as $V^\dagger V$ one finds for the transformation of $W$
\begin{align}
W \to G_L W G_R^\dagger,
\end{align}
from which the gauge-covariance of $\overline{V}$ follows.\medskip

Next we describe the projection method used for HYP smearing.   This method
is based on seeking a matrix $U_\text{max}$ for which \cite{Bali:1992ab}
\begin{align}
\label{eq:MaxReTrDef}
U_\text{max} \in SU(3) \big| \text{Re}\,\text{Tr}(U_\text{max} V^\dagger )\,
\text{is maximal} \,.
\end{align}
The maximizing matrix $U_\text{max}$ is found iteratively by decomposing it
into its $SU(2)$ subgroups as outlined in Refs.~\cite{Bali:1992ab,
Bonnet:2001rc}.  Although, by the compactness of $SU(3)$, a global maximum
exists, it is not guaranteed that this iterative procedure converges to the
true maximum.  Furthermore, the matrix obtained by the iterative procedure depends on the details of the iteration
algorithm.  In practice, however, one finds  that $U_\textrm{max}$ is
numerically very close to $\overline V$, the $SU(3)$ projection of the
matrix $V$ obtained from the unit circle projection, \cite{Kamleh:2004xk},
and hence the iteratively found maximum is close to the true maximum.  Thus
we choose
\begin{align}
\tilde V = U_\text{max}
\end{align}
as the $SU(3)$ projection of the complex matrix $V$ .  This projection is
idempotent because for $V \in SU(3)$, unitarity and the triangle inequality
imply the bound $\text{Re Tr}(\tilde VV^{\dagger}) \leq 3$ which is uniquely
saturated by $\widetilde{V} = V$. By construction this projection is also
gauge-covariant. \medskip

Finally, we show that although the two projection methods are generally not equivalent they differ only at second order. For this discussion we assume that the iterative method converges to the global maximum and that the unprojected matrix $V$ is approximately unitary and unimodular (which is true in the weak coupling limit). 

In order to compare the two projection schemes, we first write the polar decomposition of $V$ as
\begin{equation}
V = e^{i\theta} \overline{V} H ,
\label{eq:PolarDecomp}
\end{equation}
where $\theta \in (-\pi/3,\pi/3]$, $H$ is a Hermitian matrix with eigenvalues $\lambda_1, \lambda_2, \lambda_3>0$, and $\overline{V}$ is the $SU(3)$-projected matrix given by Eq.~(\ref{eq:PolarDecomposition2}).  We then diagonalize $H = M \Lambda M^{\dagger}$ [$\Lambda =\text{diag}(\lambda_1, \lambda_2, \lambda_3)$ and $M \in SU(3)$] and substitute Eq.~(\ref{eq:PolarDecomp}) into Eq.~(\ref{eq:MaxReTrDef}). The resulting function to be maximized is 
\begin{equation}
\label{eq:diagonalmax} \text{ReTr}(e^{-i\theta}U'\Lambda) =
\text{Re}(e^{-i\theta}(\lambda_1 u'_{11} + \lambda_2 u'_{22} +
\lambda_3 u'_{33})),
\end{equation}
where $u'_{ij}$ is an element of $U' \equiv M^{\dagger}\overline{V}^{\dagger}U_\text{max} M \in SU(3)$. If the maximization is carried out over $U(3)$ rather than
$SU(3)$, we could choose $u'_{11} = u'_{22} = u'_{33} = e^{i\theta}$. This satisfies the bound $\text{Re Tr}(e^{-i\theta}U'\Lambda)$ derived from the triangle inequality. If we maximize over $SU(3)$ and $\bar V$ is already an element of $SU(3)$ then it follows $U' = \openone$ and there exists a unique maximum for $\theta = 0$ such that $\widetilde{V} = \overline{V}$.
%However, when maximizing over $SU(3)$,  the only solution is $\theta = 0$ from which follows that $U' = \openone$ is the unique maximum and thus $\widetilde{V} = \overline{V}$.

For the generic case maximizing over $SU(3)$, the constraint $\det(U') = 1$ forces a maximum at $U' \neq \openone$, and hence $\widetilde{V} \neq \overline{V}$. Nevertheless, when $V$ is approximately unitary and unimodular we can write $\theta = \delta \theta$ with $|\delta \theta| \ll 1$. Then $\lambda_k = 1 + \delta \lambda_k$ and $|\delta \lambda_k| \ll 1$. From that follows $u'_{kk} = 1 + i\delta_k$ with $|\delta_k| \ll 1 ~~(k = 1, 2, 3)$, and we can work to first order in all ``small'' quantities.  Although $U'$ can in principle
develop small, off-diagonal elements, the constraint $U' \in SU(3)$ always implies that $\delta_k \in \mathbb{R}$ and $\delta_1 + \delta_2 + \delta_3 = 0$.  Therefore we must maximize
\begin{equation}
\sum_{k = 1}^{3} \text{Re}\left(1 + \delta \lambda_k + i \delta_k -
i \delta \theta\right),
\end{equation}
subject to the constraint $\delta_1 + \delta_2 + \delta_3 = 0$.  Because this is independent of $\delta \theta$, this is equivalent to the case $\theta = 0 $  discussed above. Consequently, to first order $\widetilde{V} = \overline{V}$. 
This explains why the two projection schemes lead to numerically close results as reported in \cite{Kamleh:2004xk}.

%=================================================
\section{Continuum QCD to HQET matching and HQET running}
\label{app:cont_match}
%=================================================

In Eqs.~(\ref{eq:C_A}) and~(\ref{eq:ZDB2}) we separated the matching
coefficients of the operators into three contributions:  the QCD to HQET matching factor at scale $m_b$, the HQET running from $m_b$ to $m_c$, and the HQET running from $m_c$ to $\mu$. Here we collect the results for these factors and present them for general numbers of flavors $N_f$ and colors $N_c$. The Casimir factors that appear in the expressions are
\begin{align}
C_F =  \frac{N_c^2 -1}{2N_c}, \qquad C_A = N_c, \qquad \text{and}\qquad T_F = \frac{1}{2}.
\end{align}

The one-loop QCD to HQET matching factor for the axial current operator renormalized in the $\overline{\mbox{MS}}$ scheme with naive dimensional regularization is calculated in Refs.~\cite{Eichten:1989zv, PT_Oa} 
\begin{equation}
\tilde C_A(m_b) = 1- 2 C_F \frac{\alpha_s(m_b)}{4\pi} + \CO(\alpha_s^2),
\end{equation}
while the renormalization group (RG) evolution factors entering Eq.~(\ref{eq:C_A}) are given by \cite{Ji:1991pr,Broadhurst:1991fz} 
\begin{equation}
U_A^{(N_f)}(\mu',\mu) =
 \left(\frac{\alpha_s^{(N_f)}(\mu')}{\alpha_s^{(N_f)}(\mu)}\right)^{d_A^{(N_f)}}
 \left(1+J_A^{(N_f)}\frac{\alpha_s^{(N_f)}(\mu')-\alpha_s^{(N_f)}(\mu)}{4\pi}\right) + \CO(\alpha_s^2) .
\end{equation}
The coefficients in the evolution factor are
\begin{eqnarray}
d_A^{(N_f)} &=& \gamma_A^{(0)}/2\beta_0 , \\
J_A^{(N_f)} &=& \gamma_A^{(1)}/2\beta_0-\gamma_A^{(0)}\beta_1/2\beta_0^2 ,
\end{eqnarray}
with
\begin{align}
\beta_0  &= \frac{11}{3}C_A-\frac{4}{3}T_F N_f,  \\
\beta_1  &=  \frac{34}{3} C_A^2-4C_F T_F N_f-\frac{20}{3}C_AT_FN_f, 
\end{align}
and the one- and two-loop anomalous dimensions
\begin{eqnarray}
\gamma_A^{(0)} & = & -3 C_F,\\
\gamma_A^{(1)} & = & -16C_F \left(\frac{49}{96}C_A - \frac{5}{32}C_F -  \frac{5}{24} T_F N_f - ( \frac{1}{4}C_A - C_F) \frac{\pi^2}{6}  \right).
\end{eqnarray}

The one-loop QCD to HQET matching factors for the $\Delta B=2$ four quark operator are calculated in Refs.~\cite{Flynn:1990qz, Buchalla:1996ys}:
\begin{eqnarray}
\tilde Z_1(m_b,m_b) & = & 1 -\frac{8 N_c^2 + 9 N_c - 15}{2 N_c}\, \frac{\alpha_s(m_b)}{4\pi} + \CO(\alpha_s^2) ,\\
\tilde Z_2(m_b,m_b) & = &    - 2 (N_c + 1)\, \frac{\alpha_s(m_b)}{4\pi} + \CO(\alpha_s^2) . \label{eq:Z2_mb_mb}
\end{eqnarray}
The operators $O_L$ and $O_S$ mix under RG evolution (see Eq.~(\ref{eq:ZDB2})), such that the evolution factors for the matching coefficients can be written as a $2 \times 2$ matrix:
\begin{equation}
U_L^{(N_f)} = \left( \begin{array}{cc}
		U_{11}^{(N_f)} & 0\\
		U_{21}^{(N_f)} & U_{22}^{(N_f)} 
\end{array} \right) . \label{eq:U_B2}
\end{equation}
Note that the matrix element $U_{12}^{(N_f)}$ is zero to all-orders in perturbation theory due to the fact that heavy-quark spin symmetry prohibits mixing from $O_S$ into $O_L$.  Because $\tilde Z_2$ in Eq.~(\ref{eq:Z2_mb_mb}) has no tree-level contribution, the two-loop expression for $U_{11}^{(N_f)}$ and one-loop expressions for $U_{21}^{(N_f)}$ and $U_{22}^{(N_f)}$
are sufficient to determine the $\CO(\alpha_s)$ matching coefficients. 
The matrix elements in Eq.~(\ref{eq:U_B2}) are given in  Refs.~\cite{Gimenez:1992is,Ciuchini:1996sr,Buchalla:1996ys}:
\begin{eqnarray}
U_{11}^{(N_f)}(\mu',\mu) & = &
 \left(\frac{\alpha_s^{(N_f)}(\mu')}{\alpha_s^{(N_f)}(\mu)}\right)^{d_1^{(N_f)}}
 \left(1+J_{11}^{(N_f)}\frac{\alpha_s^{(N_f)}(\mu')-\alpha_s^{(N_f)}(\mu)}
  {4\pi}\right) + \CO(\alpha_s^2) ,\\
U_{21}^{(N_f)}(\mu',\mu) & = &- \frac{1}{4}\left(
 \left(\frac{\alpha_s^{(N_f)}(\mu')}{\alpha_s^{(N_f)}(\mu)}\right)^{d_1^{(N_f)}}
 -\left(\frac{\alpha_s^{(N_f)}(\mu')}{\alpha_s^{(N_f)}(\mu)}\right)^{d_2^{(N_f)}}
				   \right) + \CO(\alpha_s),\\
U_{22}^{(N_f)}(\mu',\mu) & = & 
 \left(\frac{\alpha_s^{(N_f)}(\mu')}{\alpha_s^{(N_f)}(\mu)}\right)^{d_2^{(N_f)}} + \CO(\alpha_s),
\end{eqnarray}
with the coefficients
\begin{eqnarray}
	d_i^{(N_f)} &=& \gamma_{ii}^{(0)}/2\beta_0 , \\
	J_{11}^{(N_f)} &=& \gamma_{11}^{(1)}/2\beta_0-\gamma_{11}^{(0)}\beta_1/2\beta_0^2 ,
\end{eqnarray}
and the anomalous dimensions 
\begin{eqnarray}
\gamma^{(0)} &=& \left( \begin{array}{ccc}
		-6 C_F && 0\\
		1+ 1/N_c && -6 C_F + 4 (1+ 1/N_c) 
\end{array} \right) ,  \\
\gamma_{11}^{(1)} &=& -\frac{N_c-1}{12N_c} \Big( 127N_c^2 + 143N_c + 63 - 57/N_c \nonumber \\ \newline && \qquad+ 8\pi^2 (N_c^2 - 2N_c + 4/N_c) - N_f (28N_c + 44) \Big). 
\end{eqnarray}
%

%=================================================
\section{Finding matrix elements from Green's functions}
\label{app:Norman}
%=================================================

In this appendix we provide a derivation of Eqs.~(\ref{eq:PhiUnRen}) and (\ref{eq:box_Mq}) which 
give the relationship between the Euclidean space correlation 
functions which are the direct results of our simulations and the 
matrix elements between normalized Hilbert space states which we wish 
to determine.  In this appendix, all quantities are written in
lattice units.  The derivation of Eq.~(\ref{eq:PhiUnRen}) follows the standard steps 
of inserting a complete set of energy eigenstates into the Green's 
functions ${\cal C}^{LW}(t,t_0)$ and ${\cal C}^{WW}(t,t_0)$ which 
are defined in Eqs.~(\ref{eq:two_lw}) and (\ref{eq:two_ww}) and appear in the numerator and 
denominator of Eq.~(\ref{eq:PhiUnRen}):
\begin{eqnarray}
{\cal C}^{LW}(t,t_0) &=& \sum_n \sum_{\vec x \in V}\langle 0|A_0^L(\vec x,0)|n\rangle
                                \langle n|A_0^W(0)^\dagger|0\rangle e^{-E_n(t-t_0)}
\label{eq:p0_numer} \\
{\cal C}^{WW}(t,t_0) &=& \sum_n |\langle n|A_0^W(0)^\dagger|0\rangle|^2  e^{-E_n(t-t_0)},
\label{eq:p0_denom}
\end{eqnarray}
where $E_n$ is the energy of the state $|n\rangle$, a state with unit 
normalization.  Because of the translational invariance of either 
the wall source operator $A_0^W$ or the sum of the local operator 
$A_0^L$ over a temporal hyperplane, the intermediate state $|n\rangle$
must have zero momentum.  There is then a single, finite-volume state, 
$|B_q(\vec p = 0)_V\rangle$, separated from the other excited states by
a non-zero energy gap, which will dominate as the time separation
$t-t_0$ becomes large.  Thus, in the limit of large $t-t_0$, 
Eqs.~(\ref{eq:p0_numer}) and (\ref{eq:p0_denom}) become:
\begin{eqnarray}
{\cal C}^{LW}(t,t_0) 
    &=& L^3 \langle 0|A_0^L(0)|B_q(\vec p = 0)_V\rangle
    \langle B_q(\vec p = 0)_V|A_0^W(0)^\dagger|0\rangle e^{-m_{B_q}^*(t-t_0)}
\label{eq:p0_numer_1} \\
{\cal C}^{WW}(t,t_0) 
    &=& |\langle B_q(\vec p = 0)_V|A_0^W(0)^\dagger|0\rangle|^2  e^{-m_{B_q}^*(t-t_0)},
\label{eq:p0_denom_1}
\end{eqnarray}
where we have used translational symmetry to replace the sum over
$\vec x$ in Eq.~(\ref{eq:p0_numer}) with the factor $L^3$.  

Finally, Eq.~(\ref{eq:PhiUnRen}) follows by taking the ratio of Eq.~(\ref{eq:p0_numer_1}) 
and the square root of Eq.~(\ref{eq:p0_denom_1}) to remove the unwanted
factor $\langle 0|A_0^L(0)|B_q(\vec p = 0)\rangle$ and recognizing that 
in the limit of large volume the unit-normalized state 
$|B_q(\vec p = 0)_V\rangle$ in Eq.~(\ref{eq:p0_numer_1}) and the
covariant, delta-function normalized state $|B_q(\vec p = 0)\rangle$ 
in Eq.~(\ref{eq:fB_def}) are related by:
\begin{equation}
\sqrt{2m_{B_q}L^3}|B_q(\vec p = 0)_V\rangle 
              \rightarrow |B_q(\vec p = 0)\rangle.
\label{eq:finite_inf_volume}
\end{equation}
This discussion is standard and has been repeated here to provide a 
familiar background for the derivation of Eq.~(\ref{eq:box_Mq}) where a new approach, 
special to the static approximation, is required.  The complication in 
Eq.~(\ref{eq:box_Mq}) arises because of the use of a box source which is not 
translationally invariant and which produces a superposition of $B$-meson 
states with various momenta.  In the static limit, all of these states 
are degenerate since their energy no longer depends on their momenta.  
Thus, we cannot assume that the large time limit, $t_f \gg t \gg t_0$ 
will project onto a unique ground state.  However, as is worked out below,
we can use an additional symmetry of the static approximation to show
that the normalization of the box source cancels between the numerator 
and denominator of Eq.~(\ref{eq:box_Mq})~\cite{Christ:2007cn}.

Fortunately, this large set of degenerate states resulting from the 
momentum independence of the energy of the heavy-light meson, can
be distinguished by a new conservation law which becomes exact in the
static limit: the local conservation of heavy-quark number.  The
absence of spatial derivatives in heavy-quark actions shown in Eqs.~(\ref{eq:hqa})
and (\ref{eq:hqa_fat}) implies that the total number of heavy quarks at each spatial 
site is separately conserved.  This conservation law results from the 
invariance of the heavy-quark action under the phase rotation of the 
heavy-quark Grassmann variables:
\begin{eqnarray}
h(\vec x) &\rightarrow& e^{i\theta(\vec x)}h(\vec x) \\
\overline{h}(\vec x) &\rightarrow& e^{-i\theta(\vec x)}\overline{h}(\vec x)
\end{eqnarray} 
in which a different phase $\theta(\vec x)$ can be used for each spatial site 
$\vec x$.  In the quantum mechanical Hilbert space this symmetry corresponds to a
family of operators $N_h(\vec x) =  \overline{h}(x)\gamma^0 h(\vec x)$  which commute 
with the Hamiltonian (continuum theory) or the transfer matrix (lattice theory).  
Thus, when interpreting our Green's functions we can introduce a complete 
set of $B$-meson states $|\widetilde{B}_q(\vec x)\rangle$ which are 
eigenstates of both the Hamiltonian and the number operators 
$N_h(\vec x\,^\prime)$ with a standard normalization:
\begin{eqnarray}
N_h(\vec x\,^\prime)|\widetilde{B}_q(\vec x)\rangle
           &=& -\delta_{\vec x, \vec x\,^\prime}|\widetilde{B}_q(\vec x)\rangle \\
\langle \widetilde{B}_q(\vec x) |\widetilde{B}_q(\vec x\,^\prime)\rangle
           &=& \delta_{\vec x, \vec x\,^\prime}.
\end{eqnarray}
Here the Kronecker delta, $\delta_{\vec x, \vec x\,^\prime}$, used above 
is appropriate for the discrete positions that appear in the lattice 
theory.  The energy eigenstate $|\widetilde{B}_q(\vec x)\rangle$ can be 
thought of as composed of a static quark fixed to the position $\vec x$ 
together with a light-quark bound state centered at $\vec x$.  The tilde 
notation is chosen to suggest the relation of these new states to our 
earlier energy and momentum eigenstates, $|B(\vec p)_V\rangle$, which 
are standard superpositions of these $N_h(\vec x)$ eigenstates:
\begin{equation}
|B_q(\vec p)_V\rangle = \sum_{x \in V} \frac{e^{-i\vec p \cdot \vec x}}{L^{3/2}}
                           |\widetilde{B}_q(\vec x)\rangle.
\end{equation}

Using these $N_h(\vec x)$ eigenstates and the conservation of $N_h(\vec x\,^\prime)$ 
for all $\vec x\,^\prime \in V$ it is now straightforward to derive Eq.~(\ref{eq:box_Mq}).  Substituting
a sum over the complete set of degenerate ground states $|\widetilde{B}_q(\vec x)\rangle$
into the correlation function $C^{BB}(t,t_0)$ which appears in the 
denominator of Eq.~(\ref{eq:box_Mq}) and is defined in Eq.~(\ref{eq:two_bb}) we find:
\begin{eqnarray}
C^{BB}(t,t_0) &=& \sum_{\vec x \in \Delta V} \Biggl\{
    \langle 0 |
\overline h(\vec x)
      \gamma_0\gamma_5 
\Bigl(\sum_{\vec y \in \Delta V}  q(\vec y) \Bigr)
|\widetilde{B}_q(\vec x)\rangle
    \label{eq:box-box_corr} \\
    && \hskip 1.0in \cdot
    \langle \widetilde{B}_q(\vec x)|
\Bigl(\sum_{\vec y\,^\prime \in \Delta V} \overline q(\vec y\,^\prime)\Bigr) 
 \gamma_0\gamma_5 h(\vec x)
|0\rangle
    \Biggr\}e^{-m_{B^*}(t-t_0)} ,
    \nonumber
\end{eqnarray}
where we have used the conservation of local heavy-quark number to keep
only terms where all of the heavy quarks are located at the same spatial
position $\vec x$.  

Similarly, we can evaluate $C^B_{O_i}(t_f,t,t_0)$ which appears in the numerator 
of Eq.~(\ref{eq:box_Mq}) and is defined in Eq.~(\ref{eq:three_bb}):
\begin{eqnarray}
C^B_{O_i}(t_f,t,t_0) &=& \sum_{\vec x \in \Delta V} \Biggl\{
    \langle 0 |
\Bigl(\sum_{\vec y \in \Delta V} \overline q(\vec y,t)\Bigr)
\gamma_0\gamma_5 
h(\vec x, t)
    |\widetilde{\overline B}_q(\vec x)\rangle
    \label{eq:box-box_3pt} \\
    && \hskip 0.25in \cdot \langle \widetilde{\overline B}_q(\vec x)|
                    O_i(\vec x)|\widetilde{B}_q(\vec x)\rangle\cdot
    \langle \widetilde{B}_q(\vec x)|
\Bigl(\sum_{\vec y\,^\prime \in \Delta V} \overline q(\vec y\,^\prime,t)\Bigr)
\gamma_0\gamma_5 h(\vec x, t)    
 |0\rangle
    \Biggr\}e^{-m_{B^*}(t_f-t_0)} \nonumber \\
               &=& C^{BB}(t_f,t_0)\langle \widetilde{\overline B}_q(\vec 0)|
                  O_i(\vec 0)|\widetilde{B}_q(\vec 0)\rangle,
\end{eqnarray}
where translational invariance of the matrix element 
$ \langle \widetilde{\overline B}_q(\vec x)|
                    O_i(\vec x)|\widetilde{B}_q(\vec x)\rangle$
has been used to remove
it from the sum over $\vec x$ causing that sum to assume the same form 
which appears in the box-box correlator $C^{BB}(t,t_0)$ given in 
Eq.~(\ref{eq:box-box_corr}), after an application of charge conjugation 
symmetry.  Equation (\ref{eq:box_Mq}) is then easily recognized from ratio of 
Eqs.~(\ref{eq:box-box_3pt}) and (\ref{eq:box-box_corr}):
\begin{eqnarray}
M_{O_i} &=& \frac{1}{m_{B_q}}\langle \overline B_q(\vec p = 0)|
                  O_i(\vec 0)|B_q(\vec p = 0)\rangle \label{eq:M_p} \\
         &=& 2L^3 \langle \overline B_q(\vec p = 0)_V|
                            O_i(\vec 0)|B_q(\vec p = 0)_V\rangle \\
         &=& 2 \langle \widetilde{\overline B}_q(\vec 0)|
                            O_i(\vec 0)|\widetilde{B}_q(\vec 0)\rangle \\
         &=& 2 \frac{C^B_{O_i}(t_f,t,t_0)}{C^{BB}(t_f,t_0)} \\
         &=& 2 \frac{C^B_{O_i}(t_f,t,t_0)e^{m_{B_q}^*(t_f-t_0)/2}}
                  {\sqrt{C^{BB}(t_f,t)C^{BB}(t,t_0)}} 
\end{eqnarray}
where Eq.~(\ref{eq:finite_inf_volume}), relating our two normalization
conventions for momentum eigenstates, has been used to obtain the second 
equation.

This use of localized sources while making the static approximation may 
become more important as larger spatial volumes are used and the overlap
between the translationally invariant wall sources and the physical states
of interest becomes smaller.  Of course, this same method can be used for
localized sources with different spatial distributions such as Gaussian or
atomic wave functions.

%=================================================
\section{Chiral perturbation theory for decay constants and mixing parameters}
\label{app:ChPT}
%=================================================

In this section we present the NLO HM$\chi$PT expressions needed to extrapolate $N_f = 2+1$ domain-wall lattice data for heavy-light meson decay constants and mixing parameters to the physical quark masses and the continuum.  Although we label the formulae with the subscript ``$B$", we note that these functions can also be used to extrapolate $D$-meson decay constants and mixing matrix elements with the caveat that the low-energy constants (except for the light-light meson tree-level parameters $f$ and $B$) are different for the case of $B$-mesons and $D$-mesons. 
We first show the $SU(3)$ HM$\chi$PT formulae in Sec.~\ref{app:SU3ChPT}; we then take the appropriate limits of the $SU(3)$ expressions to obtain those in $SU(2)$ HM$\chi$PT in Sec.~\ref{app:SU2ChPT}.  %Throughout this section we follow the notation of Ref.~\cite{Laiho:2005ue}.

%=================================================
\subsection{$SU(3)$ HM$\chi$PT expressions}
\label{app:SU3ChPT}
%=================================================

The tree-level mass-squared of a meson composed of two domain-wall valence quarks with flavors $x$ and $y$ is
\beq
	m^2_{xy} = \textcolor{black}{B} ( m_x + m_y+ 2 m_\textrm{res} ) , 
\eeq
where $\textcolor{black}{B}$ is a continuum low-energy constant and $m_\textrm{res}$ is the residual quark mass.  

The NLO result for $\Phi_{B_x} = f_{B_x} \sqrt{m_{B_x}}$ in the partially-quenched domain-wall theory with 2+1 flavors of sea quarks is~\cite{Arndt:2004bg,Aubin:2005aq}:
\begin{align}\label{eq:fB_PQChPT}
\Phi_{B_x} & =  \phi_0 \Bigg\{ 1 - \frac{1}{16\pi^2f^2} \frac{1+3g_{B^*B\pi}^2}{2} \sum_{f=l,l,h}  \ell(m_{xf}^2) \nonumber\\
 + &\,  \frac{1}{16\pi^2f^2} \frac{1+3g_{B^*B\pi}^2}{6} \bigg[ R^{[2,2]}_X (\{  M_X \};\{ \mu \}) \tilde{\ell} (m^2_X) -\!\!\! \sum_{j \in \{ M_X\} }  \frac{\partial}{\partial m^2_{X}} \left( R^{[2,2]}_{j} (\{ M_X \}; \{ \mu \}) \right) \ell (m^2_j) \bigg]  \nonumber \\*+&\,
       c_\textrm{sea} (2m_l + m_h) + c_\textrm{val} m_x + c_a a^2 \Bigg\}\,, 
\end{align}
where $f \approx 130.4$~MeV is the tree-level pion decay constant.  The NLO expression for $M_{B_x} = 8/3~m_{B_x}f^2_{B_x}B_{B_x}$ is similar~\cite{Arndt:2004bg,Detmold:2006gh}:
\bea
	M_{B_x} & = & \beta_0 \Bigg\{ 1 - \frac{1 + 3 g_{B^*B\pi}^2}{16 \pi^2 f^2} \sum_{f=l,l,h} \ell(m^2_{xf}) - \frac{1 - 3 g_{B^*B\pi}^2}{16 \pi^2 f^2} \ \ell(m^2_X) \nonumber\\ 
	& + &  \frac{1}{24 \pi^2 f^2} \bigg[ R^{[2,2]}_X (\{  M_X \};\{ \mu \}) \tilde{\ell} (m^2_X) - \sum_{j \in \{ M_X\} }  \frac{\partial}{\partial m^2_{X}} \left( R^{[2,2]}_{j} (\{ M_X \}; \{ \mu \}) \right) \ell (m^2_j) \bigg] \nonumber\\ 
	& +  & d_\textrm{sea} (2 m_l + m_h) + d_\textrm{val} m_x + d_a a^2 \Bigg\} \ .
\label{eq:MB_PQChPT}
\eea 
In both the decay constant and the mixing matrix element, the only effect of the nonzero lattice spacing is a new analytic term proportional to $a^2$.  These results agree with the continuum calculation of Sharpe and Zhang in the limit of three degenerate sea quarks~\cite{Sharpe:1995qp}.  We note that one is free to multiply the above expressions by arbitrary powers of the heavy-light meson mass $m_{B_x}$ without modifying the chiral logarithms.  This is because the chiral logarithm contributions to heavy-light meson masses are suppressed by $1/m_b$ and therefore of higher-order than we consider~\cite{Jenkins:1992hx}.

In the above expressions for $\Phi_{B_x}$ and $M_{B_x}$, the functions $\ell$ and $\tilde{\ell}$ are one-loop chiral logarithms:
\bea\label{eq:logs}
	\ell(m^2) & \equiv & m^2 \ln \left( \frac{m^2}{\Lambda_\chi^2} \right) \,, \\
	\tilde{\ell}(m^2) & \equiv & - \ln \left( \frac{m^2}{\Lambda_\chi^2} \right) - 1 \ .
\eea
The function $R^{[n,k]}_j$ is due to single poles in the partially-quenched propagator:
\begin{eqnarray} \label{eq:residues}
	R^{[n,k]}_j(\{m\},\{\mu\})  &\equiv & \frac{\prod_{a=1}^k (\mu^2_a-m^2_j)}{\prod_{i=1,i\neq j}^n (m^2_i-m^2_j)} \,
%	D^{[n,k]}_{j, l}(\{m\},\{\mu\})  &\equiv & -\frac{d}{dm^2_l}R^{[n,k]}_j(\{m\},\{\mu\}) \ ,
\end{eqnarray}
and the sets of flavor-singlet masses that appear in the residue functions are
\bea\label{eq:mass_sets}
	\{ \mu \} & = & \{m^2_L,m^2_H\}\ ,\\
 	\{ M_X \} & = & \{m_X^2, m_{\eta}^2\}\ ,
\eea
where $m^2_X$ is the mass-squared of a meson composed of two $x$ valence quarks, 
$m^2_{L(H)}$ is the mass-squared of a meson composed of two $l(h)$ sea quarks, and $m^2_\eta = (m^2_L + 2m^2_H)/3$ for 2+1 flavors of sea quarks.

For completeness, we also include the expressions for the decay constant and mixing matrix element at the unitary points.  In the full QCD and isospin limits, the above expression for $\Phi_x$ becomes
\begin{eqnarray}
\label{eq:PhiD}
\Phi_{B_l} & = & \phi_0 \Bigg\{ 1 - \frac{1}{16\pi^2f^2}
 	\left(\frac{1+3g_{B^*B\pi}^2}{2}\right)
 	\bigg[ \frac{3}{2}\ell(m_{\pi}^2) + \ell(m_{K}^2) + \frac{1}{6}\ell(m^2_{\eta}) \bigg] \nonumber \\*&&{}+
       c_\textrm{sea} (2m_l + m_h) + c_\textrm{val} m_l + c_a a^2 \Bigg\}  \ , \label{eq:fBd_ChPT}   \\
\label{eq:PhiS}
\Phi_{B_h} & = & \phi_0 \Bigg\{ 1 - \frac{1}{16\pi^2f^2} \left(\frac{1+3g_{B^*B\pi}^2}{2}\right) \bigg[  2\ell(m_{K}^2) + \frac{2}{3}\ell(m^2_{\eta}) \bigg]   \nonumber \\*&&{}+
       	c_\textrm{sea} (2m_l + m_h) + c_\textrm{val} m_h + c_a a^2 \Bigg\} \ , \label{eq:fBs_ChPT}
\end{eqnarray}
where $m_\pi^2 = m_L^2$ and $m_K^2 = \left(m_L^2+ m_H^2\right)/2$. Similarly, the expression for $M_x$ becomes
\bea
\label{eq:MD}
	M_{B_l} & = & \beta_0 \Bigg\{ 1 - \frac{1 + 3 g_{B^*B\pi}^2}{16 \pi^2 f^2} \left[ 2\ell(m_\pi^2) + \ell(m_K^2)  \right] - \frac{1 - 3 g_{B^*B\pi}^2}{16 \pi^2 f^2} \ \ell(m^2_\pi) \nonumber\\ &&{}+
\frac{1}{48 \pi^2 f^2} \left[ 3\ell(m_\pi^2) - \ell(m_\eta^2) \right] + d_\textrm{sea} (2 m_l + m_h) + d_\textrm{val} m_l + d_a a^2 \Bigg\} \ , \\
\label{eq:MS}
	M_{B_h} & = & \beta_0 \Bigg\{ 1 - \frac{1 + 3 g_{B^*B\pi}^2}{16 \pi^2 f^2} \left[2\ell(m_K^2) + \ell(m_S^2)  \right] - \frac{1 - 3 g_{B^*B\pi}^2}{16 \pi^2 f^2} \ \ell(m^2_S) \nonumber\\ &&{}+
 \frac{1}{24 \pi^2 f^2} \left[ 3 \ell(m_S^2) -2 \ell(m_\eta^2) \right] + d_\textrm{sea} (2 m_l + m_h) + d_\textrm{val} m_h + d_a a^2 \Bigg\} \ .
\eea
The expressions for $\Phi_l$ and $\Phi_h$ agree with those derived by Goity in Ref.~\cite{Goity:1992tp}.

One can account for lattice finite volume effects at NLO in $\chi$PT by turning the one-loop integrals to sums.  This yields an additive correction to the chiral logarithms \cite{Bernard:2001yj}:
\bea\label{eq:FV_logs}
\ell(m^2)=m^2\left(\ln \frac{m^2}{\Lambda^2_{\chi}} +
\delta^{FV}_1(m\textrm{L})\right), \ \ \ \
\delta^{FV}_1(m\textrm{L})=\frac{4}{m\textrm{L}}\sum_{\vec{r} \neq
0} \frac{K_1(|\vec{r}|m\textrm{L})}{|\vec{r}|}, \\
\label{eq:FV_logs_2}
\widetilde{\ell}(m^2)=-\left(\ln \frac{m^2}{\Lambda^2_{\chi}}+1
\right) + \delta^{FV}_3(m\textrm{L}), \ \ \ \
\delta^{FV}_3(m\textrm{L})=2\sum_{\vec{r} \neq
0}K_0(|\vec{r}|m\textrm{L}),
\eea
where $\delta_i^{FV}(m\textrm{L})$ is the finite volume correction to the infinite volume result and $K_0$ and $K_1$ are modified Bessel functions of imaginary argument.

%=================================================
\subsection{$SU(2)$ HM$\chi$PT expressions}
\label{app:SU2ChPT}
%=================================================

The NLO $SU(2)$ HM$\chi$PT expressions can easily be obtained from the $SU(3)$ results in the previous subsection by integrating out the strange valence and sea quarks.  After this procedure, however, the expressions for the decay constant and mixing matrix element differ for $B_d$-type mesons and $B_s$-type mesons.    

First we consider $B_d$-type mesons composed of a $b$-quark and a light valence quark with mass $m_x$.  In this case, we take the limits of Eqs.~(\ref{eq:fB_PQChPT}) and~(\ref{eq:MB_PQChPT}) assuming   
\begin{equation}
	\frac{m_x}{m_h}, \frac{m_l}{m_h} \ll 1 .
\end{equation}
The resulting expression for $\Phi_{B_x}$ at NLO in the partially-quenched domain-wall theory with two degenerate light sea quarks is:
\begin{eqnarray}
\label{eq:fBx_SU2_ChPT}
\Phi_{B_x} & =  & \phi_{0}^{(2)} \Bigg\{ 1 - \frac{1+3\big(g_{B^*B\pi}^{(2)}\big)^2 }{\left(4\pi f^{(2)}\right)^2} \  \ell(m_{xl}^2) +  \frac{1+3\big(g_{B^*B\pi}^{(2)}\big)^2 }{\left(4\pi f^{(2)}\right)^2} \bigg( \frac{1}{4} \bigg) \bigg[   (m_L^2 - m_X^2) \tilde\ell(m_X^2) +  \ell(m_X^2) \bigg]  \nonumber\\
	&+& c_\textrm{sea}^{(2)} m_l + c_\textrm{val}^{(2)} m_x + c_a^{(2)} a^2 \Bigg\} \,, \quad
\end{eqnarray}       
where we use the superscript ``(2)" to distinguish the $SU(2)$ low-energy constants from their $SU(3)$ analogs in the previous section.  The NLO expression for $M_{B_x}$ is similar:
\begin{eqnarray}       
	\label{eq:MBx_SU2_ChPT}
	M_{B_x} & = & \beta_0^{(2)} \Bigg\{ 1 -  \frac{1 + 3 \big(g_{B^*B\pi}^{(2)}\big)^2 }{\left(4\pi f^{(2)}\right)^2} \ 2 \ell(m^2_{xl}) - \frac{1 - 3 \big(g_{B^*B\pi}^{(2)}\big)^2 }{\left(4\pi f^{(2)}\right)^2} \ \ell(m^2_X) \nonumber\\ 
	& + &  \frac{1}{\left(4\pi f^{(2)}\right)^2} \bigg[ (m_L^2 - m_X^2) \tilde\ell(m_X^2) + \ell(m_X^2) \bigg] + d_\textrm{sea}^{(2)} m_l + d_\textrm{val}^{(2)} m_x + d_a^{(2)} a^2 \Bigg\} \ .
\end{eqnarray}
These results agree with the continuum partially-quenched calculation of Sharpe and Zhang in the limit $a \to 0$~\cite{Sharpe:1995qp}.  The unitary QCD expressions can easily be obtained by the replacement $m_x \to m_l$.  In the $SU(2)$ theory, the effects of the dynamical strange quark are fully contained in the values of the low-energy constants, e.g. $\phi_0^{(2)}(m_h)$.  For simplicity of notation, however, we suppress the functional dependence of the coefficients on $m_h$.  

Next we consider $B_s$-type mesons composed of a $b$-quark and a heavy valence quark with mass $m_y$.  Because the $SU(2)$ chiral effective Lagrangian includes only two light quark flavors, this requires an extension of $SU(2)$ $\chi$PT to the kaon sector, and the resulting theory is sometimes called Kaon $SU(2)$ Chiral Perturbation Theory (K$\chi$PT)~\cite{Roessl:1999iu,Allton:2008pn}.  In this case, we take the limits of Eqs.~(\ref{eq:fB_PQChPT}) and~(\ref{eq:MB_PQChPT}) assuming   
\begin{equation}
	\frac{m_l}{m_y}, \frac{m_l}{m_h}  \ll 1 .
\end{equation}
The resulting NLO expressions for $\Phi_{B_y}$ and $M_{B_y}$ are:
\begin{eqnarray}
	\Phi_{B_y} & =  & \phi_{0}^{(s)} \Bigg\{ 1 + {c}_\textrm{sea}^{(s)} m_l +  {c}_a^{(s)} a^2 \Bigg\}\,,  \label{eq:fBs_SU2_ChPT} \\
	M_{B_y} & = &  \beta_0^{(s)} \Bigg\{ 1 +  {d}_\textrm{sea}^{(s)} m_l +  {d}_a^{(s)} a^2 \Bigg\} \ , \label{eq:MBs_SU2_ChPT}
\end{eqnarray}
where we use the superscript ``(s)"  to distinguish the coefficients from those in Eqs.~(\ref{eq:fBx_SU2_ChPT}) and~(\ref{eq:MBx_SU2_ChPT}).  Because the valence quark has been integrated out, there are no longer any chiral logarithms.  The effects of both the valence and sea strange quarks are encapsulated in the values of the low-energy constants, e.g. $\phi_0^{(s)}(m_y,m_h)$.

%=================================================
%The bibliography
%=================================================

\bibliography{B_meson}

\end{document}